%% file: ms.tex
\documentclass[12pt,preprint]{aastex}

\usepackage{natbib}
\newcommand{\msini}{\ensuremath{M \sin{i}}}
\newcommand{\feh}{\ensuremath{[\mbox{Fe}/\mbox{H}]}}
\newcommand{\teff}{\ensuremath{T_{\mbox{\scriptsize eff}}}}
\newcommand{\persec}{\ensuremath{\mbox{s}^{-1}}}
\newcommand{\mps}{\mbox{m} \persec}
\newcommand{\mjup}{\ensuremath{\mbox{M}_{\mbox{Jup}}}}
\newcommand{\mearth}{\ensuremath{\mbox{M}_{\earth}}}
\newcommand{\Mjupsmall}{\mbox{\scriptsize M}_{\mbox{\tiny Jup}}}

\def\astrosun {\mbox{$\odot$}}
\newcommand{\Msol}{\ensuremath{\mbox{M}_{\astrosun}}}
\newcommand{\multfrac}{14\%} 
\newcommand{\trendfrac}{14\%} 
\newcommand{\totfrac}{28\%} 
\newcommand{\multsystems}{28} 
\newcommand{\totmult}{67} 
\newcommand{\alltrend}{36}
\newcommand{\tottrend}{28} 
\newcommand{\allsys}{205} 
\newcommand{\singleone}{88} 
\newcommand{\singleseven}{38} 
\newcommand{\multone}{39} 
\newcommand{\closefrac}{43} 

\begin{document}
\pagestyle{empty}
\title{Ten New and Updated multiplanet Systems, and a Survey of
  Exoplanetary Systems} 
\author{J. T. Wright} \affil{Department of
  Astronomy, 226 Space Sciences Building, Cornell University, Ithaca,
  NY 14853\\jtwright@astro.cornell.edu} 
\author{S. Upadhyay,
  G. W. Marcy} \affil{Department of Astronomy, 601 Campbell Hall,
  University of California, Berkeley, CA
  94720-3411}
\author{D. A. Fischer} \affil{Department of Physics and Astronomy, San
  Francisco State University, San Francisco, CA
  94132}
\author{Eric B. Ford} \affil{Department of Astronomy, University of
  Florida, 211 Bryant Space Science Center, P.O. Box 112055,
  Gainesville, FL 32611-2055}
\and
\author{John Asher Johnson} \affil{Institute for Astronomy, University of
  Hawai'i, Honolulu, HI 96822 \\NSF Postdoctoral Fellow}
\begin{abstract}
We present the latest velocities for ten multiplanet systems, including a re-analysis of archival Keck and Lick 
data, resulting in improved velocities that supersede our previously published measurements. We derive updated 
orbital fits for ten Lick and Keck systems, including two systems (HD 11964, HD 183263) for which we 
provide confirmation of second planets only tentatively identified elsewhere, and two others (HD 187123 and 
HD 217107) for which we provide a major revision of the outer planet's orbit. We compile orbital elements from 
the literature to generate a catalog of the 28 published multiple-planet systems around stars within 200 pc. 
From this catalog we find several intriguing patterns emerging: 
\begin{itemize}
\item Including those systems with long-term 
radial velocity trends, at least 28\% of known planetary systems
appear to contain multiple planets;
\item  Planets in multiple-planet systems have somewhat smaller
  eccentricities than single planets; and 
\item The 
distribution of orbital distances of planets in multiplanet systems and single planets are inconsistent: single- 
planet systems show a pileup at $P \sim$ 3 days and a jump near 1 AU, while multiplanet systems show a 
more uniform distribution in log-period. 
\end{itemize}
In addition, among all planetary systems we find the following:
\begin{itemize}
\item There may be an emerging, positive correlation between stellar mass and giant-planet semimajor axis. 
\item Exoplanets more massive than Jupiter have eccentricities broadly distributed across $0 < e < 0.5$, while lower mass exoplanets 
exhibit a distribution peaked near e = 0.
\end{itemize}
\end{abstract}
\keywords{planetary systems}
\section{Introduction}\label{introsec}

\subsection{The Detection of Multiple-Planet Systems}

The first exoplanetary system known to comprise multiple
planets\footnote{Prior to this, \citet{Wolszczan92} detected
  three extraordinary planets orbiting the pulsar PSR
  1257+12.  Here, we restrict the discussion to systems orbiting
  nearby, ordinary stars.} was the triple system 
$\upsilon$ And \citep{Butler_upsand}, detected by the radial velocity
method just four years after the first confirmed
exoplanet, 51 Peg $b$ \citep{Mayor_queloz}.  The subsequent discovery of a second
planet orbiting 47 UMa \citep{Fischer_47uma} and the resonant pair of
planets orbiting GJ 876 \citep{Marcy_876} foreshadowed the discovery
of more than two dozen more systems.  Today, \multfrac\ of known host
stars of exoplanets within 200 pc are known to be multiple-planet systems, and
another \trendfrac\ show significant evidence of multiplicity in the form of
long-term radial velocity trends.

The recent proliferation of multiple-planet systems is due to the
increase in both the velocity precision and duration of the major planet
search programs.  The increased time baseline has led to the detection of
long-period outer companions, the first being the $P = 1270$ d planet
in the multiple system $\upsilon$ And.  Today, the 10$+$ year baseline
of high-precision ($< 5$ m/s) radial velocity planet searches means
that most planets with $a < 3$ AU now have multiple complete orbits
observed, improving their detectability. Even planets with $a > 4$ AU
which have not yet completed a single orbit can sometimes have
well constrained minimum masses ($\msini$), as in the case
of HD 187123 $c$ \citep{Wright07}.
 
\begin{figure} [t!]
  \vspace{-0.5in}
  \plotone{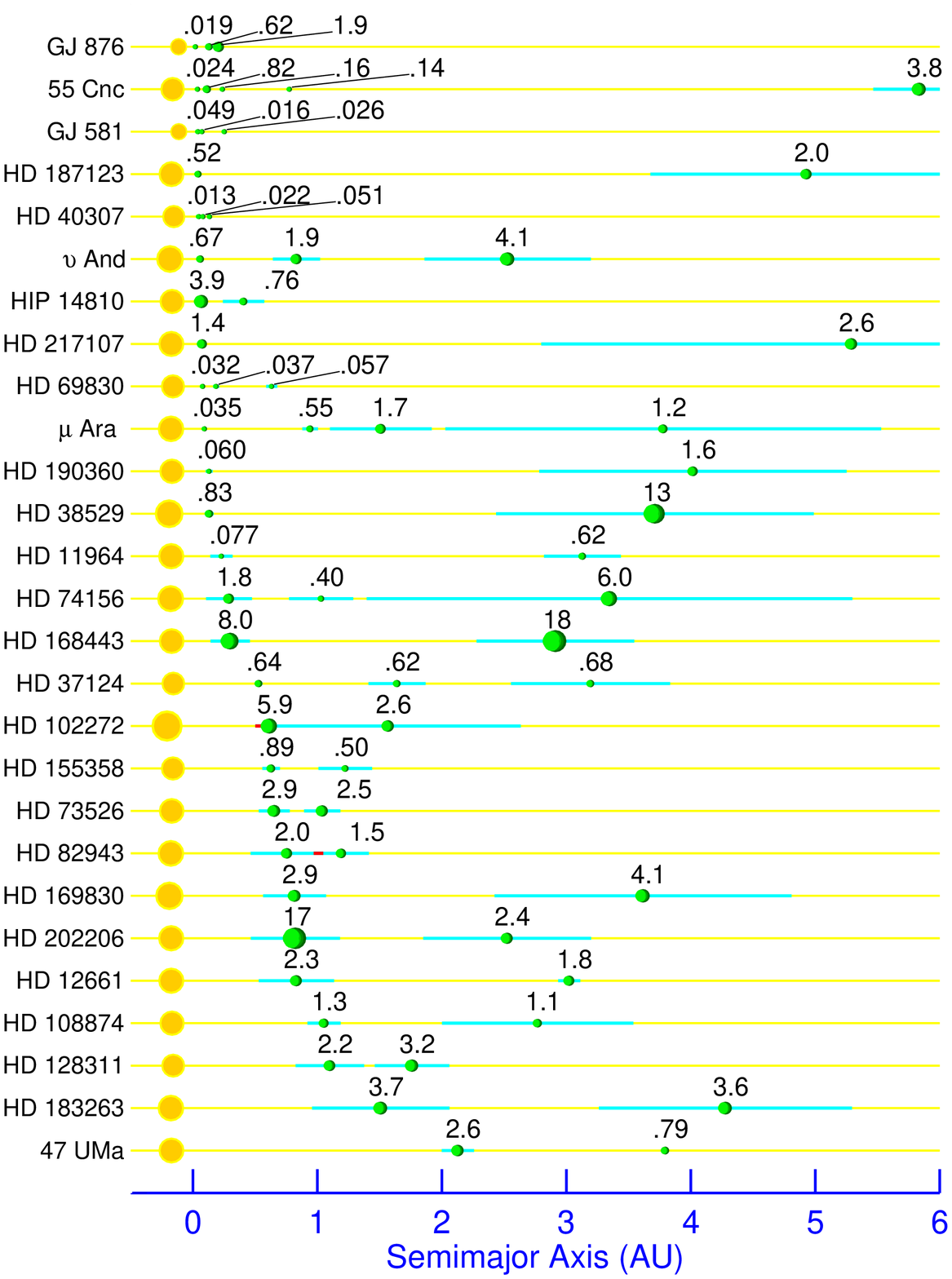}
  \caption{Chart of semimajor axes and minimum masses for the \multsystems\ known
    multiplanet systems.  The diameters depicted for planets are
    proportional the cube root of the planetary \msini. The periapse
    to apoapse excursion is shown by a horizontal line intersecting
    the planet. The diameters depicted for stars are proportional the
    cube root of the stellar mass.}\label{chart}
\end{figure}

Radial velocity precision has steadily improved toward and below 1 m/s with
the High Resolution Echelle Spectrometer (HIRES) at the Keck Observatory and with the HARPS spectrograph at La Silla
\citep{Pepe03,Wright07}.  This has allowed for the detection of ever weaker
signals and led to the discovery of some of the lowest-mass planets
known, including (among others): 55 Cnc $e$ \citep{McArthur04}, GJ 876$d$
\citep{Rivera05}, the triple system HD 69830 \citep{Lovis06}, and
$\mu$ Ara $d$ \citep{Santos04b}.  We list all published multiple-planet systems within 200pc in Table~\ref{Catalog}, a list
which benefits from updated and improved velocities of 10
of the systems provided in Table~\ref{velocities}.  We present a
graphical overview of these multiplanet systems in Fig.~\ref{chart}.

\begin{deluxetable}{rlllllllllllll}
\tablecaption{List of Exoplanets in Multiplanet Systems} 
\tablewidth{8in}
\rotate
\tablecolumns{14} 
\tabletypesize{\footnotesize}
\tablehead{ \colhead{Name}& & \colhead{Per} & \colhead{K} & \colhead{e\tablenotemark{a}}
 &\colhead{$\omega$\tablenotemark{a}} & \colhead{$T_p$} &
  \colhead{$\msini$} &\colhead{a} & \colhead{r.m.s.}  &
  \colhead{$\sqrt{\chi_\nu^2}$} &\colhead{$N_{obs}$}& \colhead{Note} &
  \colhead{ref\tablenotemark{b}}\\ & &\colhead{(d)} &
  \colhead{(\mps)}& &\colhead{($\arcdeg$)}
  &\colhead{(JD-2440000)}&\colhead{($\Mjupsmall$)} &\colhead{(AU)}
  &\colhead{(\mps)} & & & & } 

\startdata
\label{Catalog}
\input{tab1.tex} 
\enddata
\tablecomments{For succinctness, we express uncertainties using
  parenthetical notation, where the least significant digit of the
  uncertainty, in parentheses, and that of the quantity are understood
  to have the same place value.  Thus, ``$0.100(20)$'' indicates
  ``$0.100 \pm 0.020$'', ``$1.0(2.0)$'' indicates ``$1.0 \pm 2.0$'',
  and ``$1(20)$'' indicates ``$1 \pm 20$''.}  \tablenotetext{a}{When
  the uncertainty in $e$ is comparable to $e$, uncertainties in
  $\omega$ and $e$ become non-Gaussian.  See \citet{Butler06} for
  details.}  \tablenotetext{b}{References indicate which orbital
  parameters are taken from the literature as follows:
  Be8: \citet{Bean08}; Bu6: \citet{Butler06}; Cc7: \citet{Cochran07};
  Cr5: \citet{Correia05}; Ds8: \citet{Desort08};
  Fi2: \citet{Fischer02}; Fi8: \citet{Fischer08}; Le6: \citet{Lee06};
  Lv6: \citet{Lovis06}; My4: \citet{Mayor04}; My8: \citet{Mayor08};
  Ni8: \citet{Niedzielski08b}; Nf4: \citet{Naef04}; 
  Pp7: \citet{Pepe07}; R5: \citet{Rivera05}; T6: \citet{Tinney06};
  U7: \citet{Udry07}; Vo5: \citet{Vogt05}; Wr7: \citet{Wright07}.  All other orbital solutions are new Keplerian
  (kinematic) fits to the data in Table~\ref{velocities}.}
\tablenotetext{c}{The fit for this
  system includes a trend of $-0.51 \pm 0.1$ m \persec yr$^{-1}$.}
\tablenotetext{d}{The planets in HD 73526
  are in a 2:1 mean motion resonance, and planet-planet interactions
  are important, rendering Keplerian elements inadequate.  In addition
  to the elements reported here, \citet{Tinney06} report a mean
  anomaly to be $86\arcdeg\pm13$ and $82\arcdeg\pm 27$ at a Julian
  Date of 2451212.1302.  There is considerable degeneracy between $K$
  and $e$ because the orbital period of HD 73526$c$ differs from one
  year by only 12 d.}  
\tablenotetext{e}{\citet{Barnes08} found that the orbit presented in
  \citet{Bean08} for the $d$ component is unstable, and provide multiple
  stable solutions without uncertainties in the orbital parameters.}
\tablenotetext{f}{Planet-planet interactions
  are strong in 55 Cnc.  The osculating orbital elements here are
  from the dynamical fit at Julian Date 2447578.730 of
  \citet{Fischer08}.  That work puts no errors on these parameters,
  however, so the errors quoted here are those from the Keplerian
  (kinematic) fit there.}  
\tablenotetext{g}{The
  exoplanets in HD 82946 have significant interactions, which
  renders Keplerian orbital elements inadequate for describing their
  orbits, since these elements are time-variable.  \citet{Lee06}
  report the mean anomaly of the inner and outer planets to be
  $353\arcdeg$ and $207\arcdeg$, respectively, at a Julian Date of
  2451185.1.}
\tablenotetext{h}{This solution includes a linear trend with magnitude
  $-3.08 \pm 0.16 $ m \persec\ yr$^{-1}$.}
\tablenotetext{i}{The exoplanets in HD 202206 have significant
  interactions, which renders Keplerian orbital elements inadequate
  for describing their orbits, since these elements are time-variable.
  \citet{Correia05} report the mean longitude to be $266.23\arcdeg \pm
  0.06$ and $30.59\arcdeg \pm 2.84$ for the inner and outer planets,
  respectively, at a Julian Date of 2452250.}  
\tablenotetext{j}{The outer two
  exoplanets GJ 876 have significant interactions, which
  renders Keplerian orbital elements inadequate for describing their
  orbits, since these elements are time-variable.  \citet{Rivera05}
  report the mean anomaly of the planets to be
  $M_d = 309.5\arcdeg\pm5.1\arcdeg$, $M_c=308.5\arcdeg\pm1.4\arcdeg$, and $M_b =
  175.5\arcdeg\pm6.0\arcdeg$ respectively, at a Julian Date of
  2452490.  The solution quoted here assumes $i=90\arcdeg$.}

\end{deluxetable}
\begin{deluxetable}{lrrrc}
\tablecaption{Updated RV Data for Multiple-Planet Systems}
\tablecolumns{5} \tablewidth{0pc} \tablehead{ {Star} & {Time} &{Radial
    Velocity} & {Unc.} & {Telescope}\\ &(JD-2440000)
  &{(m/s)}&{(m/s)}&}

\startdata
\label{velocities}
\input{tab2.tex} \enddata
\tablecomments{A full version of this table is available in the
  electronic version of the {\it Journal} 
This is only a sample. }
\end{deluxetable}

\subsection{Multiple-Planet Systems and Planet Formation Theory}

multiple-planet systems are of special interest to test theoretical
models of planet formation, dynamics, and final architectures. To
date, over 230 planets have been discovered orbiting \allsys\ main sequence
stars within 200 pc, \multfrac\ of which harbor multiple
planets with well constrained masses and periods. With \totmult\ such
planets members of multiplanet systems, we may now make statistically
significant comparisons between properties of planets in single planet
systems with those in multiplanet systems. 

The known planet population is remarkably diverse with properties that
bear on planet formation theory. Theories to explain the semimajor
axis distributions and eccentricity distributions of planets have
especially benefitted from the constraints that multiplanet systems
impose. Most planet formation theories are based on the
core accretion model that begins with a disk of dust and gas where
the dust particles collide and grow to form rock-ice planetary cores
\citep{Aarseth93, Kokubo02, Levison98}. If a core becomes massive
enough while gas remains in the disk, it gravitationally accretes the
nearby gas and rapidly increases in mass \citep{Bodenheimer03,
  1996Icar..124...62P}. Such gas planets should form preferentially
beyond the ``ice line'' (near 3 AU for solar-type stars),
where ices can participate in the initial planetary cores.  This
appears inconsistent with the observation that about 20\% of known
exoplanets orbit within 0.1 AU, where there should be too little ice
in the protoplanetary disk for massive cores to form quickly. 

Thus it
appears that short-period planets form farther out and migrate inwards
to their final semimajor axis \citep{Paploizou05, Tanaka04,
  Trilling02}. The discovery of at least five systems in or near
mean-motion resonance (viz. GJ 876, 55 Cnc, 
HD 82943, HD 73526, and HD 128311) may lend support to the migration
hypothesis \citep{Nelson02, Lee06,Marzari05}.  Such resonances are
difficult to explain if planets form in situ, but hydrodynamical
simulations and n-body simulations with externally applied damping
\citep{Kley05b, Bryden01, Chiang02, Dangelo03, Ida04b} show that
resonance capture occurs if planets undergo significant migration at
different rates, passing through mean-motion resonances.  Also, although there are confirmed 2:1 mean
motion resonances, other period ratios may occur \citep{KleyResonance,
  Nelson02, Laughlin02}.  Future discoveries of multiplanet systems
may shed further light on the resonance capture processes and on planet
formation dynamics.

multiplanet systems also provide hints about the wide distribution of
exoplanet eccentricities and we are beginning to identify different
processes that drive some planets to large eccentricities, while
damping others to moderate and low eccentricities \citep{Murray2003,
  thestability}. \citet{Ford06} suggests that planet-planet scattering
augments eccentricities, and that after the era of strong planet-planet
scattering interactions with the remaining planetesimals will damp these
eccentricities to the observed distribution. One may hope to detect
some signature of these process in exoplanet data. Indeed, the
configuration of the $\upsilon$ Andromedae three-planet system may be
explained by a planet-planet scattering event that ejected a fourth
planet from the vicinity, as this system carries a signature of sudden
perturbation \citep{Ford05}. HD 128311, a two-planet system, also
appears to carry a signature of similar scattering \citep{SandorScatter}.

Models of planet formation in protoplanetary disks point to the disk
viscosities and lifetimes that are required to produce some observed
systems. There is some inconsistency between the required disk
behavior and that predicted by hydrodynamical theories
\citep{KleyResonance}. Simulations that include multiple
protoplanets interacting with a protoplanet swarm show that most
protoplanets will eventually accrete on to the star, leading to a
planet occurrence rate lower than that observed \citep{Cresswell06}.  

The increasing number of characterized multiplanet systems should
help constrain these theories.  We present the current catalog of
multiple-planet systems with the most recent radial velocity data in
hopes of offering theorists a thorough overview current state of
observations, and an opportunity to propose new tests of planet
formation theories as the number of known multiple-planet systems grows
and our knowledge of them improves.

\subsection{Plan}
In \S\ref{Survey} we present the sample of the \multsystems\ published nearby
multiple-planet systems, including updated orbital parameters and
radial velocities for ten of these systems.  In \S\ref{Stats} we present some 
empirical correlations among the properties of multiple-planet systems
and difference between the distributions of orbital parameters among
multiple-planet and single-planet systems.  In \S\ref{discussion} we
briefly summarize our findings and discuss our principle conclusions.

\section{A Survey of the Known Multiple-Planet Systems}
\label{Survey}
\subsection{New and updated multiplanet systems}
We present updated velocities and orbital parameters for ten
multiple-exoplanet systems in Tables~\ref{Catalog} and
\ref{velocities}.  In six systems, the updated orbits are
substantially similar to the published orbits.  In four cases there
are major updates to previously published orbital parameters, which we
detail below. 
\subsubsection{HD 183263}
\begin{figure} 
  \plotone{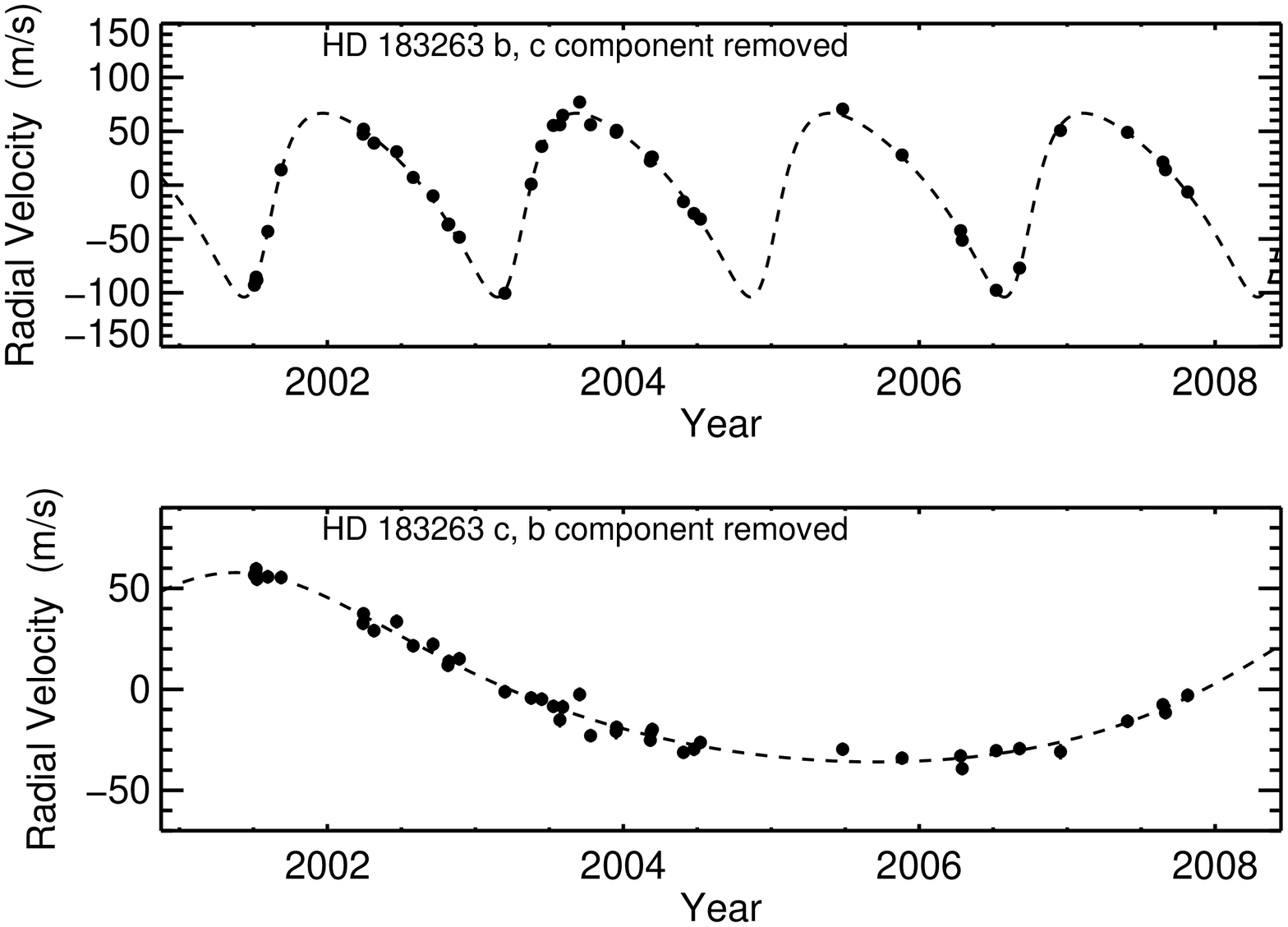}
  \caption{RV curves for HD 183263. The data are from Keck Observatory
    and show (top) the inner planet with $P = 1.7$ yr and (bottom)
    $\msini=3.7 \mjup$ the outer planet with $P\sim 8$ yr and $\msini = 3.6$
    \mjup.\label{183263bc}}
\end{figure}
\begin{figure}
  \plotone{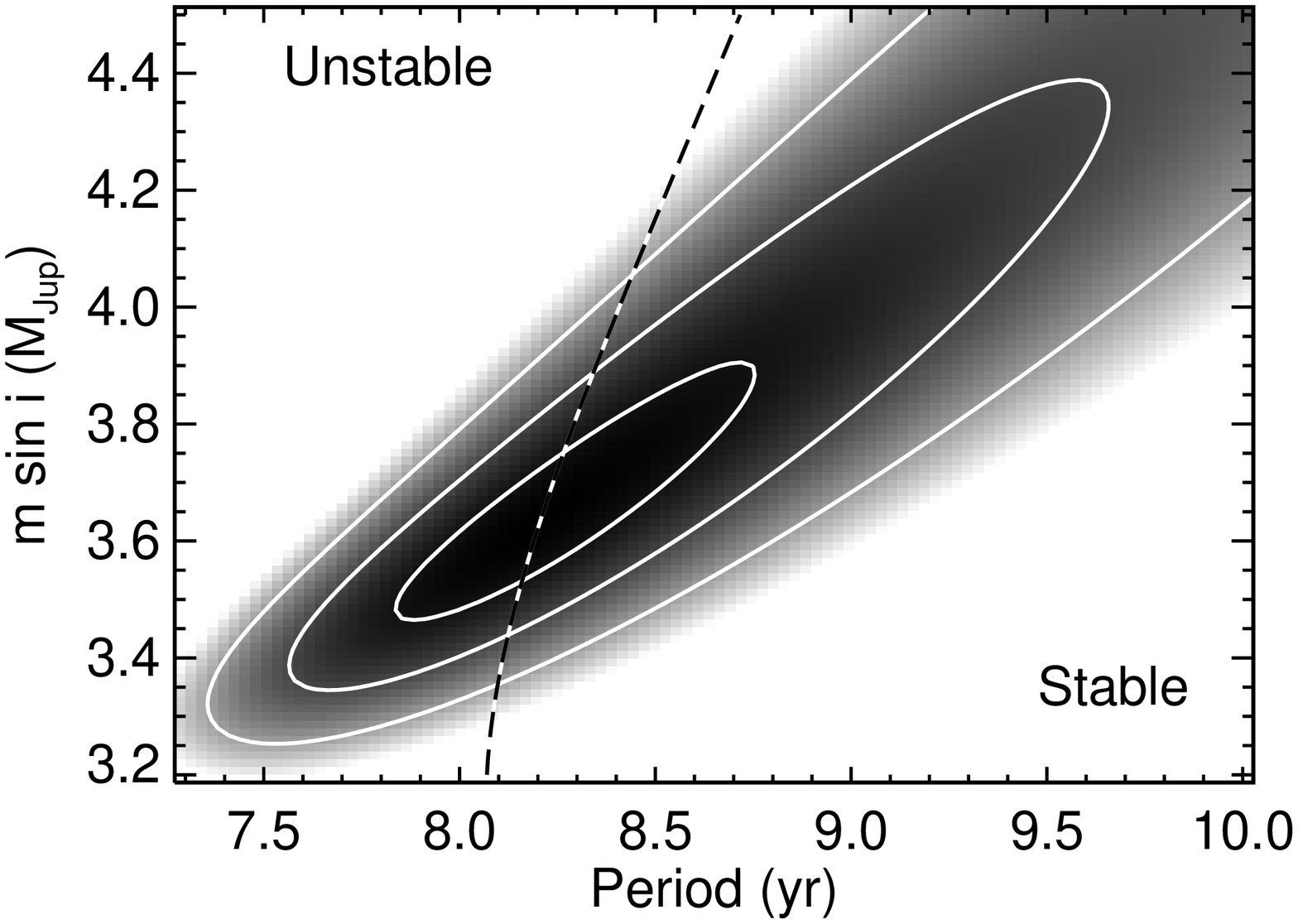}
  \caption{Contours of $\chi^2$ in $P_c-m_c\sin{i}$ space of best-fit orbits
  to the RV data of HD 183263 (Fig.~\ref{183263bc}), with $\chi^2$ in grayscale.  The solid contours mark
  the levels where $\chi^2$ increases by 1, 4 and 9 from the
  minimum.  The dashed curve marks the approximate dividing line
  between stable and unstable orbital solutions, as determined by an
  ensemble of n-body integrations of the system assuming edge-on,
  coplanar orientations (cf. Figure 13 of
  \citet{Wright07}).  \label{183263map}}  
\end{figure}

\citet{Marcy05} announced an $\msini = 3.7 \mjup$ planet in a 634-day,
eccentric ($e=0.4$) orbit around HD 183263, a G2 IV star, and pointed
out a strong residual linear trend of 32 m \persec yr$^{-1}$.
\citet{Wright07} showed that by 2007 the residuals had significant
curvature, but could not constrain the minimum mass of the outer companion.
Since then, we have obtained observations on an additional six nights.  These
data have dramatically constrained the orbital fit to the outer
planet, and we can now confidently measure the orbital elements of a
3--4 $\mjup$ planet under the assumption that there is not a third
companion contributing a detectable linear trend \citep[see][for a
  discussion of the difficulties of constraining planets with only
  partially-observed orbits]{Wright07}.  HD 183263$c$ has a 8.4$\pm$0.3
yr orbital period, orbits at 4.3$\pm0.4$ AU, and has $e=0.24\pm0.06$.  We
present the full set of newly determined orbital parameters in
Table~\ref{Catalog}, and the latest radial velocities in
Table~\ref{velocities} and Fig.~\ref{183263bc}.
Figure~\ref{183263map} shows $\chi^2$ in $P-\msini$ space,
demonstrating that, despite our having only one observed orbit, the
minimum mass and orbital period of the planet are well constrained.

In Table~\ref{Catalog} we report a stable solution near the $\chi^2$
minimum with errors estimated from the sample of stable orbits 
found through error bootstrapping \citep{Wright07}.  We tested each of
the 100 bootstrapping trials for stability, and found a rough dividing
line between stable and unstable solutions for this system such that orbital
solutions with $e_c > 10^{-4} ((P_c/\mbox{days})-3000)$ are generally
unstable.  We have mapped this line into $P_c-m_c\sin i_c$ space in
Figure~\ref{183263map} demonstrating that many solutions consistent
with the data are, in fact, unstable.   The true uncertainties are thus
asymmetric about the nominal values because the $\chi^2$ minimum is so
close to the boundary of stability.

For these long-term stability tests, we applied direct n-body
integrations on each of the orbital solutions generated in the error
bootstrapping.  The radial velocity parameters were converted into
initial conditions using a Jacobi coordinate system \citep{Lee03}.
Unless otherwise specified, we assumed edge-on, 
coplanar orbits.  We held the stellar mass fixed, adopting values from
\citet{Takeda07}.  We integrated for at least $10^{8}$ yr using
the hybrid integrator in {\tt Mercury} \citep{Chambers99}.  For the
majority of each integration, {\tt Mercury} uses a mixed-variable
symplectic integrator \citep{Wisdom91} with a time step equal
to a hundredth of the Keplerian orbital period calculated at a
semimajor axis equal to the pericenter distance of the closest
planet.  During close encounters, {\tt Mercury} uses a Bulrich-Stoer
integrator with an accuracy parameter of $10^{-10}$.  We identified each
set of initial conditions as an unstable system if: 1) two planets
collide, 2) a planet is accreted onto the star (astrocentric distance
less than 0.005AU), or 3) a planet is ejected from the system
(astrocentric distance exceeds 100AU).  We manually verified that for
the vast majority of systems not identified as unstable, the final
orbits are qualitatively similar to the initial conditions. 

\subsubsection{HD 187123}
\label{187123}
\begin{figure} 
  \plotone{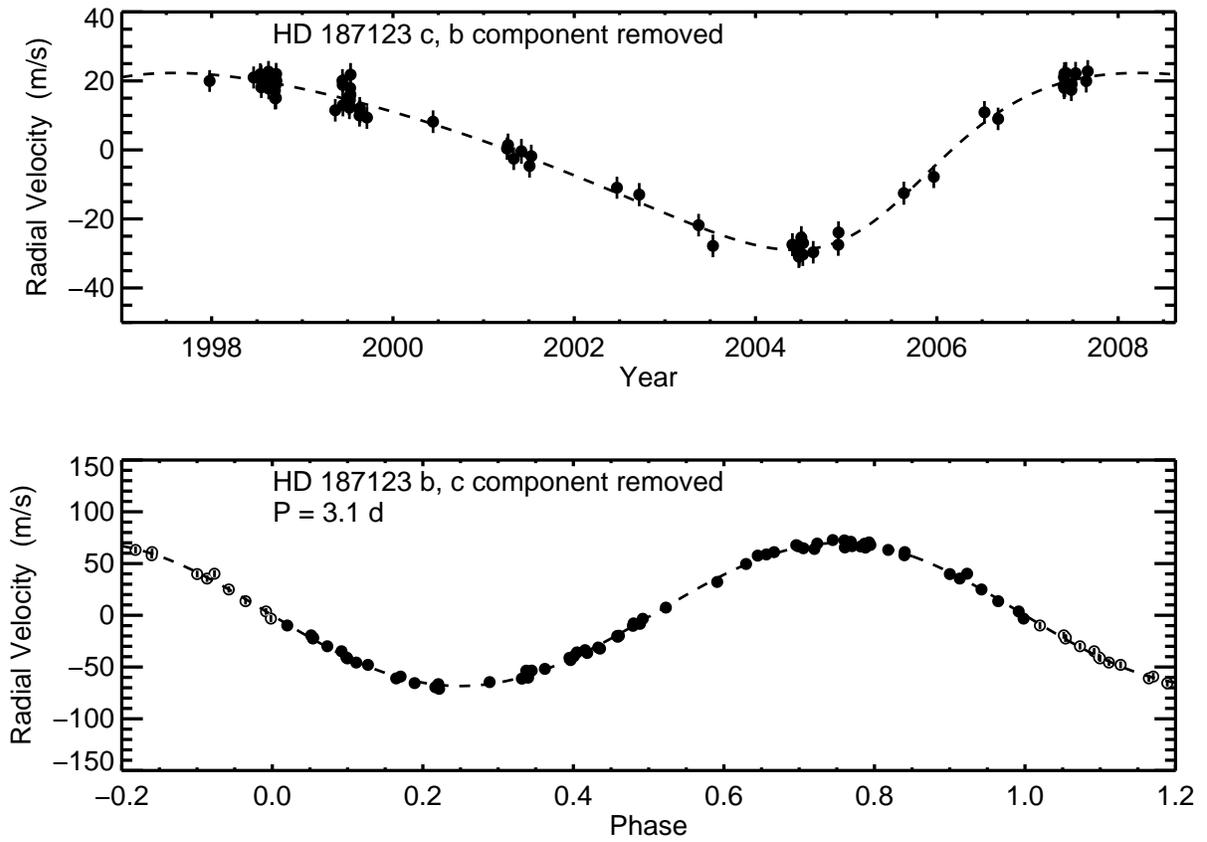}
  \caption{RV curves for HD 187123. The data are from Keck Observatory
    and show the inner planet (top) with $P = 3.1$ d and $\msini=0.5 \mjup$
    and the outer planet (bottom) with $P=10.7$ yr and $\msini =2$
    \mjup.\label{187123bc}}
\end{figure}
\begin{figure}
  \plotone{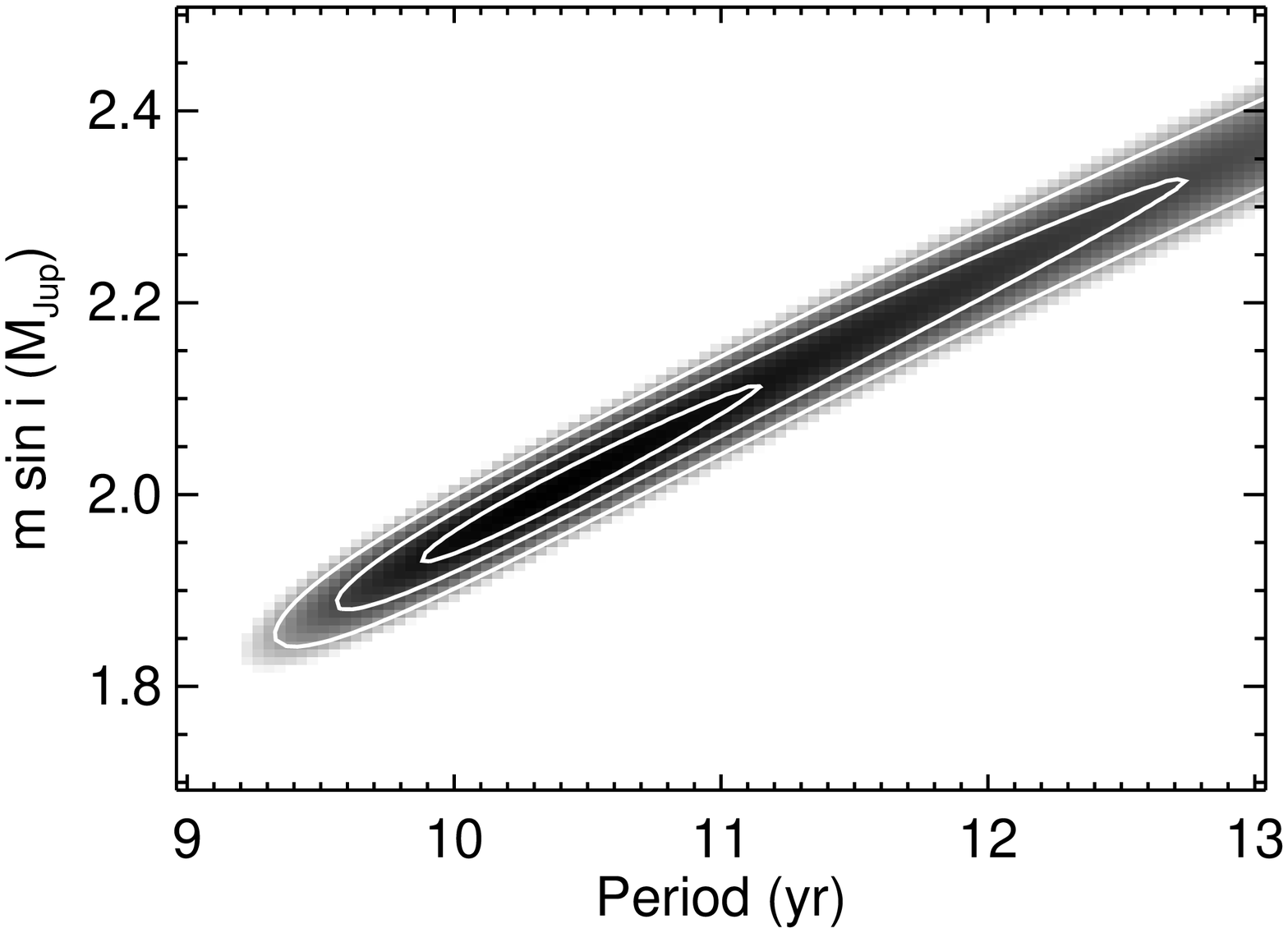}
  \caption{Contours of $\chi^2$ in $P_c-m_c\sin{i_c}$ space of best-fit orbits
  to the RV data of HD 187123 (Fig.~\ref{187123bc}), with $\chi^2$ in
  grayscale (cf. Figure~15 of \citet{Wright07}).  The solid contours
  mark the levels where $\chi^2$ increases by 1, 4 and 9 from the
  minimum.  Although the orbit is still imprecise, the orbital period
  and minimum mass are now constrained to $\sim 20$\%.}  \label{187123map}
.
\end{figure}

\citet{Butler98} announced an $\msini = 0.5 \mjup$ planet in a 3-day
orbit around HD 187123, a close solar analog ($M_* = 1.1\Msol$, $\feh
= +0.1$, $\teff = 5810$).  \citet{Wright07} announced the existence of
an outer companion with orbital period $> 10$ yr.  At that time, the
radial velocity history was too incomplete to fully determine the
orbit, which had not yet closed. \citet{Wright07} were nonetheless
able to constrain the minimum mass of this 'c' component to be
planetary ($1.5 \mjup < \msini < 10 \mjup$).  Since then we have
obtained additional observations of HD 187123, and have found that the
orbit has closed just recently.

Assuming that there is no linear trend or detectable third planet in the
system, the data constrain HD 187123$c$ to have $P=10.4\pm1.2$ yr,
$e=0.25\pm0.03$, and $\msini = 2.0\pm0.3 \mjup$.  We present the newly
determined orbital parameters in Table~\ref{Catalog} and the latest
radial velocities in Table~\ref{velocities} and Figure~\ref{187123bc}.  Figure~\ref{187123map}
shows $\chi^2$ in $P-\msini$ space, demonstrating that, despite our
having observed somewhat less than one full orbit, the minimum mass and orbital period of the
planet are constrained to $\sim 20$\%.

We have checked this orbit for long-term stability in the same manner
as for HD 183263 and find that the range of solutions shown in
Figure~\ref{187123map} are well within the stable regime.

\subsubsection{HD 11964}
\begin{figure} 
  \plotone{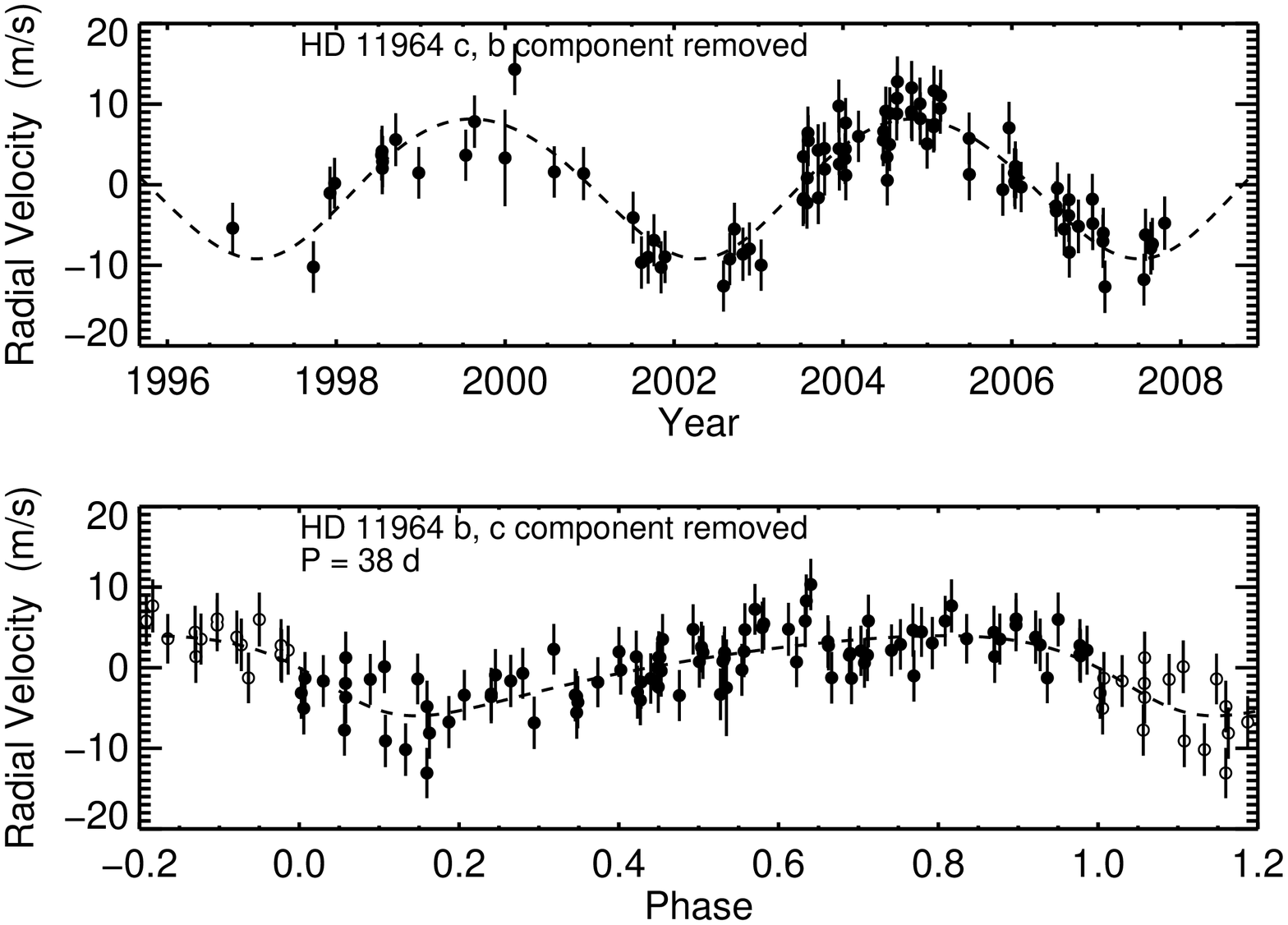}
  \caption{RV curves for HD 11964. The data are from Keck Observatory
    and show the inner planet (bottom) with $P = 38.9$ d and $\msini=23
    \mearth$, and the outer planet (top) with $P=5.5$ yr and $\msini =0.6$
    \mjup.\label{11964bc}}
\end{figure}

HD 11964 (= GJ 81.1A) is a metal-rich ($\feh = +0.12$) slightly evolved
(2 mag.\ above the main sequence) G star with a nearby (sep $\sim
30\arcsec$), K dwarf companion.  \citet{Butler06} announced the planet
HD 11964$b$, a Jovian ($\msini = 0.6 \mjup$) planet in a 5.5 yr,
circular orbit, and noted a weak, residual trend in the velocities.
An analysis by \citet{Wright07} showed that a trend was probably not
the proper interpretation of the residuals, and that they were
consistent with a low amplitude ($K = $5.6 m\persec), 38 d signal (FAP
$< 2\%$).  \citet{Wright07} cautioned that the low amplitude of this
prospective 38 d planet meant that it would require more observations
for confirmation, especially given the higher levels of jitter seen in
subgiant stars \citep{Wright05,Johnson07}.  \citet{Gregory07} also
noted the 38 d period in a Bayesian periodgram of the published
velocities (as well as a 360 d signal which is not apparent in the
rereduced data presented here).

  We have now obtained 24 additional observations
  of the star, and the 38 d signal has strengthened (FAP $< 1\%$),
  allowing us to confirm a low-mass ($\msini = 23 \mearth$)
  planet, HD 11964$c$, in a 37.9 d orbit.  The low amplitude of the
  signal makes estimation of $e$ and $\omega$ difficult, but data are
  inconsistent with a circular orbit, and favor $e=0.3$.

  Private communications from our group regarding the 38 d signal,
  which has been apparent, but not convincing, since 2005, has led to
  some confusion in the literature regarding the nomenclature of these
  planets \citep[e.g.][]{Raghavan06}. Here, we follow the convention
  that planet components are ordered by the date of a formal or a public
  announcement of their existence.  We report the orbital parameters
  of the two-planet fit in Table~\ref{Catalog}, the radial velocities in
  Table~\ref{velocities}, and we show the radial velocity curves in
  Fig.~\ref{11964bc}.  

We have checked this orbit for long-term stability with the same n-body
code as for HD 11964, and find that it is well within the stable regime.

\subsubsection{HD 217107}
\label{217107}
\begin{figure} 
  \plotone{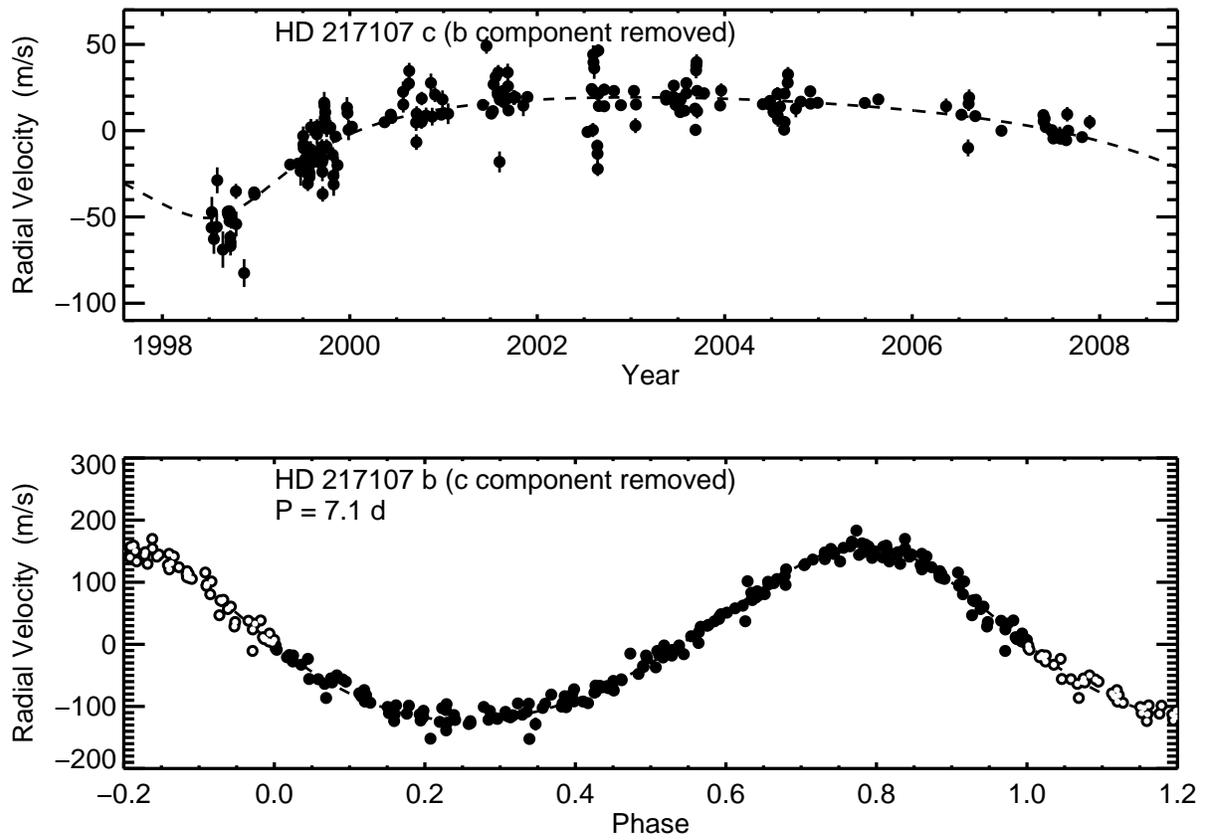}
  \caption{RV curves for HD 217107. The data are from Keck Observatory
    and Lick Observatory,
    and show the inner planet (top) with $P = 7.1$ d and $\msini=1.4 \mjup$
    and the outer planet (bottom) with $P=11.6$ yr and $\msini =2.6$
    \mjup.\label{217107bc}}
\end{figure}
\begin{figure}
  \plotone{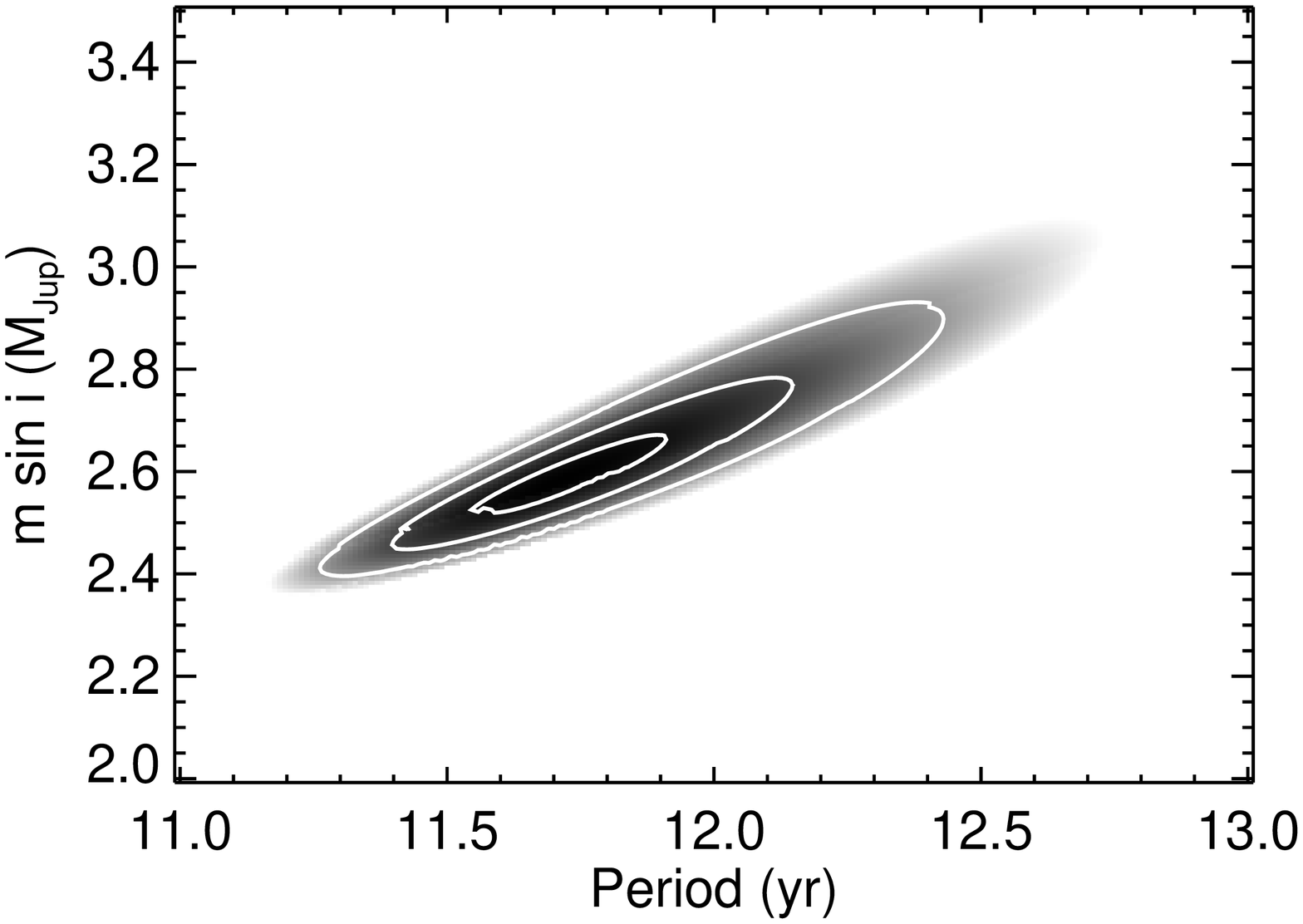}
  \caption{Contours of $\chi^2$ in $P_c-m_c\sin{i_c}$ space of best-fit orbits
  to the RV data of HD 217107 (Fig.~\ref{217107bc}), with $\chi^2$ in
  grayscale (cf. Figure~15 of \citet{Wright07}).  The solid contours
  mark the levels where $\chi^2$ increases by 1, 4, and 9 from the
  minimum.  Although the orbit is still imprecise, the orbital period
  and minimum mass are now constrained to $\sim 10$\%.}  \label{217107map}
.
\end{figure}
\citet{Fischer99} reported a 7.1 d, $\msini = 1.4 \mjup$ planet orbiting HD
217107, and \citet{Fischer01} described a linear trend superimposed on
the Lick and Keck radial velocities.  \citet{Vogt05} updated the
orbital fit, finding significant curvature in the residuals to the
inner planet fit, and estimated its orbital period, though poorly
constrained, to be $~8.5$ yr, with $\msini \sim 2 \mjup$.  

Since then, data collected at Lick and Keck continue to map out the
orbit of the outer planet.  Today, we can constrain the minimum mass
and period to within $\sim 10\%$ under the assumption that there are
no additional planets in the system.  

We present the newly determined orbital parameters in
Table~\ref{Catalog} and the latest radial velocities in
Table~\ref{velocities} and Figure~\ref{217107bc}.
Figure~\ref{217107map} shows $\chi^2$ in $P-\msini$ space,
demonstrating that, despite our having observed somewhat less than one
full orbit, the minimum mass and orbital period of the planet are constrained
to $\sim 10$\%, at $P\sim 11.7$ yr and $\msini \sim 2.6 \mjup$.

We have checked this orbit for long-term stability in the same manner
as for HD 183263, and find that the range of solutions shown in
Figure~\ref{217107map} are well within the stable regime.

\subsubsection{47 UMa}

\citet{Butler96a} announced the existence of a 1090-day planet
orbiting 47 UMa from data collected at Lick Observatory.  After
collecting an additional 6 years of data, \citet{Fischer_47uma}
announced the existence of a second, 0.46 \mjup\ long period companion
in a $\sim 2600$-day orbit.  \citet{Naef04} and \citet{Wittenmyer07},
using data from ELODIE and McDonald Observatory, respectively, have
questioned the existence of 47 UMa {\it c}.  Neither of the latter data sets, however, have
both the precision and the duration to rule out the outer planet.  The
parameters quoted here are the literature values.

\subsection{The Current Sample}

We consider here the \multsystems\ known multiple-planet systems among the \allsys\
known, normal exoplanet host stars within 200 pc.  This is the sample
of the {\it Catalog of Nearby Exoplanets} \citep[CNE, ][]{Butler06}\footnote{Available on the World Wide
  Web at http$://$exoplanets.org}.  The CNE employs a liberal
upper mass limit in its definition of an exoplanet (any companion with
$\msini < 24 \mjup$) but restricts itself to systems with high-quality
radial velocity detections around the bright stars most amenable to
confirmation and follow-up. This distance cutoff excludes the multiple
planets orbiting the pulsar PSR 1257+12 \citep{Konacki03} and the
Jupiter-Saturn analogs orbiting OGLE-2006-BLG-109 \citep{Gaudi08}
detected by microlensing.  We also exclude several speculative claims
of additional planets around known exoplanet host stars, systems where
second planets have very poorly constrained orbits, and two
announced  multiplanet systems (HD 47186 and HD 181433 by Bouchy et
al. 2009) for which orbital parameters were not available at submission time.

We provide up-to-date fits with radial velocities as
recent as 2008 June for those multiplanet systems with no
significant planet-planet interactions (see \S\ref{Newtonian}) and
for which we have Lick and Keck data from our planet search \citep[see][for
  details]{Butler06}.  We have employed a fitting
algorithm which exploits linear parameters in the Kepler problem (Wright \& Howard, 2008, ApJ submitted) to efficiently search the
high-dimensional $\chi^2$ space associated with multiple-planet
systems.  We have also updated some of our radial velocity data
reduction procedures at Lick and Keck Observatories, including a small
correction to our calculation of telescopic barycentric motion.  The radial
velocities presented here are thus more accurate and precise than our previously
published velocities for these systems.  The resulting best-fit
orbital parameters and uncertainties supersede previously published
parameters.   

Table~\ref{Catalog} contains measured properties and derived
quantities for these \multsystems\ systems.  Properties of the host stars can be
found in \citet{Butler06}.

\subsection{The Multiplicity Rate}
\label{multiplicity}
In addition to these \multsystems\ systems, \alltrend\ single-planet
systems are best fit with the addition of a linear trend
\citep[]{Butler06}.  If we conservatively exclude the eight such
cases in which the host star has a known stellar 
companion \citep[determined from a survey of the literature
  including][]{Eggenberger07}, then we are left with
\tottrend\ apparently single-star systems with an outer, 
potentially planetary companion.\footnote{Our radial velocity analysis
  is sensitive to the presence of a second set of spectral lines, and
  we estimate that we can rule such a binary companion in a
  close orbit down to $\sim 0.1\Msol$ in most cases.  Some
  contamination from as-yet undetected binary companions may still
  remain in the sample, however.}  Of the \allsys\ known nearby 
planetary systems then, \multfrac\ have multiple confirmed planets and another
\trendfrac\ show significant evidence of being multiple, meaning
the true planet multiplicity rate may be \totfrac\ or higher.  This is
consistent with the estimate of \citet{Wright07} and somewhat more
conservative than the value of $\sim 50$\% in \citet{Fischer01}.

\subsection{Kinematic vs.\ Dynamical Fits}
\label{Newtonian}
The radial velocity signature of multiple-planets is significantly
more complex than that of a single-planet system, and fitting such curves to
observed radial velocity data requires care.  Each planet has five
spectroscopic orbital parameters, most of which are neither orthogonal nor
linear, so finding a global minimum in $\chi^2$ space becomes
significantly more difficult as the number of planets grows.
Efficient algorithms are necessary to conduct a thorough search
(Wright \& Howard 2008, ApJ submitted).

Planets in mean-motion resonances can be particularly difficult to
identify from radial velocity curves because of degeneracies among the
best-fit orbital parameters.  Care must also be taken not to confuse
weak signals resulting from aliasing of orbital periods with the
observing window function with genuine planet detections
\citep{Fischer08, Tinney06}.

Most importantly, interactions between planets may require
consideration.  In many cases these planet-planet interactions are
sufficiently weak that they can be ignored, and the resulting radial
velocity signal is simply the linear superposition of multiple
Keplerian radial velocity curves (a ``Keplerian'' or ``kinematic'' fit).

When these interactions are important, however, both short term and
long term numerical $n$-body integrations of the physical system must
be performed.  In the short term, these interactions can cause 
detectable variations in the orbits of the planets
\citep{Rivera05,Fischer08}.  In these cases a set of constant
Keplerian orbital elements is insufficient to model the observed
radial velocities, and a proper fit must be driven by an $n$-body
code (a ``Newtonian'', or ``dynamical'', fit).

In some cases even good Newtonian fits to the data may 
yield orbital parameters for planets which, while stable for the
duration of the observations, are not stable on timescales comparable
to the age of the planetary system. Thus, long-term stability is an
additional constraint that multiplanet fits must satisfy.

The new fits listed in Table~\ref{Catalog} are all Keplerian fits
which have been confirmed stable with the $n$-body code described in
\S\ref{187123}.  The compiled literature fits are a mixture of
Keplerian and Newtonian fits, and in some cases a more sophisticated
Newtonian fit may be superior to the published one.  For
instance, the orbits of the planets in the HD 74156 system reported by
\citet{Bean08} are apparently unstable on $10^5$ yr timescales, but an
orbit with similar parameters is stable \citep{Barnes08}.  Our
conclusions on the statistical properties of planets are insensitive to these details, but
detailed work on planet-planet interactions and resonant dynamics
should employ the published RV data directly, rather than the
orbital elements presented here. 

\section{Statistics of Multiple-Planet Systems}
\label{Stats}
\subsection{Semimajor Axis Distributions}
\label{semimajordist}
\begin{figure} [ht!]
\begin{center}
\plotone{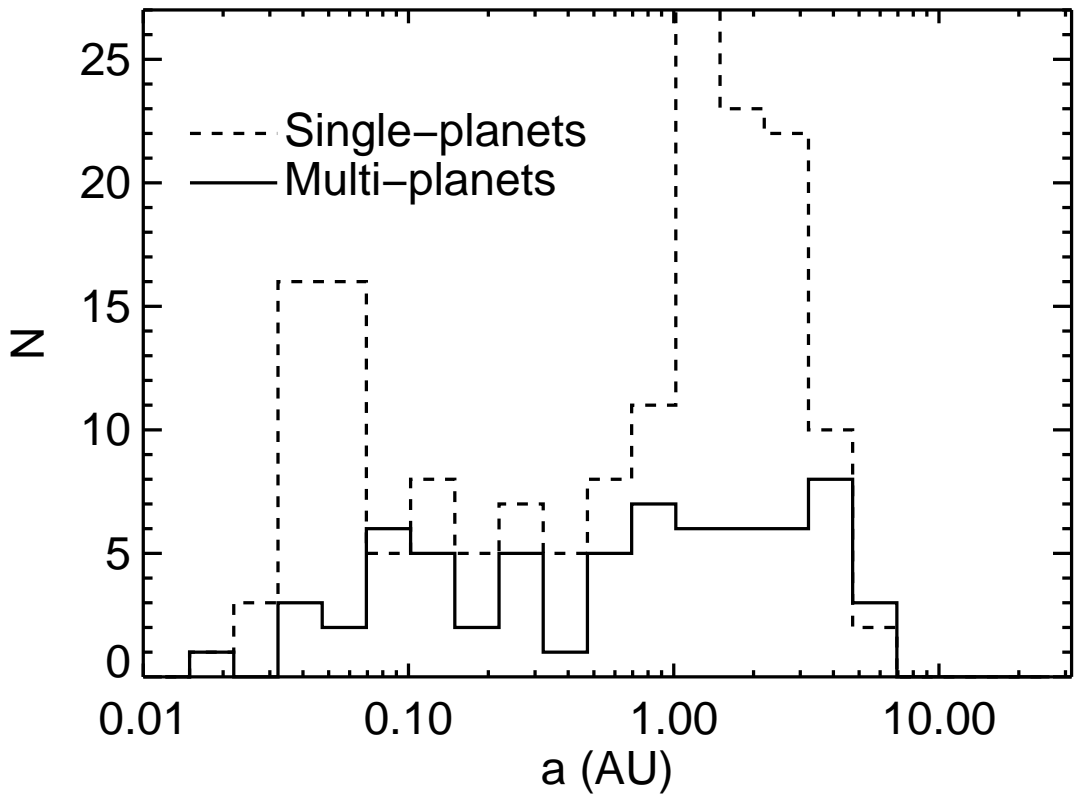}
  \caption{Distribution of semimajor axes of exoplanets for
    multiple-planet systems (solid) and apparently single systems
    (dashed).  Note the enhanced frequency of hot jupiters and the
    jump in abundance beyond 1 AU in the single-planet systems.}\label{a_hist_log}
\end{center}
\end{figure}

    Figure~\ref{a_hist_log} shows the distribution of semimajor
    axes\footnote{Following \citet{Butler06}, we calculate semimajor
      axes from the measured orbital periods using Newton's version of
      Kepler's Third Law;
     the major source of uncertainty in $a$ is usually the host star's mass.}
    for multiplanet and apparently single-planet systems. The semimajor axes
    of multiplanet systems appear to show some significant departures
    from the single-planet 
    systems. Most strikingly, the pile up of Hot-Jupiter planets
    between 0.03 AU and 0.07 AU and the jump at 1AU observed in the
    distribution for 
    single-planet systems are both absent from the corresponding
    distribution for multiplanet systems, which appears rather
    uniform.  

    It thus appears that planets in multiple systems are not
    consistent with having been drawn randomly from the population of
    apparently single-planet systems.  To provide a numerical measure of the
    difference between the two distributions, we created the
    cumulative distribution function (CDF) of the semimajor axis
    distribution (with an upper limit of $13\mjup)$ and applied a
    Monte Carlo Kolmogorov-Smirnov (K-S) test for a difference in distribution.  The K-S
    test uses the two sided D-statistic, defined as the maximum
    difference between the two CDF curves. The Appendix contains a
    complete description of our Monte Carlo K-S procedure.

\begin{figure} [ht!]
\begin{center}
\plotone{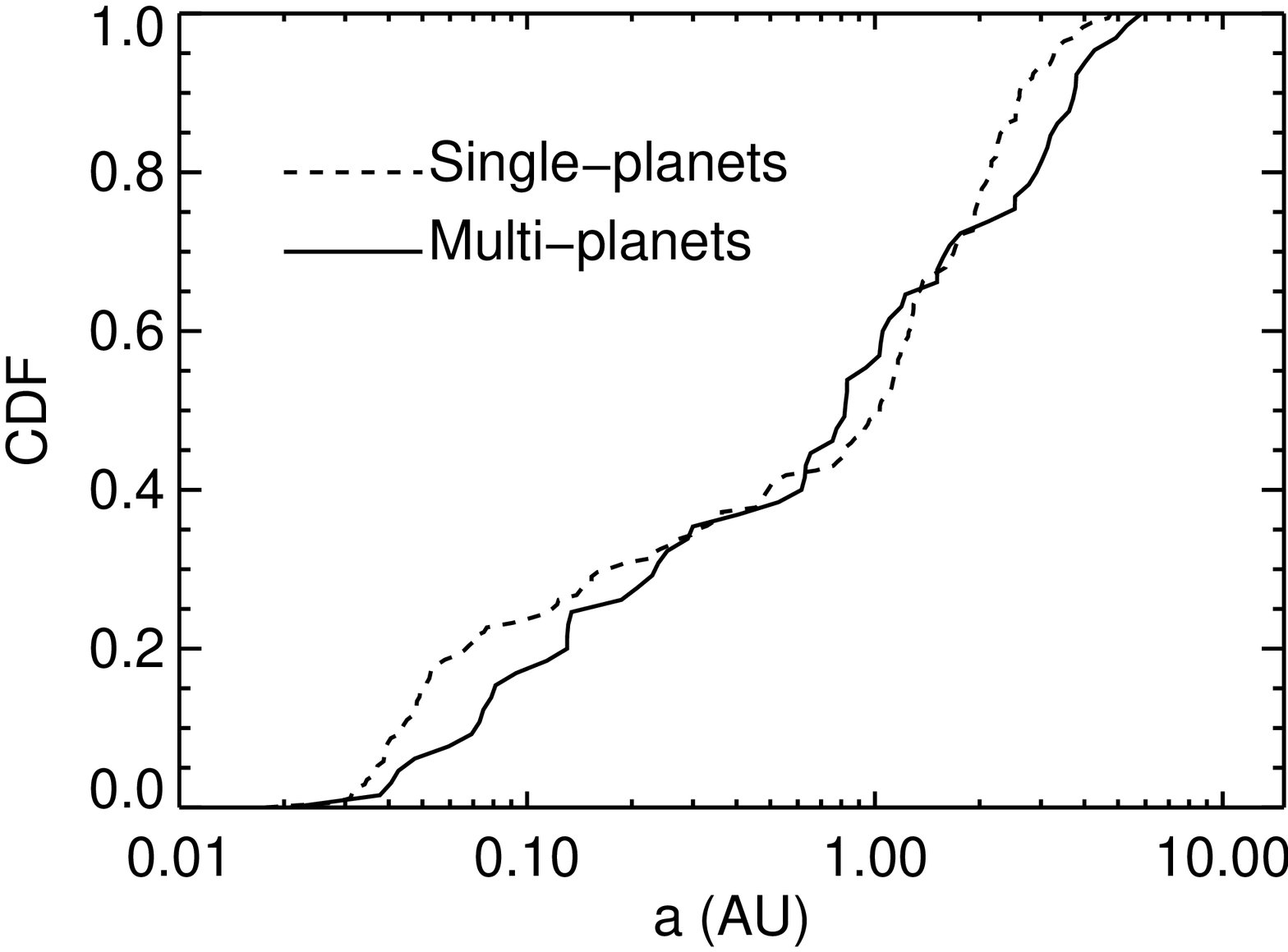}
  \caption{CDF of semimajor axis for known multiplanet systems (solid) and
    apparently single systems (dashed).}\label{KS_a}
\end{center}
\end{figure}

Figure~\ref{KS_a} shows the CDF for
    these two distributions.  A K-S test of
    the two distributions shows that we can reject the null hypothesis
    that the two samples were  drawn from the same distribution with 90\%
    confidence. 

This is somewhat surprising, since
    presumably many apparently single-planet systems have as-yet
    undetected low-mass planets and are, in fact, multiple systems.
    Presumably, then the difference is due to the presence of
    multiple {\it giant} planets within 5 AU.

    In particular, close-in planets ($a < 0.07$ AU) are not as common
    in multiplanet systems 
    as in apparently single-planet systems.  Among the \singleone\ single
    planets having $a < 1$ AU, \singleseven\ of them (\closefrac\%)
    have $a <0.07$ AU.  A 
    similar rate for the \multone\ planets in multiple systems with $a < 1$
    AU would yield 15 planets within $0.07$ AU, but only
    seven are known.  

    The 1 AU jump in the single-planet sample may relate to the
    location of the ``ice line'' beyond which ices can participate in
    planet formation in the protoplanetary nebula.  If this peak
    therefore represents planets which have not experienced
    significant migration, then the lack of such a peak in the
    multiple-planet distribution suggests that planets in multiple
    planet systems generally undergo more significant migration.

    Alternatively, the 1AU jump may be indicative of a pile-up of
    planets at the typical orbital distance within which migration
    becomes inefficient due to either the presence of the
    ice line or a ``dead zone'' near 1AU \citep[e.g.][]{Ida08}.   For instance, in a
    model where planet migration is driven by planetesimal scattering,
    the disk surface density of planetesimals is large enough to drive
    migration beyond the ice line, but not inside it.   



\subsection{\msini}
\label{msini}
\begin{figure} [ht!]
\begin{center}
  \plotone{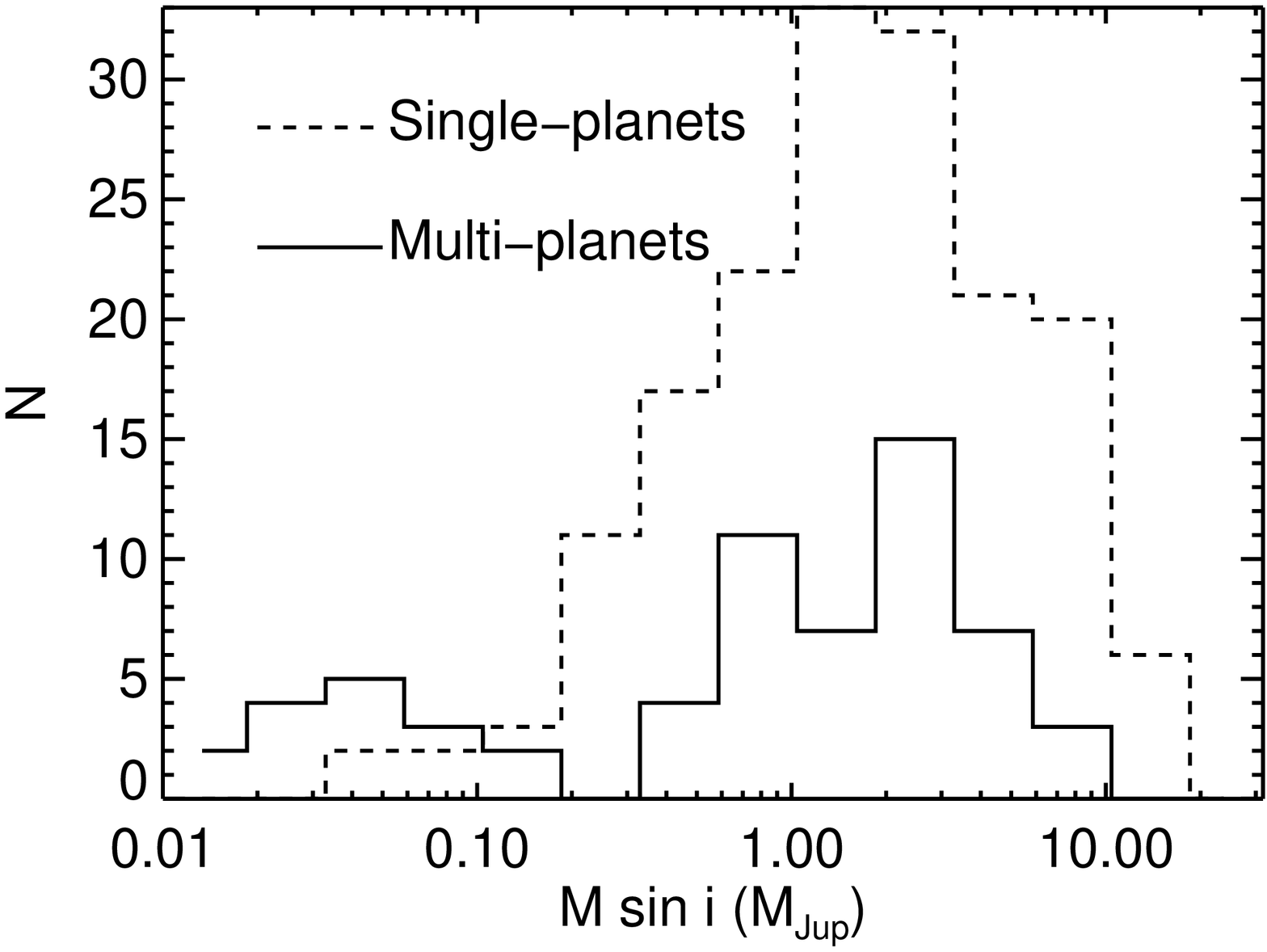}
  \caption{Distribution of \msini\ of exoplanets, with an upper limit
    of $13\mjup$ for known multiple-planet systems (solid) and apparently
    single systems (dashed)}\label{msini_hist_log}
\end{center}
\end{figure}

 The distribution of minimum masses (\msini) of planets in multiple
 planet systems are shown in Figure~\ref{msini_hist_log}, along with that for
 single planets. multiplanet systems exhibit an apparent
 overabundance of planets with $\msini$ between 0.01 to 0.2 \mjup, but
 this may be amplified by a selection effect.  When we find a planet
 around a star we tend to observe that star more frequently --- 
 making it more likely that we will find another planet that was not
 detectable beforehand.  This appears to explain the detection of very
 low mass planets around 55 Cnc, GJ 876, and $\mu$ Ara. In these
 systems, more massive planets were known in advance, and an
 especially large number of observations were made to refine their
 orbits. The lowest-mass planets were found in the course of these
 detailed observations.

\begin{figure} [ht!]
\begin{center}
  \plotone{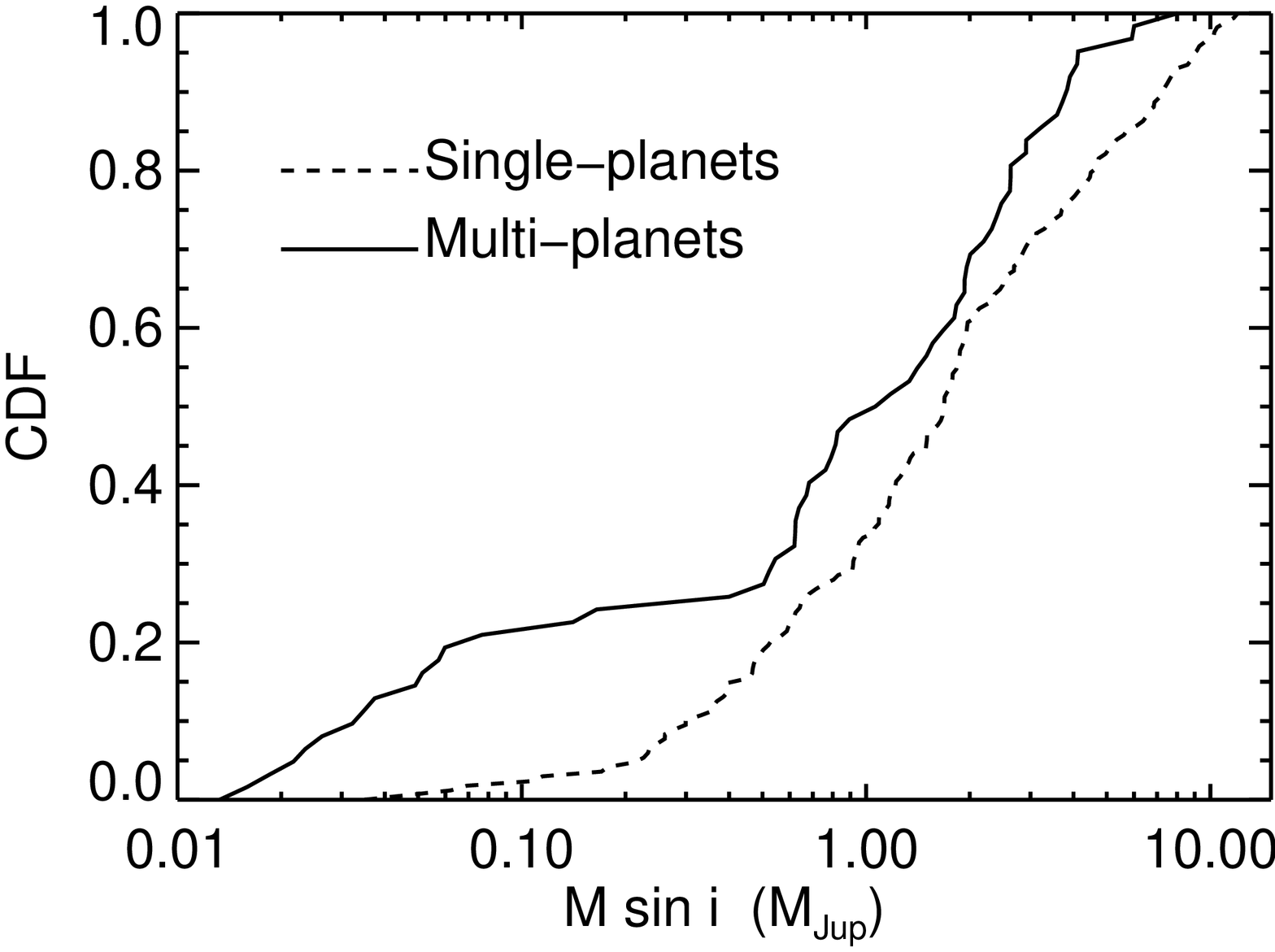}
  \caption{CDF of \msini for known multiplanet systems (solid) and
    apparently single systems (dashed) Note the enhanced frequency of
    known low mass planets in multiplanet systems.}\label{ks_msini_13}
\end{center}
\end{figure}

The \msini\ CDF of single-planet
systems (Figure~\ref{ks_msini_13}) is relatively featureless with the
logarithmic \msini\ axis, but the CDF of multiplanet systems deviates
markedly from its single-planet counterpart.  We calculate
$D^+ = 1.36$ and $D^- = -0.089$, so we find $p(D^+) = 1.8\%$, so we
can reject the null hypothesis that our samples are from a common
distribution with 
$> 95\%$ confidence. 

\subsection{Eccentricity}
\label{ecc}
\begin{figure} [ht!]
\begin{center}
  \plotone{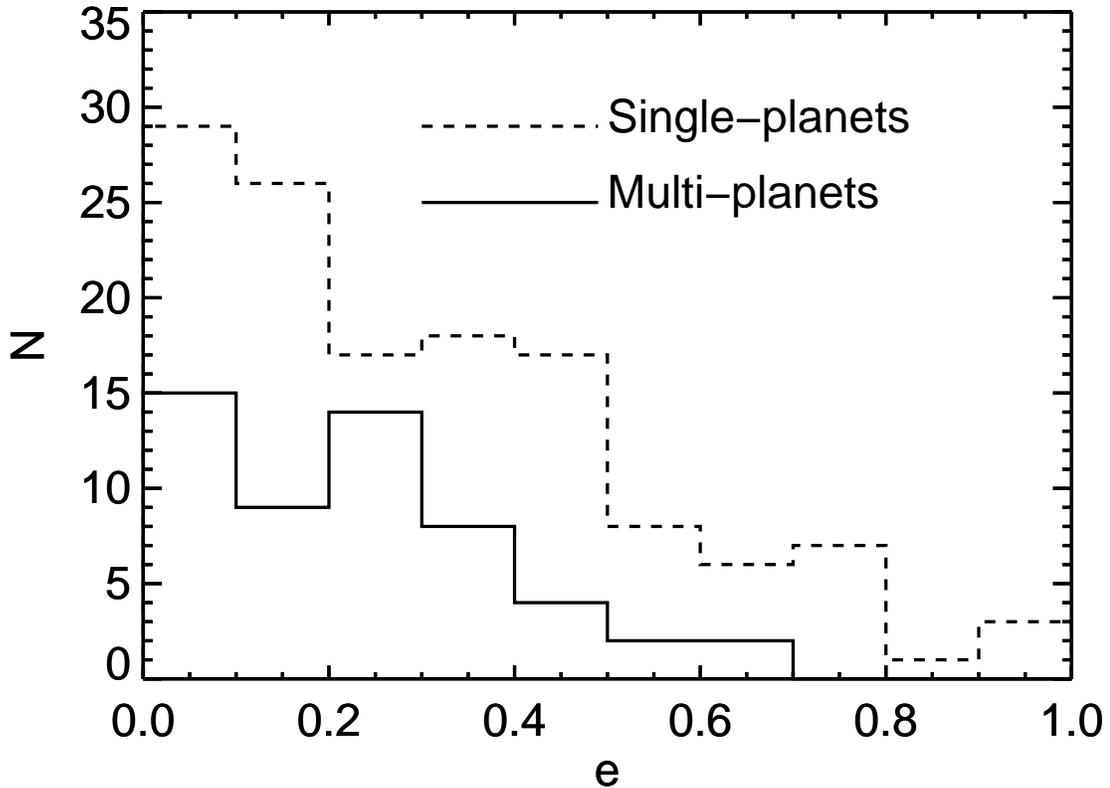}
  \caption{Distribution of eccentricities of exoplanets for known multiple-planet systems (solid) and apparently single-planet systems
    (dashed).  Note the high eccentricity orbits, $e>0.6$ occur
    predominantly in single planets.}\label{e_hist}
\end{center}
\end{figure}

The distribution of eccentricities for single and multiplanet systems
are shown in Figure~\ref{e_hist}. Note that we have excluded planets
with $a<0.1$ AU from consideration here to remove the effects of tidal
circularization on the analysis. Of the planets selected, both single
and multiplanet systems exhibit a wide range of eccentricities from
0.0 to 0.8.  The mean and
standard deviation for single planets are 0.30 and 0.24,
respectively, and 0.22 and 0.17 for multiplanet systems. Also, 11
single planets have eccentricities above 0.7 ($7\%$ of this sample),
but none of the multiplanet systems have an eccentricity above 0.7. 

\begin{figure} [ht!]
\begin{center}
  \plotone{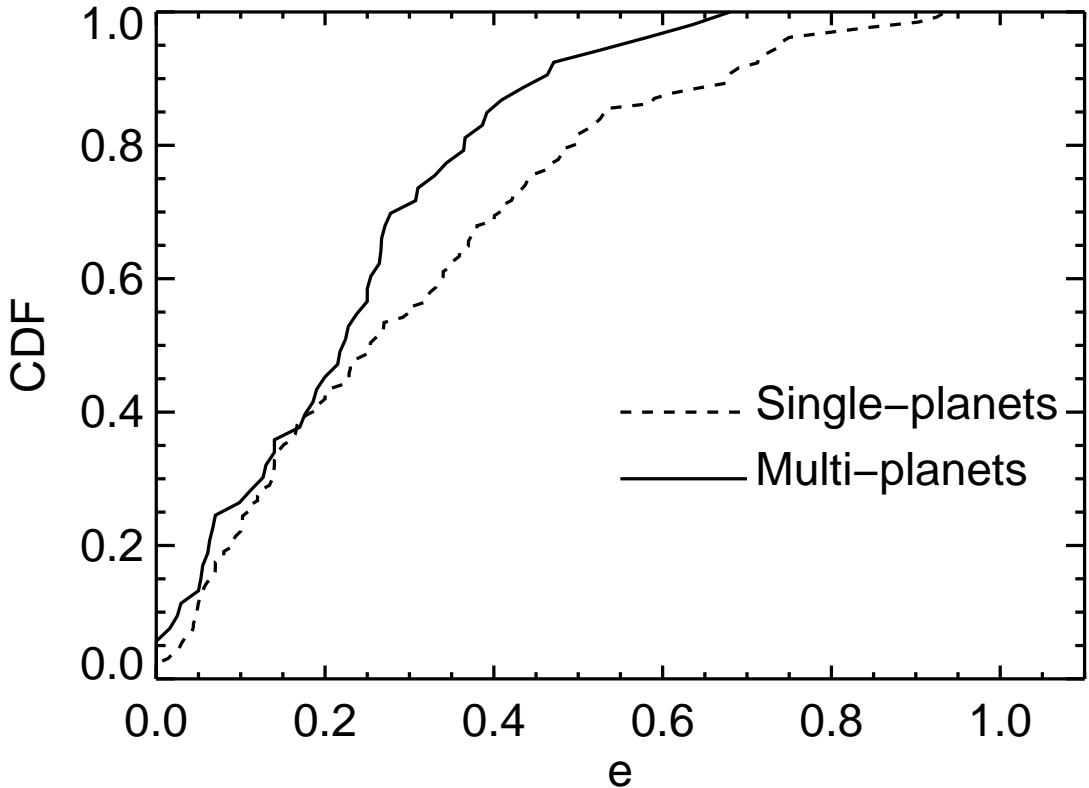}
  \caption{CDF of eccentricity for multiplanet systems (solid) and
    apparently single-planet systems (dashed). The tidally circularized hot
    jupiters have been removed.  Note that the highest eccentricities,
    $e>0.6$ occur predominantly in the single-planet
    systems.}\label{ks_e_exclude}
\end{center}
\end{figure}

To provide a numerical measure of the significance of the difference between the
two distributions, we computed the cumulative distribution function of
the eccentricities and applied a Monte Carlo K-S test for difference of
distribution (Figure~\ref{ks_e_exclude}). The eccentricity CDF for
multiplanet systems is greater than the eccentricity CDF for single-planet
systems at any given eccentricity, suggesting that multiplanets systems have systematically
lower eccentricity. The K-S statistics are $D^+ = 1.34$ and $D^- =
0.265$, and  $p(D^+) = < 1\%$, so we reject the null hypothesis that
our samples are from a common distribution with over 99\% confidence. Thus
it appears that the known multiplanet systems have systematically
lower eccentricities.  Selection effects in eccentricity do not
significantly affect the delectability of planets for $e<0.7$ \citet{Cumming08}.

It is surprising that multiplanet systems have lower orbital
eccentricities, as mutual interactions between giant planets might be
expected to excite eccentricities.   The lack of very high
eccentricities in multiplanet systems may be 
partially explained by the additional constraint in multiplanet
systems of orbital stability, which favors low-eccentricity orbits.
Conversely, some single-planet systems may exhibit high 
eccentricities as a result of a series of ejections 
of former members from the system.  Both factors can be at play
simultaneously: \cite{Ford05} explain the observed eccentricities of
the planets in the $\upsilon$ Andromedae system as the end result of
the ejection of a hypothesized fourth planet from the system.

\subsection{Metallicity}

\citet{Fischer05c}, using uniformly calculated metallicities from
\citet{SPOCS}, found that the 14 multiple-planet systems then known
had a somewhat higher average 
metallicity than single-planet systems (+0.18 versus +0.14), suggesting
that metallicity traces multiplicity in planets even more strongly
than it traces single-planet occurrence \citep[see also][]{Santos01b}.
They suggested that further discoveries of multiplanet systems could
confirm this trend.  Using the metallicities compiled\footnote{In most cases, the ultimate origin of these metallicites is
\citet{SPOCS}} in
the CNE, we find that single and multiplanet
systems have mean \feh\ values of +0.10 and +0.10, respectively,
although the median values are +0.15 and +0.18, still showing some
evidence of the disparity.

If we include systems showing long-term RV trends (not including known
binaries) among the multiple-planet systems, however, the difference
becomes slightly stronger.  Such systems have a mean $\feh$ value of +0.20,
bringing the average for apparent multiple systems overall up to
$\feh = +0.15$.  

\begin{figure} [ht!]
\begin{center}
  \plotone{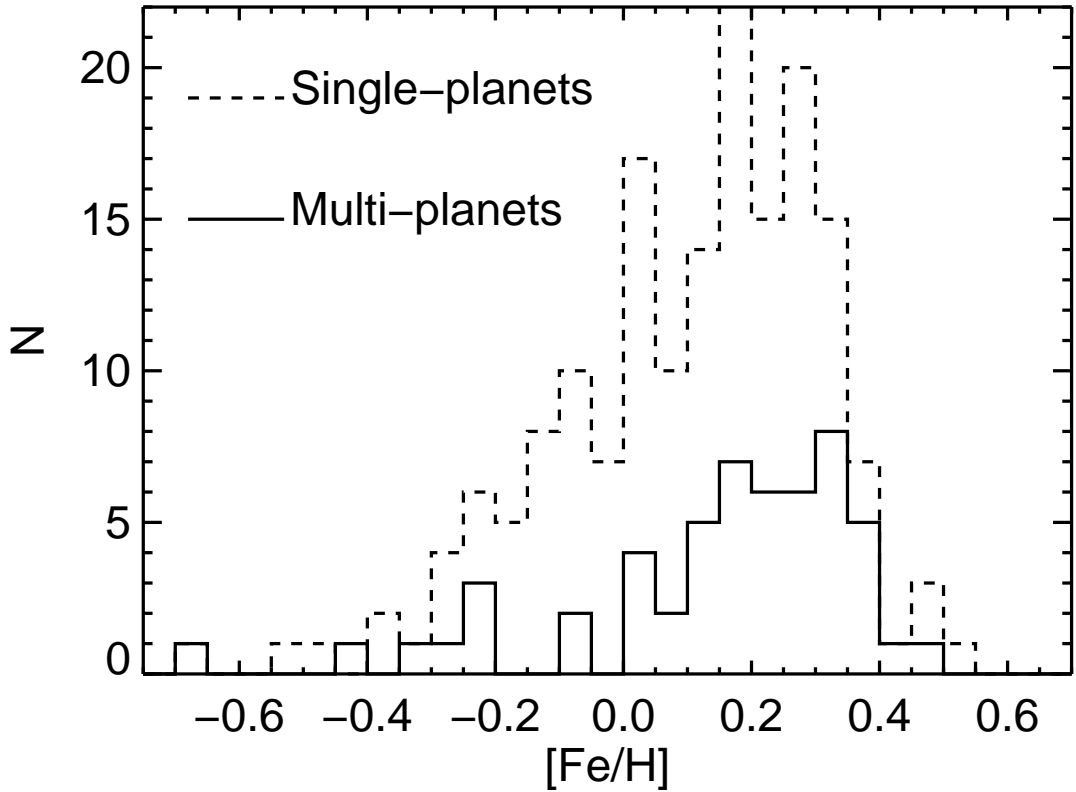}
  \caption{Distribution of \feh\ for exoplanet bearing stars harboring
    likely multiple-planet systems (including single-planet systems
    with long-term RV trends;  solid line) and apparently single systems
    (dashed).  The median \feh\ for known multiplanet systems, +0.18, is
    higher than that for the single-planet systems, +0.14. The
    multiplanet system HD 155358 has the lowest \feh\ of any system,
    with $\feh =-0.68$.}\label{feoh_hist}
\end{center}
\end{figure}

\begin{figure} [ht!]
\begin{center}
  \plotone{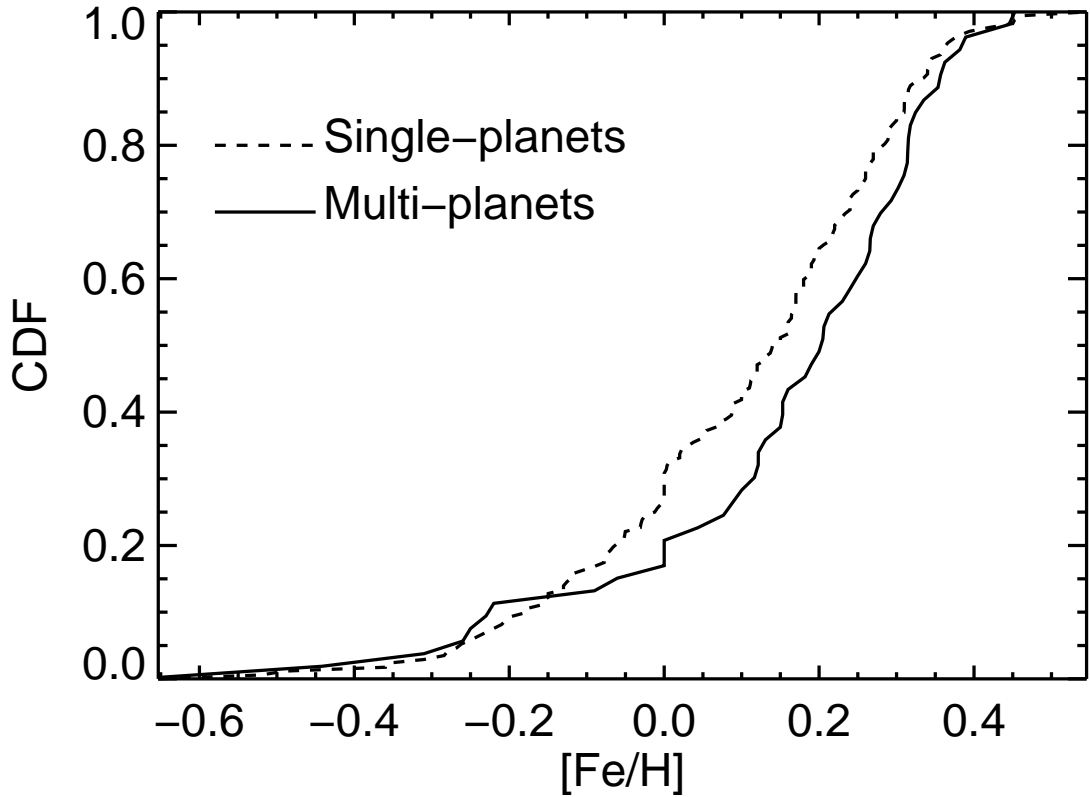}
  \caption{CDF of \feh\ for known multiple-planet systems (solid) and
    apparently single systems (dashed).  The CDF shows that the
    metallicity of the multiplanet systems is consistently higher
    than that of the single-planet systems.}\label{ks_feoh}
\end{center}
\end{figure}

We plot the two distributions in Fig.~\ref{feoh_hist}.  To test
whether these two distributions differ significantly, we have
performed a K-S test, as shown in Figure~\ref{ks_feoh}.  A K-S test
rejects, with 97\% confidence, the null hypothesis that the apparently
multiplanet systems (including systems with trends) have
metallicities drawn from the same distribution as the single-planet
systems.  It appears that metallicity 
traces not only planet occurrence rate, but multiplicity among
planet-bearing stars, as well.

\subsection{Stellar Mass}

\begin{figure} [ht!]
\begin{center}
  \plotone{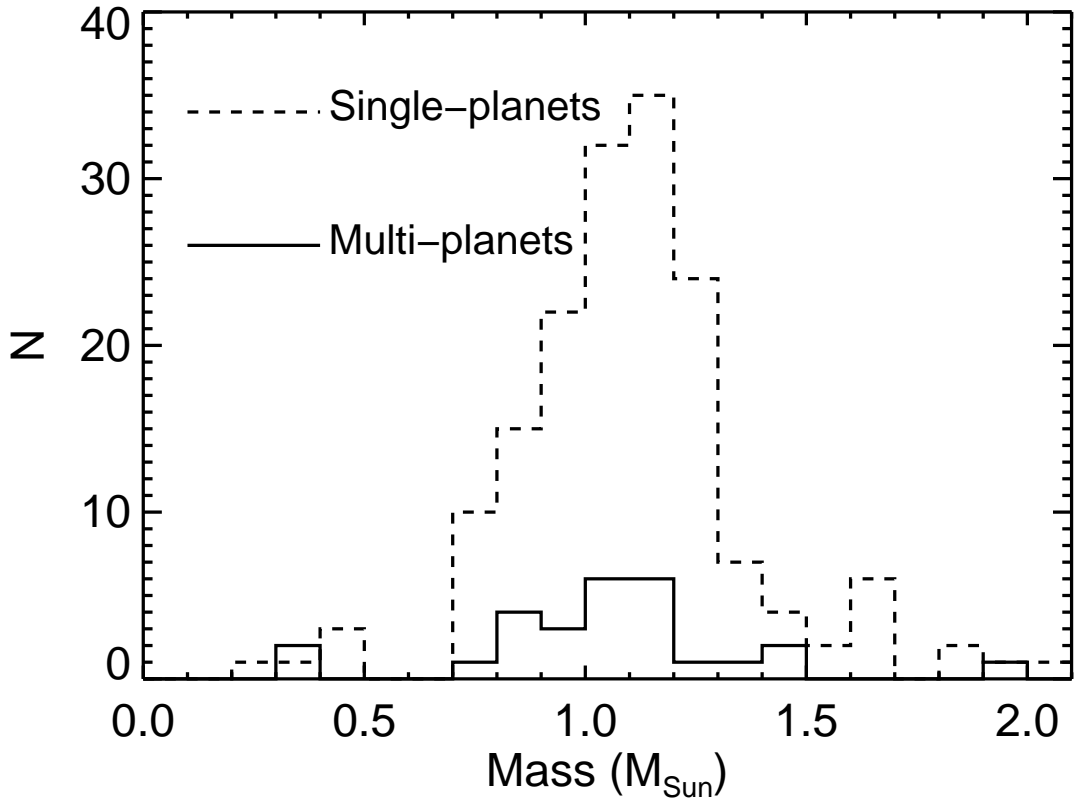}
  \caption{Mass distribution of exoplanet-bearing stars for known multiple
    planet systems (solid) and apparently single systems
    (dashed). There is no significant difference between the stellar
    masses of the single and multiplanet systems.}\label{starmass_hist}
\end{center}
\end{figure}

The distributions of stellar mass for stars hosting single planets and
stars hosting multiplanet systems are shown in
Figure~\ref{starmass_hist}. multiplanet systems have a mean stellar mass of 1.1 \Msol\
and single-planet systems have a mean mass of 1.13 \Msol. K-S tests of
the histogram in Figure~\ref{starmass_hist} and of the corresponding CDF show
no significant difference between the stellar masses of single and
multiplanet systems.  Unlike metallicity, stellar mass does not seem
to strongly trace multiplicity.

\subsection{Multiplicity vs.\ Stellar Mass}
\label{Mfreq}
Although there is no strong stellar-mass--multiplicity relation, there
may be an emerging trend regarding the M dwarfs.  Among the eight known M-dwarf
exoplanet hosts, two are well characterized multiple-planet
systems (GJ 876 and GJ 581) and two others (GJ 317, and GJ 849) show a
trend in the single-planet velocity residuals.  If we attribute these trends to planets and not 
undetected stellar or brown dwarf companions, then the total, true
multiple-planet rate is at least 50\%, which is higher
than the \totfrac\ similarly calculated for the entire CNE sample in
\S\ref{multiplicity}.  It appears that planets around M dwarfs may
be found preferentially in multiplanet 
systems rather than singly.  This trend, if it is not simply Poisson
noise and should hold up as more M-dwarf systems are discovered, is
especially surprising since M dwarfs have a lower-than-average planet
occurrence rate \citep{Endl06b,Johnson07b,Mayor08}. 

There may be subtle observational selection effects at work here,
however.  If M dwarfs in general have lower mass planets than F--K
stars (and thus require more observations before publication), then we
may simply be seeing the already-documented increase in planet
occurrence rate amongst low-mass {\it planets}.  The ongoing RV M
dwarf surveys will improve the statistics of these systems, which
should help illuminate if the effect is due to selection effects,
small numbers, or astrophysics. 

\subsection{Eccentricity vs.\ \msini}
\label{eccmsinidist}
\begin{figure} [ht!]
\begin{center}
  \plotone{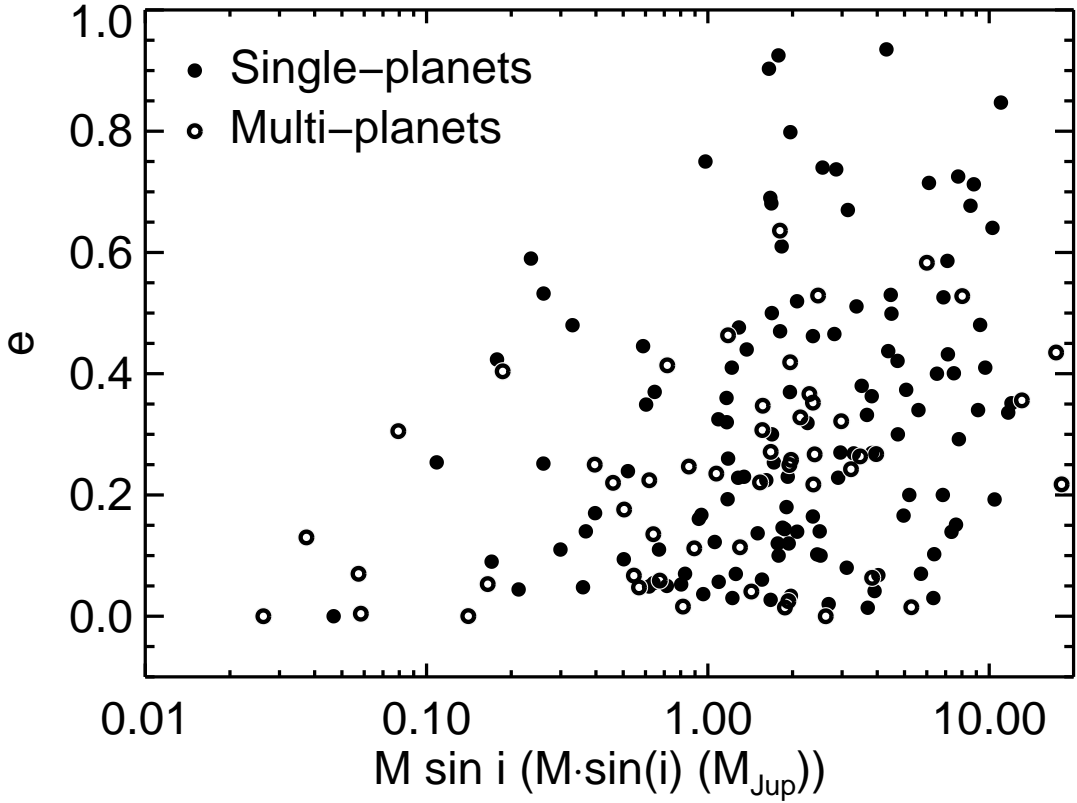}
  \caption{Plot of eccentricity vs.\ \msini\ for planets with $a>0.1$
    AU, to avoid contamination from tidal circularization. Filled
    circles are single-planet systems, and open circles represent
    multiplanet systems.  There is a slight sense of increase in the
    upper envelope in the range $0.05-1.0$ \mjup}\label{chart_ecc_vs_msini}
\end{center}
\end{figure}

Consider the plot of eccentricity vs.\ \msini\ of planets with $a>0.1$
AU, including both multi and single-planet systems
(Figure~\ref{chart_ecc_vs_msini}). We have excluded planets within 0.1 AU to remove
the effects of tidal circularization on our analysis. Planets with
minimum mass below 1.0 \mjup\ have a mean eccentricity of 0.19, while planets
above this threshold have a mean eccentricity of 0.34. 

\begin{figure} [ht!]
\begin{center}
  \plotone{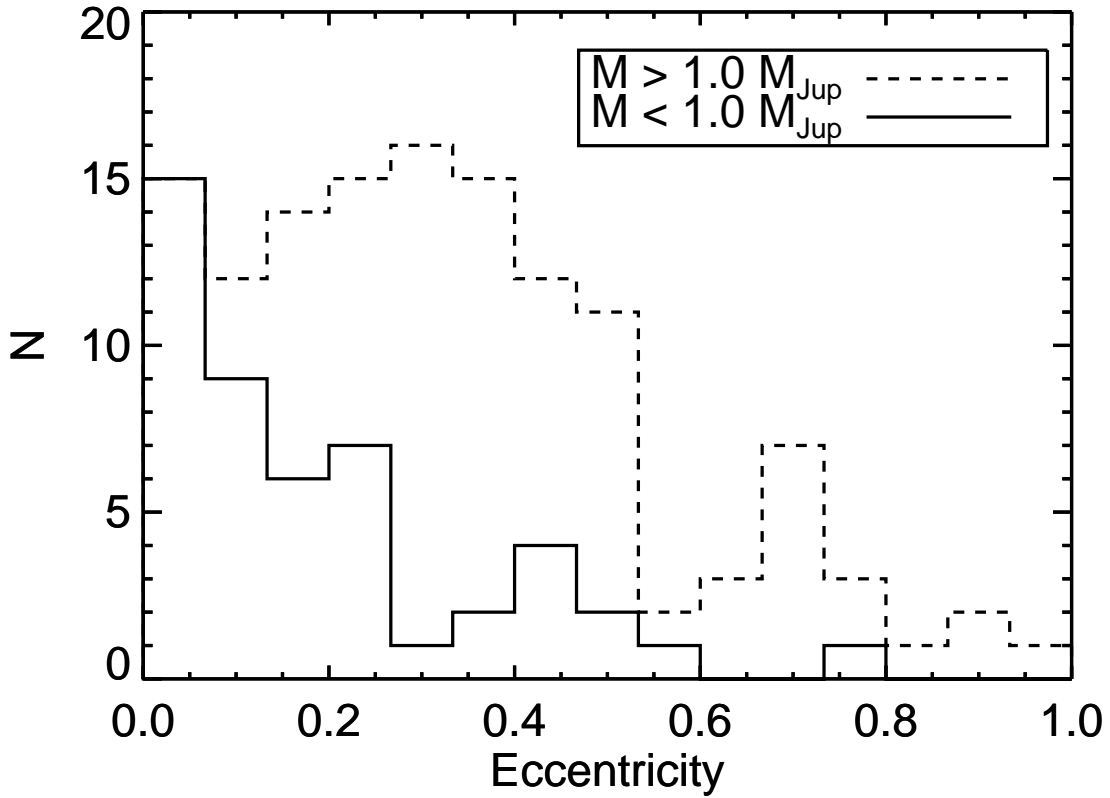}
  \caption{Distribution of eccentricities of exoplanets with $\msini <
    1.0 \mjup$ (solid) $\msini > 1.0 \mjup$ (dashed).  The tidally
    circularized hot jupiters have been removed. Note that the
    eccentricity of planets of minimum mass $ < 1.0 \mjup$ peaks at
    eccentricity $< 0.2$, while the eccentricities $e$ of planets of
    minimum mass $ > 1.0 \mjup$ are distributed broadly from
    $0.0<e<0.6$.}\label{eccmass}
\end{center}
\end{figure}

\begin{figure} [ht!]
\begin{center}
  \plotone{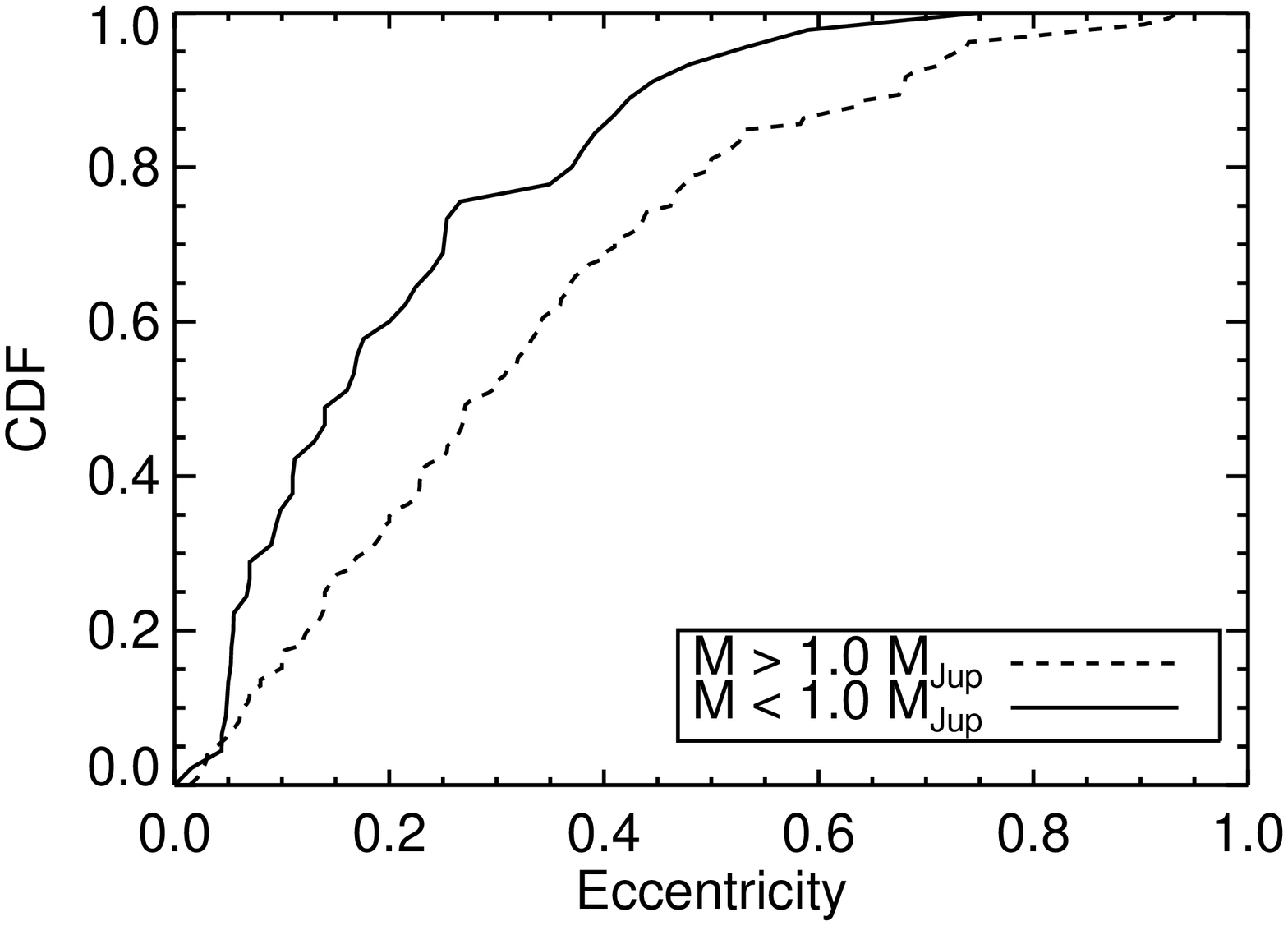}
  \caption{CDF of eccentricity for planets of minimum mass $ < 1.0
    \mjup$ (solid line) and minimum mass $ > 1.0 \mjup$ (dashed line)
    The tidally circularized hot jupiters have been removed.  Note
    that the CDF for the light planets rises steeply from eccentricity
    0.0 to 0.2, while the CDF for the heavy planets rises with a
    nearly uniform slope from eccentricity 0.0 to 0.6. }\label{KS_ecc_jupsplit}
\end{center}
\end{figure}

Figure~\ref{eccmass} shows the strong dichotomy between the
eccentricity distributions of super- and sub-Jupiters:  the 
eccentricity of sub-Jupiters peaks at $e < 0.1$, while the
eccentricity of super-Jupiters is distributed broadly from
$0.0<e<0.6$.  Figure~\ref{KS_ecc_jupsplit} shows that a K-S statistic
bears out this difference: for these two populations $D^+ = 2.06$ and
$D^- = -.034$, yielding $p(D^+) < 0.1\%$, so we reject the null
hypothesis that the samples are drawn from the same distribution, and
conclude that the apparent difference is not
a chance result of small-number statistics. 

This correlation between eccentricity and minimum planet mass provides a
valuable clue about the origins of eccentricities.  One
possibility is that conditions that encourage the formation of planets
with $\msini > 1.0 \mjup$ may contribute to greater
eccentricity pumping as well.  If so, then these would need to be
strong effects as more massive planets are relatively more difficult
to perturb.  Also possibly relevant is the work of
\citet{Goldreich03}, who describe a mechanism for modest eccentricity
pumping of a planet through interactions with a protoplanetary disk
that should be more efficient for more massive planets.  This
mechanism could create a ``seed'' eccentricity preferentially in
massive planets, which would grow through planet-planet interactions.
Alternatively, it is possible that significant eccentricities are the norm
for all systems shortly after the dissipation of the protoplanetary
disk triggers strong planet-planet interactions
\citep[e.g.][]{Ford06}, and that circularization through dynamical
friction with planetesimals is more efficient for planets with $M <
1.0 \mjup$.  Another possibility, that this is the signature of
eccentricity pumping by planet-planet scattering, is discussed in
\S\ref{theta}.  

\subsection{Eccentricity vs.\ Metallicity}

\begin{figure} [ht!]
\begin{center}
  \plotone{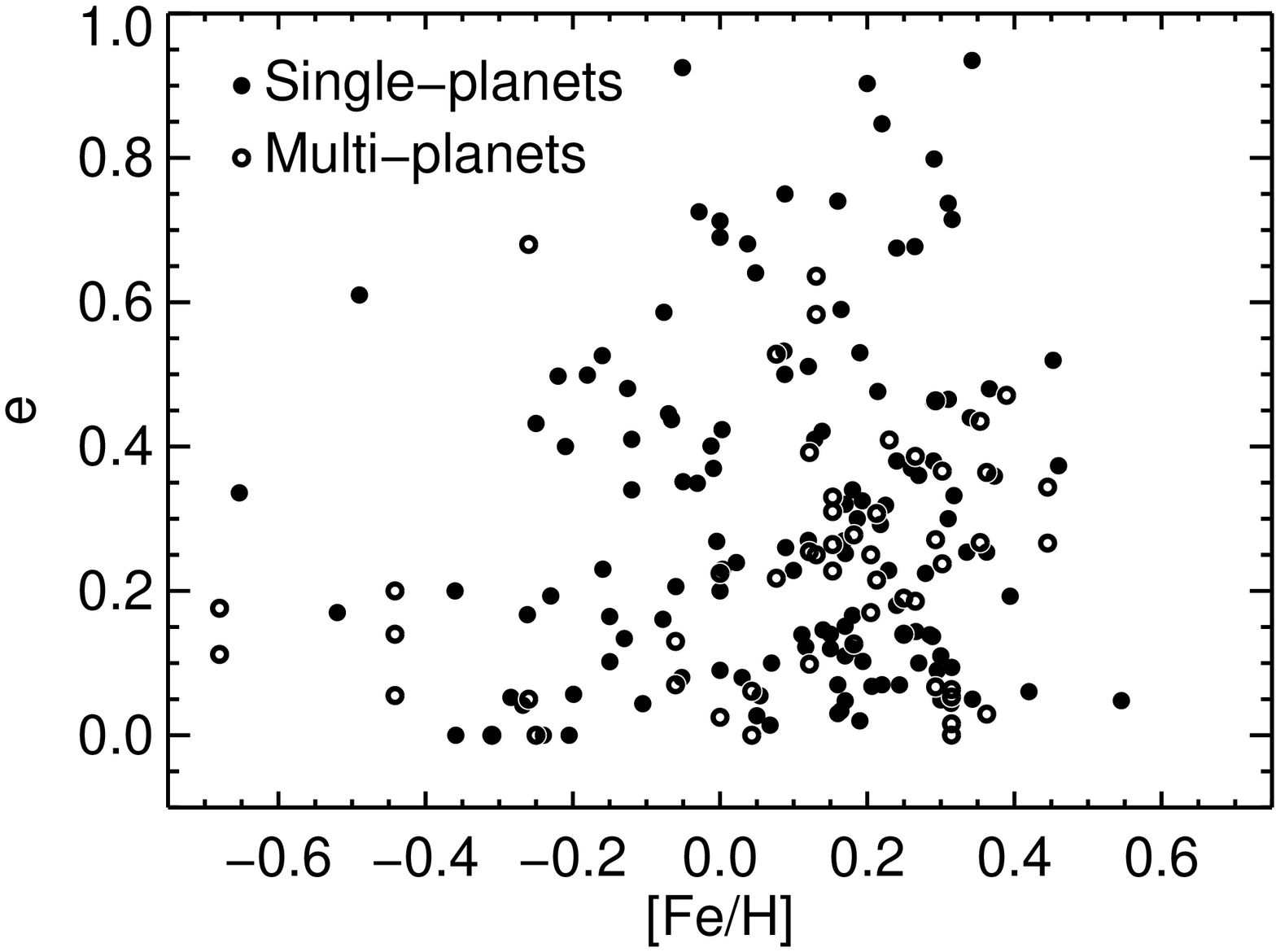}
 \caption{Plot of eccentricity vs.\ \feh for planets with $a>0.1$ AU,
   to remove bias caused by tidal circularization. Filled circles are
   single-planet systems, and open circles represent multiplanet
   systems. No correlation is apparent.}\label{chart_ecc_vs_feoh}
\end{center}
\end{figure}

We plot eccentricity vs.\ metallicity (\feh) for planets with $a>0.1$
AU, including both multi and single planet systems
(Figure~\ref{chart_ecc_vs_feoh}).  There appears to be no significant correlation
between \feh\ and eccentricities. While the systems with $\feh < -0.2$
appear to have lower eccentricities than those with $\feh > -0.2$,
this may be due to the small number of such systems.

\subsection{Eccentricity vs.\ Semimajor Axis}
\begin{figure} [ht!]
\begin{center}
  \plotone{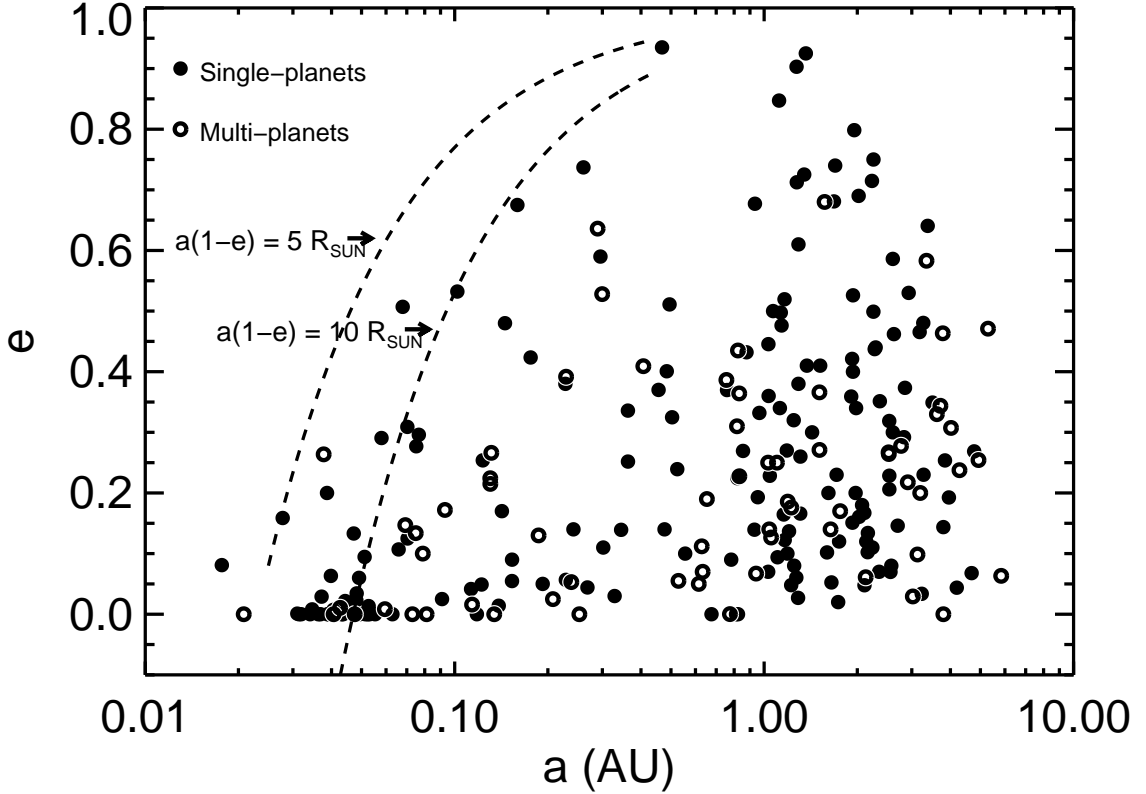}
  \caption{Plot of eccentricity vs.\ semimajor axis.  Filled circles
    are single-planet systems, and open circles represent multiplanet
    systems. Note the increasing upper envelope from 0.01 to 0.5
    AU. We have drawn the curves representing periastron passage
    distances of $a(1-e) = 5 R_{\odot}$ and $a(1-e) =
    10 R_{\odot}$ with dashed lines for reference.}\label{chart_ecc_vs_a}
\end{center}
\end{figure}

Figure~\ref{chart_ecc_vs_a} shows the eccentricity vs.\ semimajor
axis for all planetary systems. There is a clear paucity of planets
with small $a$ and large $e$, as expected from the effects of tidal
circularization \citep{Rasio96b,Ford99}.

Planets from 0.5 AU to about 3 AU have the widest range of
eccentricities, and beyond 3 AU there may be a paucity of
planets with $e>$0.6. This could easily be due to observational biases:
planets beyond 3 AU have such long orbital periods that only a small
number of orbits have necessarily been observed.  In cases with fewer
than two complete orbits, the radial velocity signature of any additional exterior
planets with orbital periods longer than the span of the observations
can sometimes be absorbed into the eccentricity term of the orbital
solution of the inner planet.  Further, since an $e > 0.6$ planet
spends a small fraction of its orbit near periastron, where its
velocity signal is largest, such a planet may not reveal itself until
it has completed nearly an entire orbit. The discovery of more long-period
planets and the observation of more complete orbits will help reveal
the true eccentricity distribution of planets with $a>$ 3 AU.

\subsection{Ratio of Escape Speeds vs.\ Eccentricity}
\label{theta}
\begin{figure}
\begin{center}
\plotone{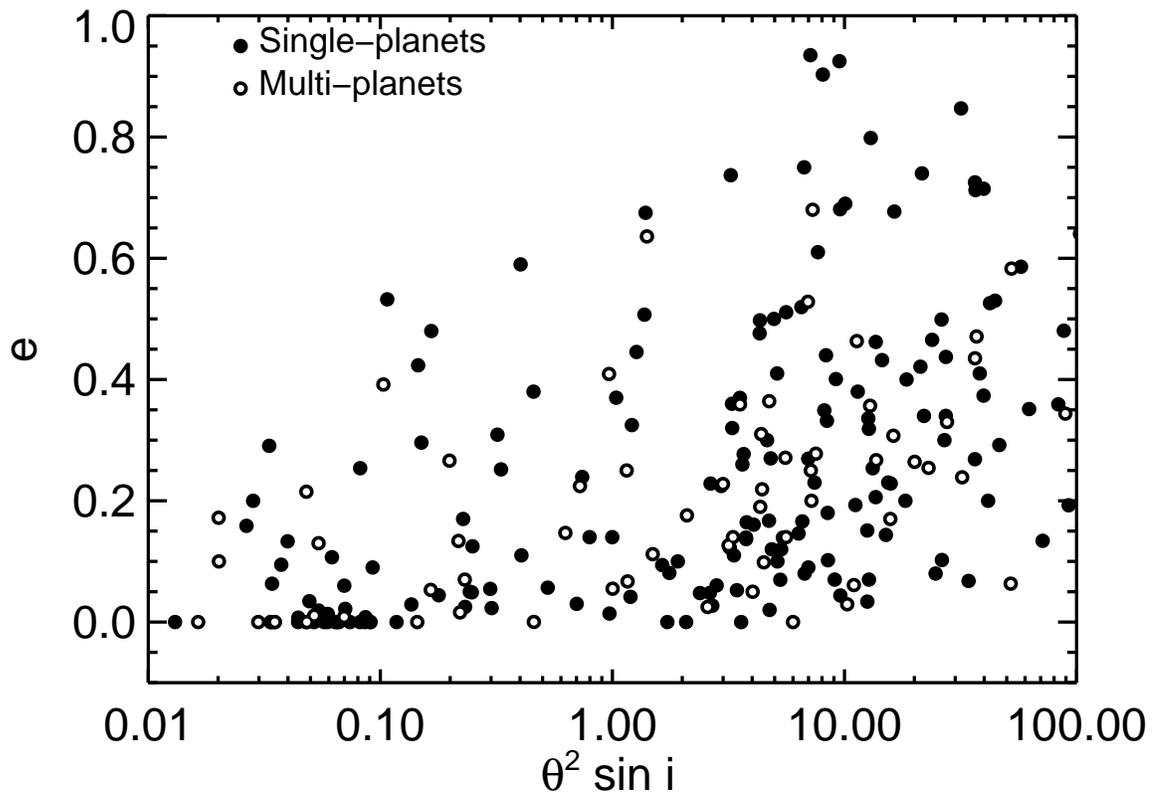}
\caption{Eccentricity vs. the ratio of the escape velocity from the
  planet to the escape velocity of the star ($\theta^2$). \label{Safronov}}
\end{center}
\end{figure}

Some theoretical models of the evolution of eccentricity through planet-planet
scattering focus on the parameter $\theta$, defined\footnote{Some
  authors refer to this quantity as the Safronov number.} as the ratio of
the escape speed from the planet to that of the planetary system \citep{Ford08}:
\begin{equation}
  \theta^2 \equiv \left(\frac{GM}{R_p}\right)\left(\frac{r}{GM_*}\right)
\end{equation}
where $M$ is the mass of the planet, $R_p$ its radius, $r$ its orbital
distance, and $M_*$ is the
mass of the host star.  Because we do not have exact masses or radii
for many of the exoplanets in our sample, and because the following
analysis is rather insensitive to the exact values of those
quantities, here we approximate
\begin{equation}
  \theta^2\sin{i} = 10 \left(\frac{\msini}{\mjup}\right) \left(\frac{\Msol}{M_*}\right)\left(\frac{R_{\mbox{Jup}}}{R_p}\right)\left(\frac{a\left(1+e\right)}{5\mbox{AU}}\right)
\end{equation}
and crudely estimate exoplanetary radii from the assumption that exoplanets
below the mass of Jupiter have similar mean densities:
\begin{equation}
  \left(\frac{R_p}{R_{\rm  Jup}}\right)^3 = \left\{ 
\begin{array}{cc}
\frac{\msini}{M_{\rm Jup}} & \msini < M_{\rm Jup}\\
 1  & \msini > M_{\rm Jup}
\end{array}
\right.
\end{equation}
When $\theta > 1$, a planet can efficiently eject bodies during close
encounters, and when $\theta < 1$ collisions are more frequent.
Figure~\ref{Safronov} shows the distribution of $\theta^2\sin i$
vs.\ eccentricity for single planets and those in multiple-planet
systems.  Consistent with \citet{Ford08}, Figure~\ref{Safronov} shows
that planets that scatter planets and planetesimals efficiently
(i.e. those with $\theta >> 1$) have a wider range of eccentricities
than inefficient scatterers.  This is consistent with the hypothesis
that planet-planet and planet-planetesimal scattering is a dominant
mechanism for the excitation of exoplanetary eccentricites.

Because $\theta^2 \propto \msini$, these results are consistent with
\S\ref{eccmsinidist} where we showed that the eccentricity
distribution of planets with $\msini < 1 \mjup$ is peaked at very
small $e$:  the eccentricity dichotomy between high- and low-mass
planets may be related to the ability of high-mass planets to more
efficiently scatter planets and planetesimals.  

\begin{figure}
\begin{center}
\plotone{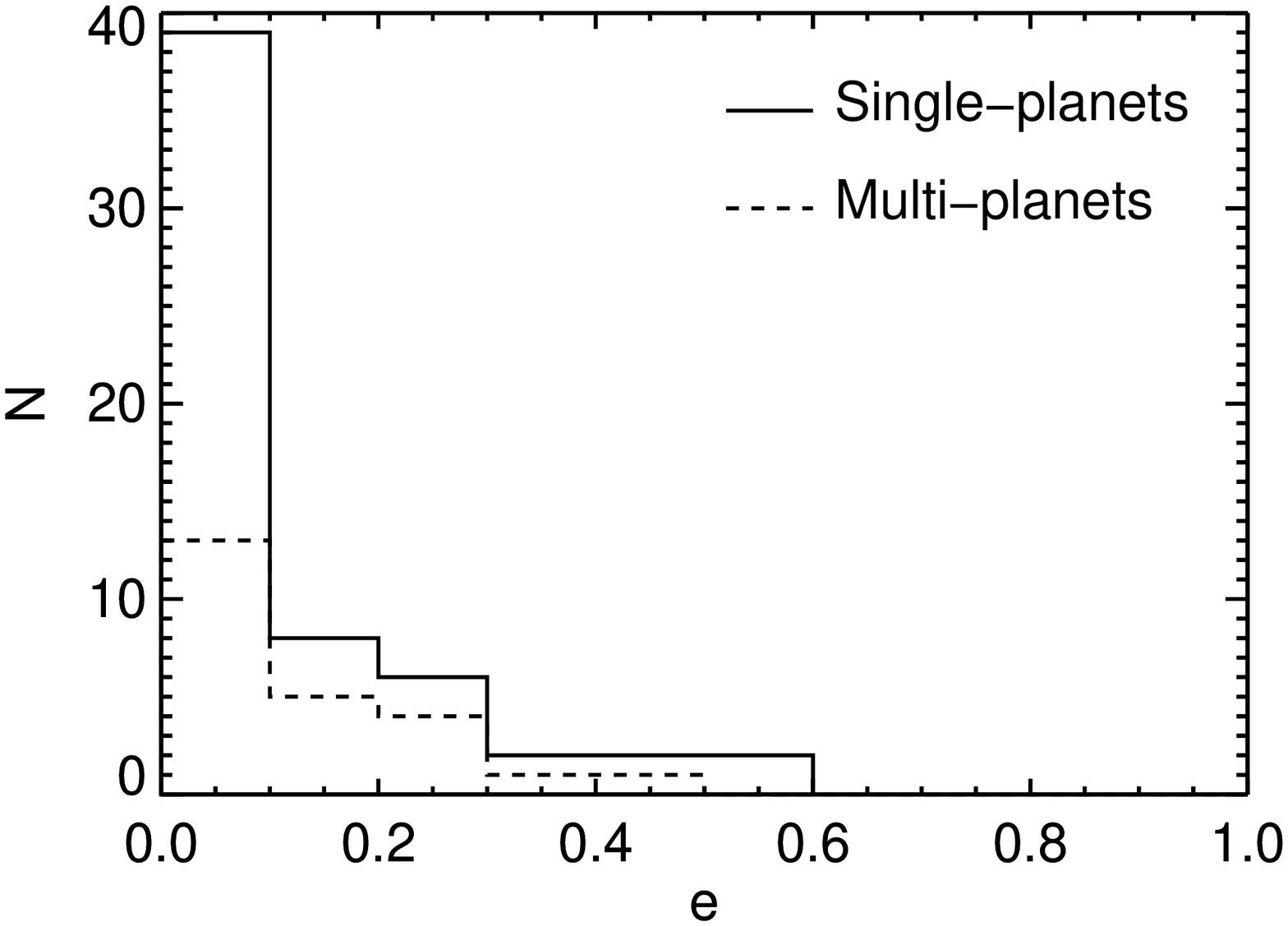}
\caption{Distribution of eccentricites for single-
  and multiple-planet systems with $\theta < 1$. \label{Safronov_hist_1}}
\end{center}
\end{figure}

\begin{figure}
\begin{center}
\plotone{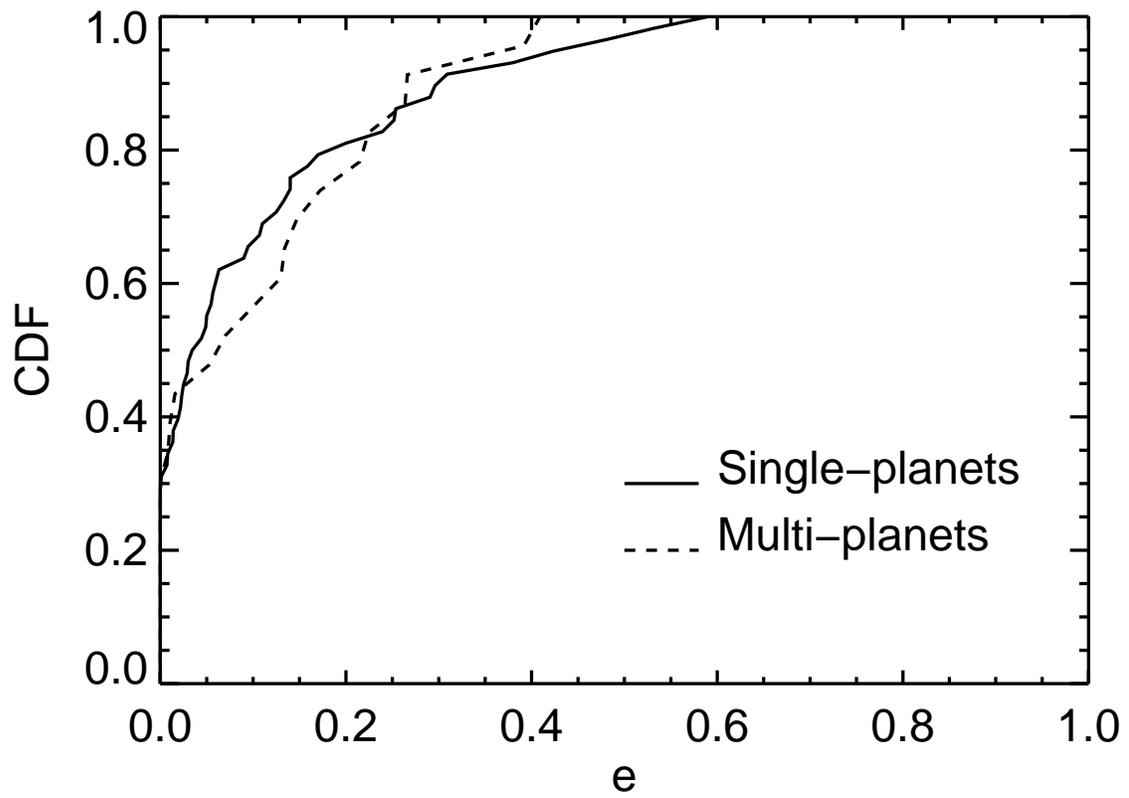}
\caption{CDF for eccentricity for $\theta < 1$ planets in single and
  multiplanet systems.  The eccentricity distributions for these
  inefficient scatterers do not differ significantly. \label{KS_Safronov_1}}
\end{center}
\end{figure}

\begin{figure}
\begin{center}
\plotone{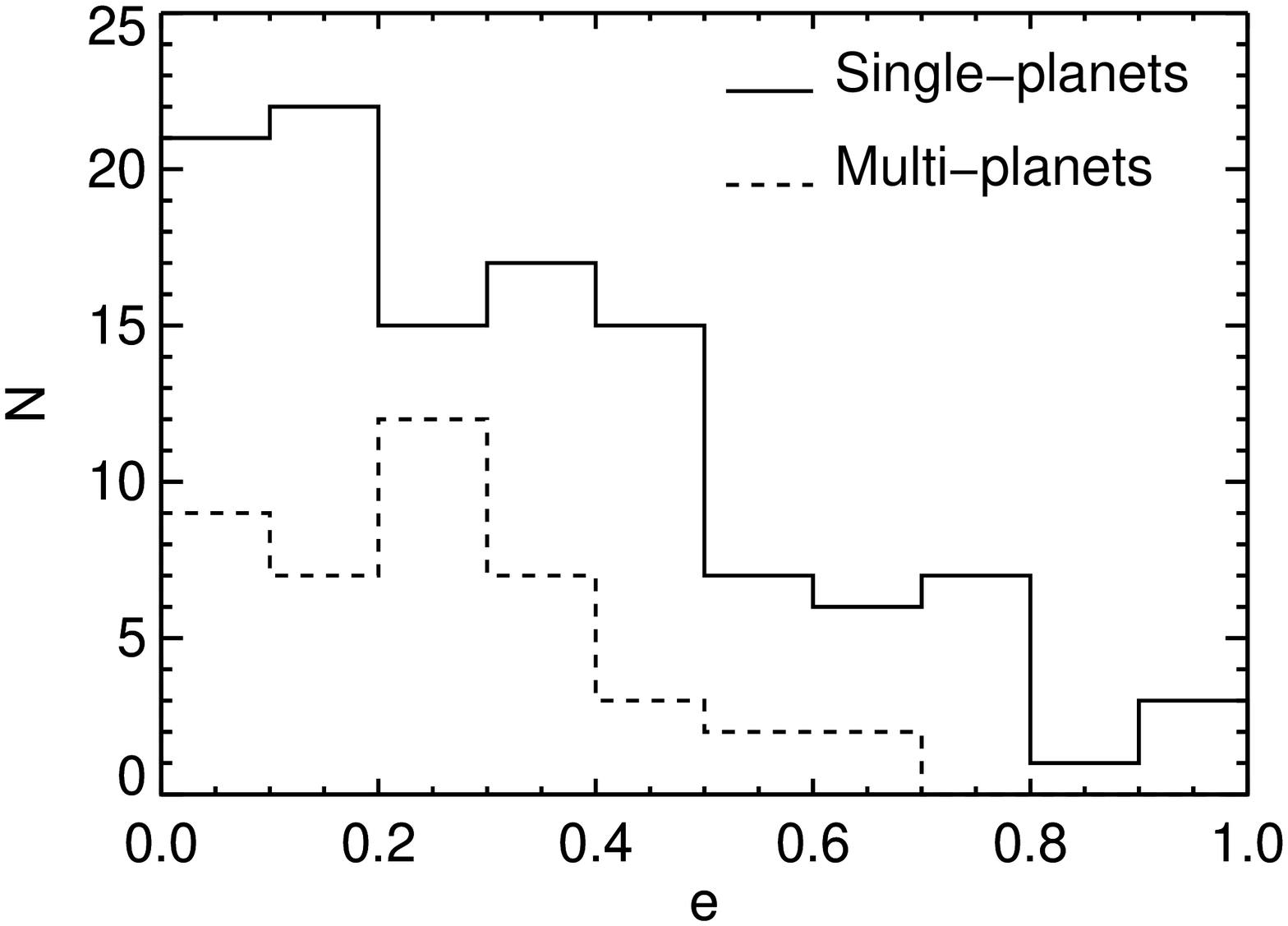}
\caption{Distribution of eccentricites for single
  and multiple-planet systems with $\theta > 1$. \label{Safronov_hist_2}}
\end{center}
\end{figure}

\begin{figure}
\begin{center}
\plotone{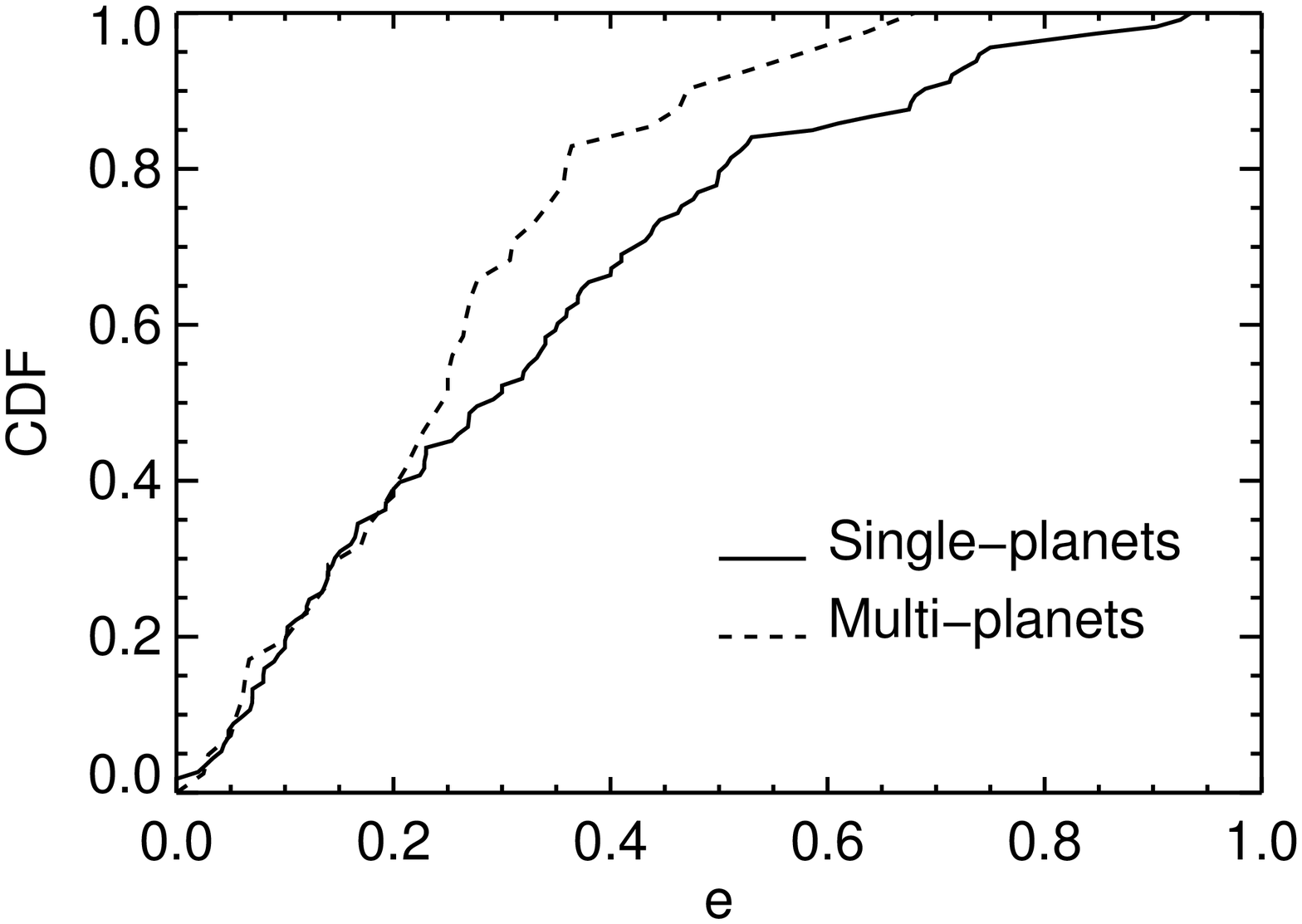}
\caption{CDF for eccentricity for $\theta > 1$ planets in single and
  multiplanet systems.  The eccentricity distributions for these
  efficient scatterers differ significantly.  It appears that
  efficient scatterers in multiplanet systems display larger
  eccentricities than those in single-planet systems. \label{KS_Safronov_2}}
\end{center}
\end{figure}

If we divide the sample into those planets with $\theta < 1$ and those
with $\theta >> 1$, we can examine whether the eccentricity
distributions of efficient scatterers in single and multiplanet
systems differ from one another.  Figures~\ref{Safronov_hist_1} \&
\ref{KS_Safronov_1} show that inefficient scaterers in single and
multiplanet systems have very similar eccentricity distributions.
Interestingly, Figures~\ref{Safronov_hist_2} \&
\ref{KS_Safronov_2} show that efficient scatterers (those for which
$\theta > 1$) in multiplanet systems appear to have higher
eccentricities than those in single-planet systems.  A K-S test for these
two populations gives $D^+ = 1.36$ and $D^- =0.11$, yielding $p(D^+)
= 2\%$,  showing that it is unlikely that these samples are drawn from
 the same distributions.  This is despite the overall tendency for
 multiplanet systems to have lower eccentricities (see \S\ref{ecc}.)


\subsection{Metallicity and Stellar Mass vs.\ \msini}

\begin{figure} [ht!]
\begin{center}
  \plotone{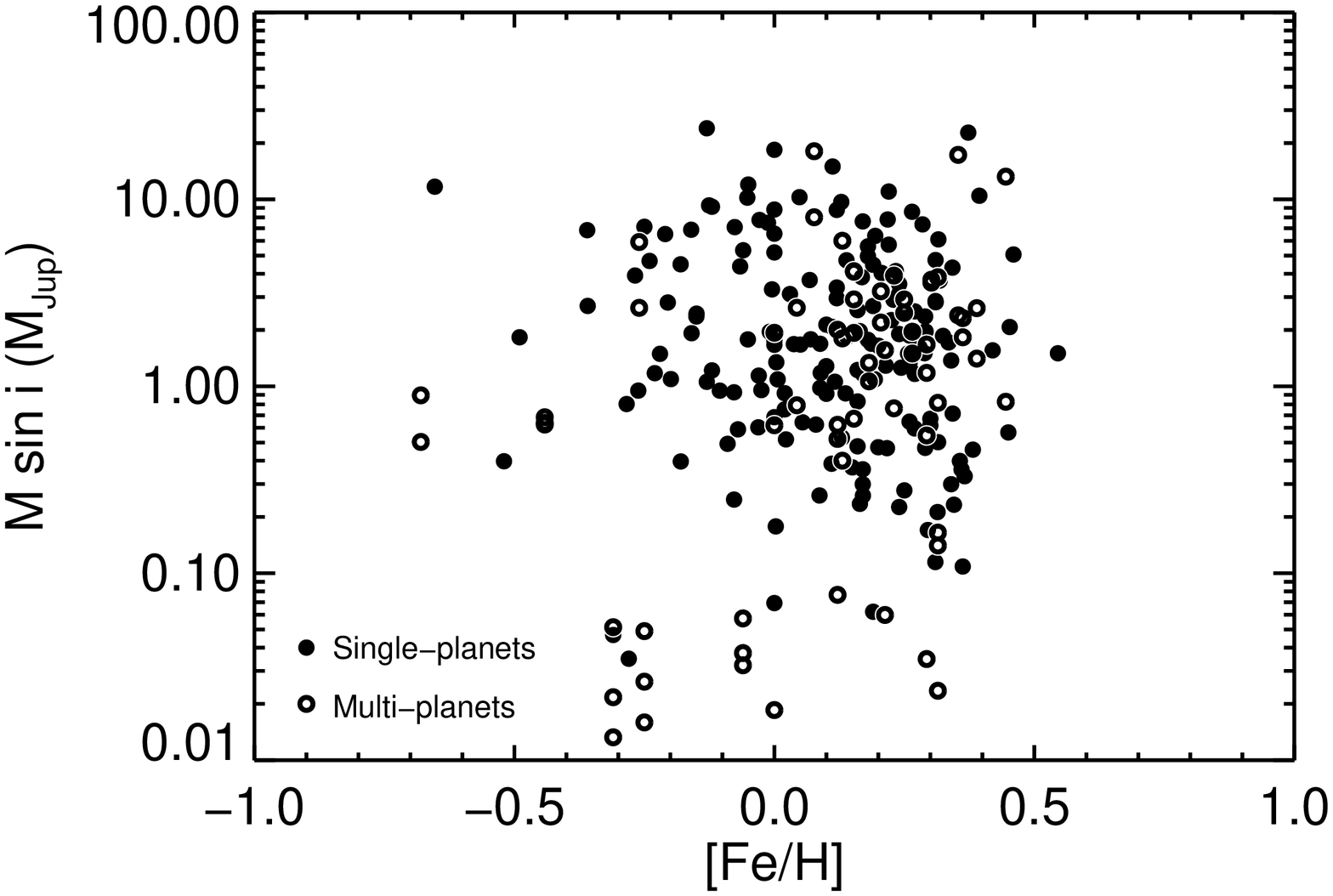}
  \caption{Plot of \feh\ vs.\ \msini. Filled circles are single-planet
    systems, and open circles represent multiplanet
    systems.}\label{chart_feoh_vs_msini}
\end{center}
\end{figure}

\begin{figure} [ht!]
\begin{center}
  \plotone{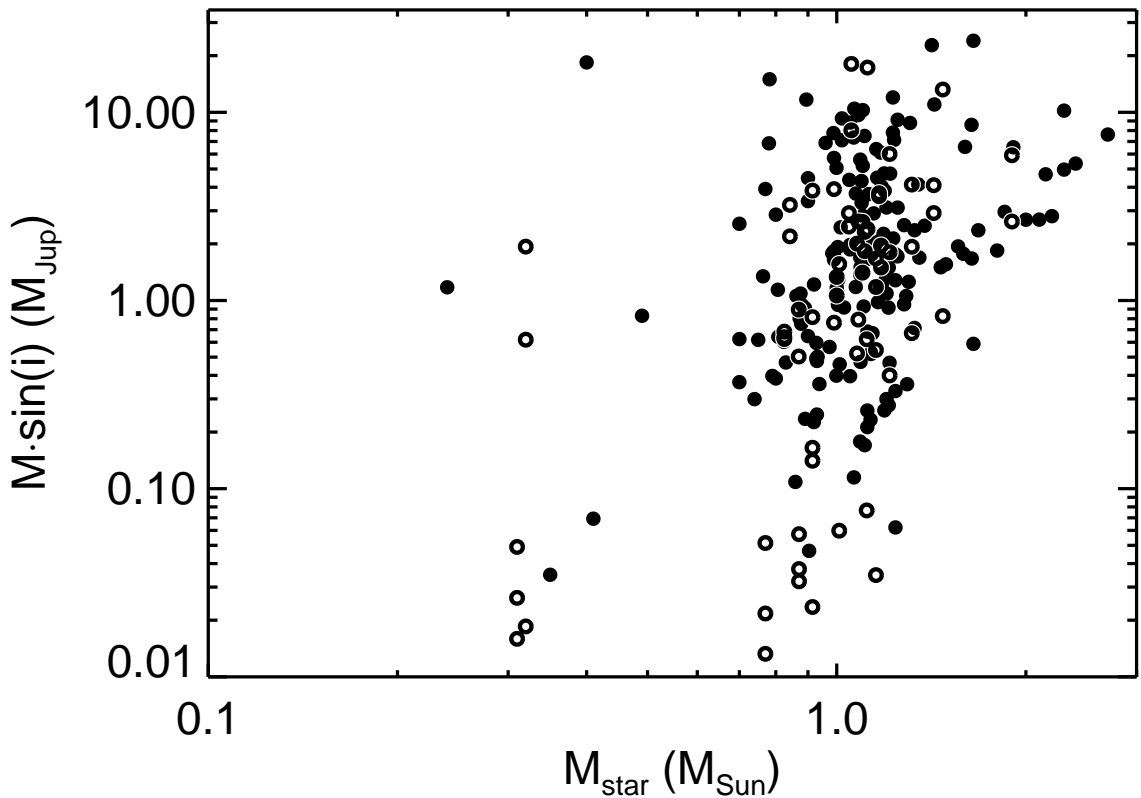}
  \caption{\msini\ vs.\ stellar mass. Filled circles are single-planet
    systems, and open circles represent multiplanet
    systems. Apparently, minimum planet mass does not correlate well with stellar
    mass in the range of 0.3 - 1.9 $M_{\odot}$.}\label{chart_mass_vs_mstar}
\end{center}
\end{figure}

Since metallicity and stellar mass both correlate with the occurrence
rate of planets, it is reasonable to check to see if either also
correlates with minimum planet mass.  In Figs.~\ref{chart_feoh_vs_msini} \& \ref{chart_mass_vs_mstar}, we
plot \msini\ versus these two quantities.  Neither figure shows a strong
correlation.  There is a dearth of low-mass planets found
around high-mass stars, but this may simply be an observational
artifact, as such planets would have a lower reflex amplitude and so
be more difficult to detect.  Likewise, the typical minimum mass of
planets orbiting M-dwarf stars appears to be lower than that around
solar mass stars, but this may simply be an artifact of the fact that
these low-mass planets are more detectable around M dwarfs, and that
M dwarfs appear to have a lower planet occurrence rate overall.

\subsection{Semimajor Axis vs.\ Stellar Mass}
\label{Mclose}
\begin{figure} [ht!]
\begin{center}
  \plotone{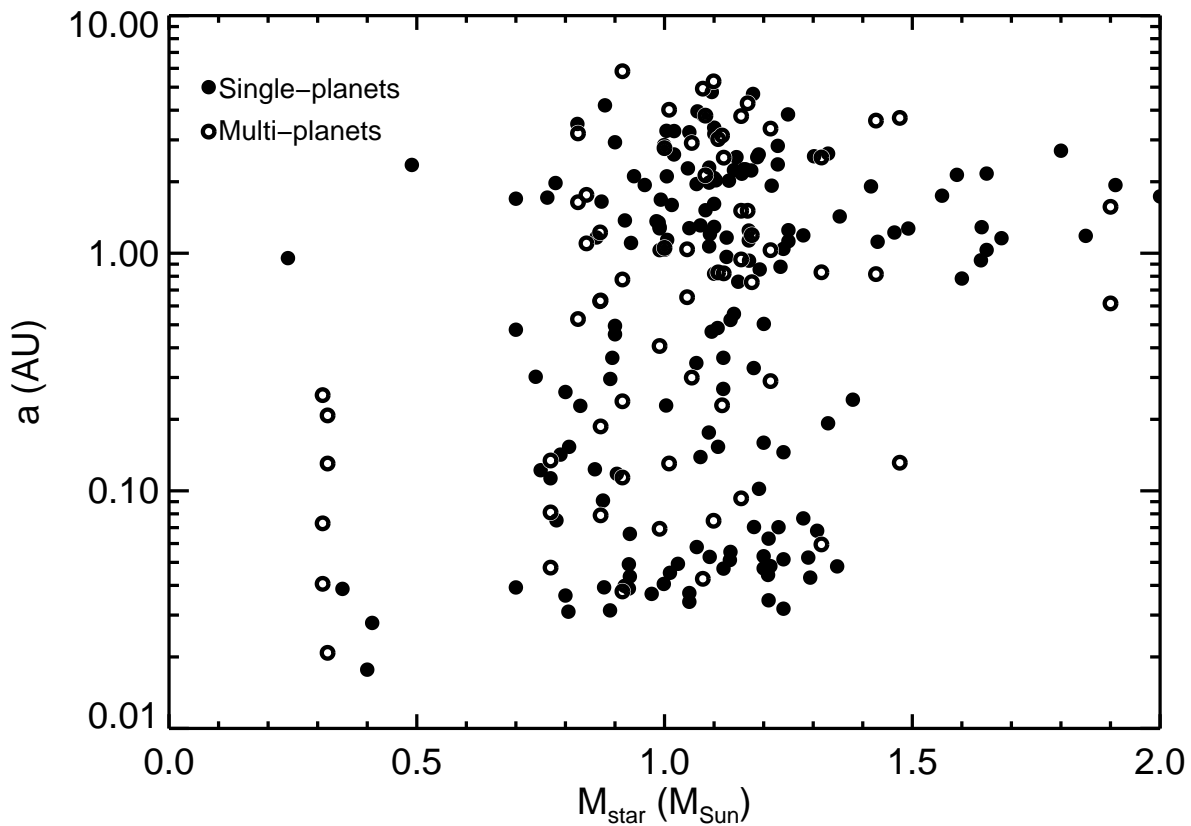}
  \caption{Plot of $a$ vs.\ stellar Mass. Filled circles are
    single-planet systems, and open circles are multiplanet
    systems. There is no strong correlation among planets orbiting
    stars with $0.8 < \Msol < 1.3$, but planets around the
    lowest-mass stars have smaller semimajor axes.}\label{chart_a_vs_mstar}
\end{center}
\end{figure}

We plot semimajor axis vs.\ stellar mass (Figure~\ref{chart_a_vs_mstar}).  Two
features are readily apparent in this plot: a lack of close-in planets
orbiting stars with $M > 1.5 \Msol$ and a lack of long-period planets
orbiting stars with $M < 0.5 \Msol$.  Thus, the semimajor axes
of giant planets correlate positively and sensitively with stellar
mass.  The first of these features has
already been noted by \citet{Johnson07b}, who find that the effect is
statistically significant.  The lack of long-period planets around
M dwarfs is puzzling, since significant numbers of M dwarfs have been
a part of the major radial velocity planet searches since at least
1995, sufficient to detect any massive long-period planets at orbital
distances of a few AU.  We are undertaking a more thorough study of
the occurrence rate of long-period planets around all of our targets
to confirm the reality of the apparent dearth of long-period planets
around low-mass stars.  Nonetheless, a correlation seems to be emerging, driven primarily by
the lack of long-period planets orbiting M dwarfs.   

\subsection{Mass Ratio}

\begin{figure} [ht!]
\begin{center}
  \plotone{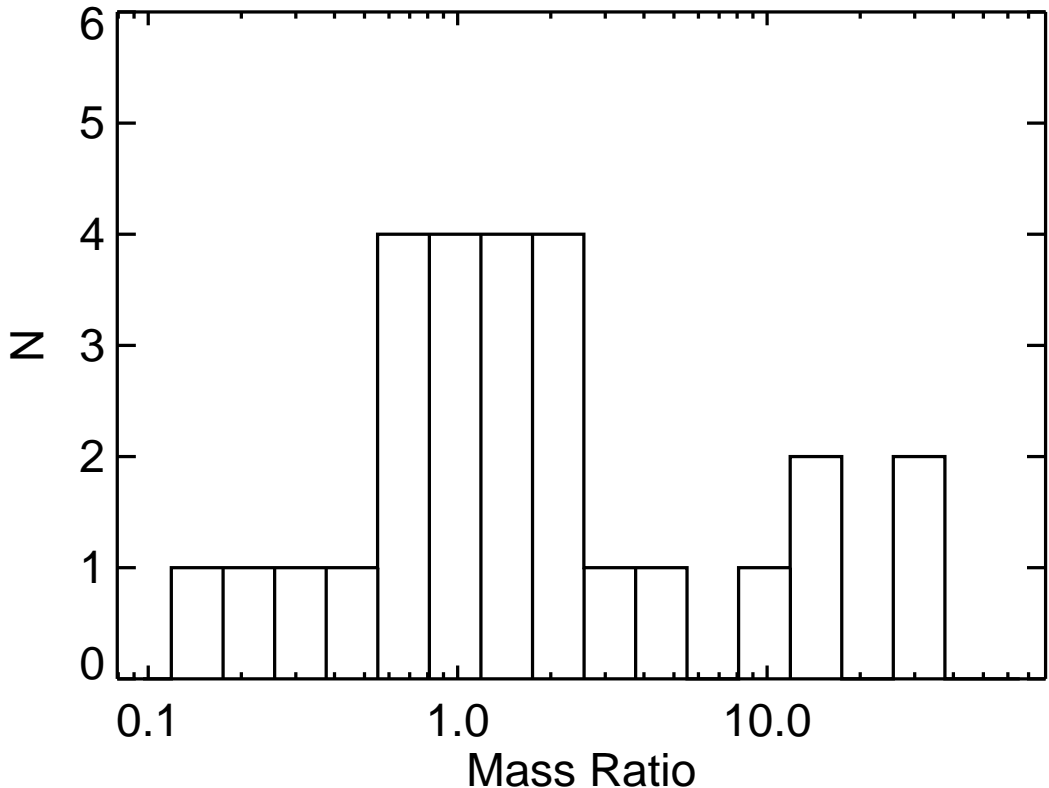}
  \caption{Distribution of \msini\ ratios computed as
    \msini(outer)/\msini(inner) for the outermost two planets in the
    20 known multiplanet systems.}\label{hist_massratio}
\end{center}
\end{figure}

For the multiplanet systems, we plotted the ratio of the minimum mass of the
outermost planet to the minimum mass of the next outermost planet
$\msini(outer)/\msini(inner)$ as shown in Figure~\ref{hist_massratio}. The
distribution peaks near 1 and appears somewhat skewed toward systems
in which the outer planet is more massive.  However, the
$a^{-\frac{1}{2}}$ dependence of orbital distance on reflex amplitude
undoubtedly plays an important role, since low-mass planets are
more easily detected closer to their parent star.  While it is true
that {\it detected} outer planets tend to be more massive, this may
not reflect the actual distribution of planet masses in
multiple-planet systems.

\section{Summary and Discussion}
\label{discussion}
A major outstanding question in planet formation remains: how common
are planetary systems with architectures similar to the solar system?
In the known planetary systems within
200 pc, the median \msini\ is 1.6 \mjup, and the median orbital
distance is 0.9 AU.  Given the observational biases, these numbers are
suggest that the solar system may yet prove to be typical.

The primary way in which the solar system appears unusual is in the nearly
circular orbits of its planets, since the median eccentricity of the
detected exoplanets is 0.25 (excluding tidally circularized planets).
Figure~\ref{chart_ecc_vs_msini}, however, shows that these high eccentricities are
generally restricted to planets with $\msini > 1 \mjup$, and that
{\it nearly circular orbits are typical in planetary systems having no
planets with $M > 1 \mjup$.}

We have also shown that {\it there are four classes of planetary systems
  with distinct orbital distance distributions} ---
1) planetary systems around stars with $M > 1.5 \Msol$, 2) those with $M <
0.5 \Msol$, 3) other apparently single-planet systems, and 4) multiple-planet
systems\footnote{There is, at present, only slight overlap between
  these classes:  GJ 581 and GJ 876 are both multiple-planet systems
  and have $M<0.5 \Msol$}.  This provides a challenge for planet
formation and migration theories to reproduce these disparate distributions.

The 1 AU jump in the orbital distance distribution (Fig.~\ref{a_hist_log})
of planets orbiting single stars may correspond to the planet
formation ice line, or at least constitute an observational
constraint on its location.  It also suggests the existence of a
large, yet unobserved population of giant  
planets beyond 1 AU.  The next decade of radial velocity planet
detection will reveal the population of giant plants from 3 to 7 AU, and
thus provide a census of giant planets which have {\it not}
experienced significant inward migration.  This provides theory with
an opportunity not only to explain the above observations, but also to
provide observational tests of those explanations by predicting the
frequency of such {\it in situ} planets orbiting various classes of
stars.

\appendix
\section{Statistical Methods: K-S tests}\label{statsec}

In this work, we used the Kolmogorov-Smirnov (K-S) statistical test,
based on the two sided D-statistic, to measure the likelihood that any
apparent differences in distributions (for a given property) were in
fact attributable to chance \citep{computerprogramming}. Considering
any independent variable, "$x$", we construct two CDFs of that variable, one for planets in
single-planet systems, $F_s(x)$, and another for planets in
multiplanet systems, $F_m(x)$. We define the D-statistic as follows:
\begin{equation}\label{dstat}
    D^+=max(F_m(x) - F_s(x))$$$$ D^-=max(F_s(x) - F_m(x))
\end{equation}
This statistic measures the maximum difference between the two CDF
graphs being compared, with $D^+$ measuring how high the multiplanet
CDF rises above the single-planet CDF, and $D^-$ measuring how high
the single-planet CDF rises above the multiplanet CDF. A large $D^+$
implies that multiplanet systems systematically have lower values of
$x$, and a large $D^-$ implies that they have higher values. We
quantify this using a Monte Carlo method to produce a K-S confidence
value for this statistic.

To establish confidence levels we count $n$ planets in single-planet
systems and $m$ planets in multiplanet systems. The "D" statistic
measures how frequently $m$ planets drawn with 
replacement from the distribution of $x$ for single-planet systems
will have $D^+_{simulated}$ and $D^-_{simulated}$ as large or larger
than the D-statistics for the actual multiplanet systems.  We
simulated 100,000 multiplanet distributions using the 
distribution for single-planet systems, and the fraction of trials
with $D^+_{simulated}$ greater than the $D^+$ for multiplanet systems
is the K-S test p-value ($p$) that the CDF for multiplanet systems
lies \textit{above} the CDF for single-planet systems - and similarly,
the fraction of trials with $D^-_{simulated}$ greater than the $D^-$
for multiplanet systems is the K-S test p-value that the CDF for
multiplanet systems lies \textit{below} the CDF for single-planet
systems. The K-S test does not assume any particular shape for the
distributions, however, we have assumed that the counts follow Poisson
statistics.

\acknowledgments

The authors acknowledge and thank R. Paul Butler and Steven S. Vogt
for their major contributions in obtaining the radial velocity
measurements herein, without which this work would not have been possible.

The authors also wish to acknowledge helpful discussions with many
astronomers for providing and refining many of the ideas of this
paper.  An incomplete list would include Alan
Boss, Eric Agol, Kristen Menou, James Kasting, Rory Barnes, and Ed
Thommes.  

The work herein is based on observations obtained at the W. M. Keck
Observatory, which is operated jointly by the University of California
and the California Institute of Technology.  The Keck Observatory was
made possible by the generous financial support of the W.M. Keck
Foundation.  We wish to recognize and acknowledge the very significant
cultural role and reverence that the summit of Mauna Kea has always
had within the indigenous Hawaiian community.  We are most fortunate
to have the opportunity to conduct observations from this mountain.

This research has made use of the SIMBAD database, operated at CDS,
Strasbourg, France, and of NASA's Astrophysics Data System
Bibliographic Services, and is made possible by the generous support
of NASA and the NSF, including grant AST-0307493.

J.T.W received support from NSF grant AST-0504874.  G.W.M. received
support from Nasa grant NNG06AH52G, and D.A.F from Nasa grant
NNG05G164G and the Cottrell Science Scholar Program.  E.B.F
acknowledges the support of NASA RSA 1326409.  J.A.J is supported by
NSF grant AST-0702821.  The authors acknowledge
the University of Florida High-Performance Computing Center for
providing computational resources and support that have contributed to
the results reported in this paper.

\end{document}

%% file: tab1.tex
$\upsilon$ And & b & 4.617136(47) & 68.2(1.1) & 0.013(16) & 51 & 14425.02(64) & 0.672(56) & 0.0595(34) & 14 & 1.6 & 284 &  & \\
 & c & 241.33(20) & 53.6(1.4) & 0.224(21) & 250.8(4.9) & 14265.57(64) & 1.92(16) & 0.832(48) & 14 & 1.6 & 284 &  & \\
 & d & 1278.1(2.9) & 66.7(1.4) & 0.267(21) & 269.7(6.4) & 13937.73(64) & 4.13(35) & 2.53(15) & 14 & 1.6 & 284 &  & \\[2pt]
HD 11964 & b & 1945(26) & 9.41(39) & 0.041(47) & 155 & 14170(380) & 0.622(56) & 3.16(19) & 3.1 & 1.1 & 119 &  & \\
 & c & 37.910(41) & 4.65(59) & 0.30(17) & 102 & 14370(380) & 0.0788(97) & 0.229(13) & 3.1 & 1.1 & 119 &  & \\[2pt]
HD 12661 & b & 262.709(83) & 73.56(56) & 0.3768(77) & 296.0(1.5) & 14152.76(87) & 2.30(19) & 0.831(48) & 5.1 & 1.0 & 107 &  & \\
 & c & 1708(14) & 30.41(62) & 0.031(22) & 165 & 16153.42(87) & 1.92(16) & 2.90(17) & 5.1 & 1.0 & 107 &  & \\[2pt]
HIP 14810 & b & 6.6742(20) & 428.3(3.0) & 0.1470(60) & 158.6(2.0) & 13694.588(40) & 3.91(32) & 0.0692(40) & 5.1 & 1.4 & 30 &  & Wr7\\
 & c & 95.2847(20) & 37.4(3.0) & 0.4091(60) & 354.2(2.0) & 13679.585(40) & 0.762(83) & 0.407(23) & 5.1 & 1.4 & 30 &  & Wr7\\[2pt]
HD 37124 & b & 154.46 & 27.5 & 0.055 & 140.5 & 10000.11 & 0.64(11) & 0.529(31) & 18 & 1.9 & 52 &  & Vo5\\
 & c & 2295.00 & 12.2 & 0.2 & 266.0 & 9606.00 & 0.683(88) & 3.19(18) & 18 & 1.9 & 52 &  & Vo5\\
 & d & 843.60 & 15.4 & 0.140 & 314.3 & 9409.40 & 0.624(63) & 1.639(95) & 5.1 & 1.1 & 52 &  & Vo5\\[2pt]
HD 38529 & b & 14.31020(81) & 57.0(1.2) & 0.244(28) & 95.4(5.8) & 14384.8(8.7) & 0.856(72) & 0.1313(76) & 12 & 1.8 & 175 &  & \\
 & c & 2146.1(5.5) & 169.0(1.5) & 0.3551(74) & 17.9(1.6) & 12255.9(8.4) & 13.1(1.1) & 3.72(21) & 12 & 1.8 & 175 &  & \\[2pt]
HD 40307 & b & 4.31150(60) & 1.97(11) & 0 & 0 & 14562.770(80) & 0.0133 & 0.0475(27) & 0.85 & 1.6 & 135 & (c) & My8\\
 & c & 9.6200(20) & 2.47(11) & 0 & 0 & 14551.53(15) & 0.0217 & 0.0811(47) & 0.85 & 1.6 & 135 & (c) & My8\\
 & d & 20.460(10) & 4.55(12) & 0 & 0 & 14532.42(29) & 0.0514 & 0.1342(77) & 0.85 & 1.6 & 135 & (c) & My8\\[2pt]
HD 60532 & b & 201.30(60) & 29.3(1.4) & 0.280(30) & -8.1(4.9) & 13987.0(2.0) & 1.03(16) & 0.759(44) & 4.4 & 2.1 & 147 &  & Ds8\\
 & c & 604.0(9.0) & 46.4(1.7) & 0.020(20) & -209(92) & 13730(160) & 2.46(36) & 1.580(93) & 4.4 & 2.1 & 147 &  & Ds8\\[2pt]
HD 69830 & b & 8.6670(30) & 3.51(15) & 0.100(40) & 340(26) & 13496.80(60) & 0.0322(45) & 0.0789(46) & 0.81 & 1.1 & 74 &  & Lv6\\
 & c & 31.560(40) & 2.66(16) & 0.130(60) & 221(35) & 13469.6(2.8) & 0.0374(52) & 0.187(11) & 0.81 & 1.1 & 74 &  & Lv6\\
 & d & 197.0(3.0) & 2.20(19) & 0.070(70) & 224(61) & 13358(34) & 0.0573(80) & 0.633(37) & 0.81 & 1.1 & 74 &  & Lv6\\[2pt]
HD 73526 & b & 188.3(0.9) & \nodata & 0.19(05) & \nodata & \nodata & 2.07(16) & 0.66(5) & 7.9 & 1.3 & 30 & (d) & T6\\
 & c & 377.8(2.4) & \nodata & 0.14(9) & \nodata & \nodata & 2.5(3) & 1.05(8) & 7.9 & 1.3 & 30 & (d) & T6\\[2pt]
HD 74156 & b & 51.643(11) & 112.0(1.9) & 0.6360(91) & 181.5(1.4) & 11981.321(91) & 1.80(26) & 0.290(17) & 11 & 1.3 & 95 & (e) & Nf4\\
 & c & 2025(11) & 104.0(5.5) & 0.583(39) & 242.4(4.0) & 10901(10) & 6.00(95) & 3.35(19) & 11 & 1.3 & 95 & (e) & Nf4\\
 & d & 346.6(3.6) & 10.5(1.2) & 0.25(11) & 167(27) & 678(44) & 0.400(86) & 1.030(60) & 6.0 & 0.94 & 242 & (e) & Be8\\[2pt]
55 Cnc & b & 14.65126(70) & 71.84(41) & 0.0159(80) & 164(30) & 7572.0(1.2) & 0.82(12) & 0.1138(66) & 7.71 & 2.012 & 636 & (f) & Fi8\\
 & c & 44.3787(70) & 10.06(43) & 0.053(52) & 57(29) & 7547.5(3.3) & 0.165(26) & 0.238(14) & 7.71 & 2.012 & 636 & (f) & Fi8\\
 & d & 5370(230) & 47.2(1.8) & 0.063(30) & 163(32) & 6860(230) & 3.84(58) & 5.84(39) & 7.71 & 2.012 & 636 & (f) & Fi8\\
 & e & 2.79674(10) & 3.73(53) & 0.264(60) & 157(38) & 7578.21590(10) & 0.0235(49) & 0.0377(22) & 7.71 & 2.012 & 636 & (f) & Fi8\\
 & f & 260.7(1.1) & 4.75(60) & 0.00(20) & 206(60) & 7488.0(1.1) & 0.141(39) & 0.775(45) & 7.71 & 2.012 & 636 & (f) & Fi8\\[2pt]
HD 82943 & b & 219.3 & 66.0 & 0.359 & 127 & \nodata & 2.01 & 0.752 & 7.9 & 1.4 & 155 & (g) & Le6\\
 & c & 441.2 & 43.6 & 0.219 & 284 & \nodata & 1.75 & 1.20 & 7.9 & 1.4 & 155 & (g) & Le6\\[2pt]
47 UMa & b & 1089.0(2.9) & 49.3(1.2) & 0.061(14) & 102 & 10356(34) & 2.63(22) & 2.13(12) & 7.4 & 1.0 & 90 &  & Fi2\\
 & c & 2594(90) & 11.1(1.1) & 0.00(12) & 127(56) & 11360(500) & 0.792(92) & 3.79(24) & 7.4 & 1.0 & 90 &  & Fi2\\[2pt]
HD 102272 & b & 127.58(30) & 155.5(5.6) & 0.050(40) & 118(58) & 12146(64) & 5.91(89) & 0.615(36) & 15 & 0.87 & 37 &  & Ni8\\
 & c & 520(26) & 59(11) & 0.680(60) & 320(10) & 14140(260) & 2.63(66) & 1.57(11) & 15 & 0.87 & 37 &  & Ni8\\[2pt]
HD 108874 & b & 394.48(60) & 37.3(1.1) & 0.128(22) & 219.4(9.3) & 14045(49) & 1.34(11) & 1.053(61) & 4.0 & 1.0 & 55 &  & \\
 & c & 1680(24) & 18.90(72) & 0.273(40) & 10(11) & 12797(49) & 1.064(99) & 2.77(16) & 4.0 & 1.0 & 55 &  & \\[2pt]
HD 128311 & b & 458.6(6.8) & 66.8(8.7) & 0.25(10) & 111(36) & 10210.9(7.6) & 2.19(27) & 1.100(65) & 18 & 1.9 &  &  & Vo5\\
 & c & 928(18) & 76.2(4.6) & 0.170(90) & 200(150) & 10010(400) & 3.22(29) & 1.76(11) & 18 & 1.9 &  &  & Vo5\\[2pt]
GJ 581 & b & 5.36870(30) & 12.42(19) & 0 & 0 & 12999.990(50) & 0.0490 & 0.0406(23) & 1.3 & 1.8 & 50 &  & U7\\
 & c & 12.9310(70) & 3.01(16) & 0 & 0 & 12996.74(45) & 0.0159 & 0.0730(42) & 1.3 & 1.8 & 50 &  & U7\\
 & d & 83.40(40) & 2.67(16) & 0 & 0 & 12954.1(3.7) & 0.0263 & 0.253(15) & 1.3 & 1.8 & 50 &  & U7\\[2pt]
HD 155358 & b & 195.0(1.1) & 34.6(3.0) & 0.112(37) & 162(20) & 13950(10) & 0.89(15) & 0.628(36) & 6.0 & 1.1 & 71 &  & Cc7\\
 & c & 530(27) & 14.1(1.6) & 0.18(17) & 279(38) & 14420(79) & 0.50(13) & 1.224(87) & 6.0 & 1.1 & 71 &  & Cc7\\[2pt]
$\mu$ Ara & b & 630.0(6.2) & 37.4(1.6) & 0.271(40) & 259.8(7.4) & 10881(28) & 1.67(17) & 1.510(88) & 4.7 & 1.1 & 108 &  & Bu6\\
 & c & 2490(100) & 18.1(1.1) & 0.463(53) & 183.8(7.9) & 11030(110) & 1.18(12) & 3.78(25) & 4.7 & 1.1 & 108 &  & Bu6\\
 & d & 9.6386(15) & 3.06(13) & 0.172(40) & 213(13) & 12991.10(40) & 0.0347(53) & 0.0930(54) & 1.7 & 1.1 & 171 &  & Pp7\\
 & e & 310.55(83) & 14.91(59) & 0.067(12) & 189.6(9.4) & 12708.7(8.3) & 0.546(80) & 0.942(54) & 1.7 & 1.1 & 171 &  & Pp7\\[2pt]
HD 168443 & b & 58.11212(48) & 475.54(88) & 0.5295(11) & 172.95(13) & 14347.728(20) & 8.01(65) & 0.300(17) & 3.5 & 0.84 & 112 & (h) & \\
 & c & 1748.2(1.0) & 298.14(61) & 0.2122(20) & 64.68(52) & 13769.768(21) & 18.1(1.5) & 2.91(17) & 3.5 & 0.84 & 112 & (h) & \\[2pt]
HD 169830 & b & 225.62(22) & 80.70(90) & 0.310(10) & 148.0(2.0) & 11923.0(1.0) & 2.92(25) & 0.817(47) & 8.9 & \nodata & 112 &  & My4\\
 & c & 2100(260) & 54.3(3.6) & 0.330(20) & 252.0(8.0) & 12516(25) & 4.10(41) & 3.62(42) & 8.9 & \nodata & 112 &  & My4\\[2pt]
HD 183263 & b & 624.8(1.2) & 86.2(1.3) & 0.378(11) & 234.6(2.1) & 12120(130) & 3.73(31) & 1.508(87) & 3.8 & 0.99 & 41 &  & \\
 & c & 3070(110) & 50.3(4.1) & 0.253(76) & 339.6(9.1) & 11910(120) & 3.57(55) & 4.35(28) & 3.8 & 0.99 & 41 &  & \\[2pt]
HD 187123 & b & 3.0965828(78) & 69.40(45) & 0.0103(59) & 25 & 14343.12(31) & 0.523(43) & 0.0426(25) & 2.5 & 0.65 & 76 &  & \\
 & c & 3810(420) & 25.5(1.5) & 0.252(33) & 243(19) & 13580.04(30) & 1.99(25) & 4.89(53) & 2.5 & 0.65 & 76 &  & \\[2pt]
HD 190360 & b & 2915(29) & 23.24(46) & 0.313(19) & 12.9(4.0) & 13542(31) & 1.56(13) & 4.01(23) & 3.1 & 0.84 & 107 &  & \\
 & c & 17.1110(48) & 4.84(51) & 0.237(82) & 5(26) & 14390(31) & 0.0600(76) & 0.1304(75) & 3.1 & 0.84 & 107 &  & \\[2pt]
HD 202206 & b & 255.870(60) & 564.8(1.3) & 0.4350(10) & 161.18(30) & \nodata & 17.3(2.4) & 0.823(48) & 9.6 & 1.5 &  & (i) & Cr5\\
 & c & 1383(18) & 42.0(1.5) & 0.267(21) & 79.0(6.7) & \nodata & 2.40(35) & 2.52(15) & 9.6 & 1.5 &  & (i) & Cr5\\[2pt]
GJ 876 & b & 60.940(13) & 212.60(76) & 0.0249(26) & 175.7(6.0) & \nodata & 1.93(27) & 0.208(12) & 4.6 & 1.2 & 155 & (j) & R5\\
 & c & 30.340(13) & 88.36(72) & 0.2243(13) & 198.30(90) & \nodata & 0.619(88) & 0.1303(75) & 4.6 & 1.2 & 155 & (j) & R5\\
 & d & 1.937760(70) & 6.46(59) & 0 & \nodata & \nodata & 0.0185(31) & 0.0208(12) & 4.6 & 1.2 & 155 & (j) & R5\\[2pt]
HD 217107 & b & 7.126816(39) & 139.20(92) & 0.1267(52) & 24.4(3.0) & 14396(39) & 1.39(11) & 0.0748(43) & 11 & 2.1 & 207 &  & \\
 & c & 4270(220) & 35.7(1.3) & 0.517(33) & 198.6(6.0) & 11106(39) & 2.60(15) & 5.32(38) & 11 & 2.1 & 207 &  & \\[2pt]

%% file: tab2.tex
upsilon And & 9680.753969 & -126.4 &6.8 & L\\
upsilon And & 9942.007812 & 17.0 &9.1 & L\\
upsilon And & 9969.979687 & -22.4 &6.5 & L\\
upsilon And & 9984.852758 & -55.0 &6.9 & L\\
upsilon And & 10032.741397 & 28.7 &8.0 & L\\
upsilon And & 10056.842667 & 47.9 &6.4 & L\\
upsilon And & 10068.585579 & -60.2 &7.1 & L\\
upsilon And & 10068.773435 & -62.9 &7.4 & L\\
upsilon And & 10069.598633 & 2 &12 & L\\
upsilon And & 10072.602539 & -91 &10 & L\\

%% file: ms.bbl
\begin{thebibliography}{}

\bibitem[{Aarseth}, {Lin}, \& {Palmer}(1993){Aarseth}, {Lin}, and
  {Palmer}]{Aarseth93}
{Aarseth}, S.~J., {Lin}, D.~N.~C., \& {Palmer}, P.~L. 1993, \apj, 403, 351

\bibitem[{Barnes} \& {Quinn}(2004){Barnes} and {Quinn}]{thestability}
{Barnes}, R., \& {Quinn}, T. 2004, \apj, 611, 494--516

\bibitem[{Barnes}, {Gozdziewski}, \& {Raymond}(2008){Barnes}, {Gozdziewski},
  and {Raymond}]{Barnes08}
{Barnes}, R., {Gozdziewski}, K., \& {Raymond}, S.~N. 2008, \apjl, 680, 1407

\bibitem[{Bean} {et~al.}(2008){Bean}, {McArthur}, {Benedict}, and
  {Armstrong}]{Bean08}
{Bean}, J.~L., {McArthur}, B.~E., {Benedict}, G.~F., \& {Armstrong}, A. 2008,
  \apj, 672, 1202--1208

\bibitem[{Bodenheimer}, {Laughlin}, \& {Lin}(2003){Bodenheimer}, {Laughlin},
  and {Lin}]{Bodenheimer03}
{Bodenheimer}, P., {Laughlin}, G., \& {Lin}, D.~N.~C. 2003, \apj, 592, 555--563

\bibitem[{Bouchy} {et~al.}(2009){Bouchy} {et~al.}]{Bouchy09}]{Bouchy}
  {et~al.}, \aa, in press

\bibitem[{Bryden} {et~al.}(2000){Bryden}, {R{\' o}{\. z}yczka}, {Lin}, and
  {Bodenheimer}]{Bryden01}
{Bryden}, G., {R{\' o}{\. z}yczka}, M., {Lin}, D.~N.~C., \& {Bodenheimer}, P.
  2000, \apj, 540, 1091--1101

\bibitem[{Butler} \& {Marcy}(1996){Butler} and {Marcy}]{Butler96a}
{Butler}, R.~P., \& {Marcy}, G.~W. 1996, \apjl, 464, L153

\bibitem[{Butler} {et~al.}(1999){Butler}, {Marcy}, {Fischer}, {Brown},
  {Contos}, {Korzennik}, {Nisenson}, and {Noyes}]{Butler_upsand}
{Butler}, R.~P., {Marcy}, G.~W., {Fischer}, D.~A., {Brown}, T.~M., {Contos},
  A.~R., {Korzennik}, S.~G., {Nisenson}, P., \& {Noyes}, R.~W. 1999, \apj, 526,
  916--927

\bibitem[{Butler} {et~al.}(1998){Butler}, {Marcy}, {Vogt}, and
  {Apps}]{Butler98}
{Butler}, R.~P., {Marcy}, G.~W., {Vogt}, S.~S., \& {Apps}, K. 1998, \pasp, 110,
  1389--1393

\bibitem[{Butler} {et~al.}(2006){Butler}, {Wright}, {Marcy}, {Fischer}, {Vogt},
  {Tinney}, {Jones}, {Carter}, and {Johnson}]{Butler06}
{Butler}, R.~P., {Wright}, J.~T., {Marcy}, G.~W., {Fischer}, D.~A., {Vogt},
  S.~S., {Tinney}, C.~G., {Jones}, H.~R.~A., {Carter}, B.~D., \& {Johnson},
  J.~A. 2006, \apj, 646

\bibitem[{Chambers}(1999){Chambers}]{Chambers99}
{Chambers}, J.~E. 1999, \mnras, 304, 793--799

\bibitem[{Chiang} \& {Murray}(2002){Chiang} and {Murray}]{Chiang02}
{Chiang}, E.~I., \& {Murray}, N. 2002, \apj, 576, 473--477

\bibitem[{Cochran} {et~al.}(2007){Cochran}, {Endl}, {Wittenmyer}, and
  {Bean}]{Cochran07}
{Cochran}, W.~D., {Endl}, M., {Wittenmyer}, R.~A., \& {Bean},
J.~L. 2007, \apj, 665, 1407

\bibitem[{Correia} {et~al.}(2005){Correia}, {Udry}, {Mayor}, {Laskar}, {Naef},
  {Pepe}, {Queloz}, and {Santos}]{Correia05}
{Correia}, A.~C.~M., {Udry}, S., {Mayor}, M., {Laskar}, J., {Naef}, D., {Pepe},
  F., {Queloz}, D., \& {Santos}, N.~C. 2005, \aap, 440, 751--758

\bibitem[{Cresswell} \& {Nelson}(2006){Cresswell} and {Nelson}]{Cresswell06}
{Cresswell}, P., \& {Nelson}, R. 2006, \aap, 450, 833--853

\bibitem[{Cumming} {et~al.}(2008){Cumming}, {Butler}, {Marcy}, {Vogt},
  {Wright}, and {Fischer}]{Cumming08}
{Cumming}, A., {Butler}, R.~P., {Marcy}, G.~W., {Vogt}, S.~S., {Wright}, J.~T.,
  \& {Fischer}, D.~A. 2008, \pasp, 120, 531--554

\bibitem[{D'Angelo}, {Kley}, \& {Henning}(2003){D'Angelo}, {Kley}, and
  {Henning}]{Dangelo03}
{D'Angelo}, G., {Kley}, W., \& {Henning}, T. 2003, \apj, 586, 540--561

\bibitem[{Desort} {et~al.}(2008){Desort}, {Lagrange}, {Galland}, {Beust},
  {Udry}, {Mayor}, and {Lo Curto}]{Desort08}
{Desort}, M., {Lagrange}, A.~., {Galland}, F., {Beust}, H., {Udry}, S.,
  {Mayor}, M., \& {Lo Curto}, G. 2008, \aap, 491,883

\bibitem[{Eggenberger} {et~al.}(2007){Eggenberger}, {Udry}, {Chauvin},
  {Beuzit}, {Lagrange}, {S{\'e}gransan}, and {Mayor}]{Eggenberger07}
{Eggenberger}, A., {Udry}, S., {Chauvin}, G., {Beuzit}, J.-L., {Lagrange},
  A.-M., {S{\'e}gransan}, D., \& {Mayor}, M. 2007, \aap, 474, 273--291

\bibitem[{Endl} {et~al.}(2006){Endl}, {Cochran}, {K{\"u}rster}, {Paulson},
  {Wittenmyer}, {MacQueen}, and {Tull}]{Endl06b}
{Endl}, M., {Cochran}, W.~D., {K{\"u}rster}, M., {Paulson}, D.~., {Wittenmyer},
  R.~A., {MacQueen}, P.~J., \& {Tull}, R.~G. 2006, \apj, 649, 436--443

\bibitem[{Fischer} \& {Valenti}(2005){Fischer} and {Valenti}]{Fischer05c}
{Fischer}, D.~A., \& {Valenti}, J. 2005, \apj, 622, 1102--1116

\bibitem[{Fischer} {et~al.}(1999){Fischer}, {Marcy}, {Butler}, {Vogt}, and
  {Apps}]{Fischer99}
{Fischer}, D.~A., {Marcy}, G.~W., {Butler}, R.~P., {Vogt}, S.~S., \& {Apps}, K.
  1999, \pasp, 111, 50--56

\bibitem[{Fischer} {et~al.}(2001){Fischer}, {Marcy}, {Butler}, {Vogt}, {Frink},
  and {Apps}]{Fischer01}
{Fischer}, D.~A., {Marcy}, G.~W., {Butler}, R.~P., {Vogt}, S.~S., {Frink}, S.,
  \& {Apps}, K. 2001, \apj, 551, 1107--1118

\bibitem[{Fischer} {et~al.}(2002a){Fischer}, {Marcy}, {Butler}, {Laughlin}, and
  {Vogt}]{Fischer_47uma}
{Fischer}, D.~A., {Marcy}, G.~W., {Butler}, R.~P., {Laughlin}, G., \& {Vogt},
  S.~S. 2002a, \apj, 564, 1028--1034

\bibitem[{Fischer} {et~al.}(2002b){Fischer}, {Marcy}, {Butler}, {Vogt}, {Walp},
  and {Apps}]{Fischer02}
{Fischer}, D.~A., {Marcy}, G.~W., {Butler}, R.~P., {Vogt}, S.~S., {Walp}, B.,
  \& {Apps}, K. 2002b, \pasp, 114, 529--535

\bibitem[{Fischer} {et~al.}(2008){Fischer}, {Marcy}, {Butler}, {Vogt},
  {Laughlin}, {Henry}, {Abouav}, {Peek}, {Wright}, {Johnson}, {McCarthy}, and
  {Isaacson}]{Fischer08}
{Fischer}, D.~A., {Marcy}, G.~W., {Butler}, R.~P., {Vogt}, S.~S., {Laughlin},
  G., {Henry}, G.~W., {Abouav}, D., {Peek}, K.~M.~G., {Wright}, J.~T.,
  {Johnson}, J.~A., {McCarthy}, C., \& {Isaacson}, H. 2008, \apj, 675, 790--801

\bibitem[{Ford}(2006){Ford}]{Ford06}
{Ford}, E.~B. 2006, In New Horizons in Astronomy: Frank N. Bash Symposium,
  S.~J. {Kannappan}, S.~{Redfield}, J.~E. {Kessler-Silacci}, M.~{Landriau}, and
  N.~{Drory}, eds., volume 352 of {\em Astronomical Society of the Pacific
  Conference Series\/}, pp. 15--+

\bibitem[{Ford} \& {Rasio}(2008){Ford} and {Rasio}]{Ford08}
{Ford}, E.~B., \& {Rasio} 2008, ApJ, accepted

\bibitem[{Ford}, {Rasio}, \& {Sills}(1999){Ford}, {Rasio}, and {Sills}]{Ford99}
{Ford}, E.~B., {Rasio}, F.~A., \& {Sills}, A. 1999, \apj, 514, 411--429

\bibitem[{Ford}, {Lystad}, \& {Rasio}(2005){Ford}, {Lystad}, and
  {Rasio}]{Ford05}
{Ford}, E.~B., {Lystad}, V., \& {Rasio}, F.~A. 2005, \nat, 434, 873--876

\bibitem[{Gaudi} {et~al.}(2008){Gaudi}, {Bennett}, {Udalski}, {Gould},
  {Christie}, {Maoz}, {Dong}, {McCormick}, {Szyma{\'n}ski}, {Tristram},
  {Nikolaev}, {Paczy{\'n}ski}, {Kubiak}, {Pietrzy{\'n}ski}, {Soszy{\'n}ski},
  {Szewczyk}, {Ulaczyk}, {Wyrzykowski}, {DePoy}, {Han}, {Kaspi}, {Lee},
  {Mallia}, {Natusch}, {Pogge}, {Park}, {Abe}, {Bond}, {Botzler}, {Fukui},
  {Hearnshaw}, {Itow}, {Kamiya}, {Korpela}, {Kilmartin}, {Lin}, {Masuda},
  {Matsubara}, {Motomura}, {Muraki}, {Nakamura}, {Okumura}, {Ohnishi},
  {Rattenbury}, {Sako}, {Saito}, {Sato}, {Skuljan}, {Sullivan}, {Sumi},
  {Sweatman}, {Yock}, {Albrow}, {Allan}, {Beaulieu}, {Burgdorf}, {Cook},
  {Coutures}, {Dominik}, {Dieters}, {Fouqu{\'e}}, {Greenhill}, {Horne},
  {Steele}, {Tsapras}, {Chaboyer}, {Crocker}, {Frank}, and
  {Macintosh}]{Gaudi08}
{Gaudi}, B.~S., {Bennett}, D.~P., {Udalski}, A., {Gould}, A., {Christie},
  G.~W., {Maoz}, D., {Dong}, S., {McCormick}, J., {Szyma{\'n}ski}, M.~K.,
  {Tristram}, P.~J., {Nikolaev}, S., {Paczy{\'n}ski}, B., {Kubiak}, M.,
  {Pietrzy{\'n}ski}, G., {Soszy{\'n}ski}, I., {Szewczyk}, O., {Ulaczyk}, K.,
  {Wyrzykowski}, {\L}., {DePoy}, D.~L., {Han}, C., {Kaspi}, S., {Lee}, C.-U.,
  {Mallia}, F., {Natusch}, T., {Pogge}, R.~W., {Park}, B.-G., {Abe}, F.,
  {Bond}, I.~A., {Botzler}, C.~S., {Fukui}, A., {Hearnshaw}, J.~B., {Itow}, Y.,
  {Kamiya}, K., {Korpela}, A.~V., {Kilmartin}, P.~M., {Lin}, W., {Masuda}, K.,
  {Matsubara}, Y., {Motomura}, M., {Muraki}, Y., {Nakamura}, S., {Okumura}, T.,
  {Ohnishi}, K., {Rattenbury}, N.~J., {Sako}, T., {Saito}, T., {Sato}, S.,
  {Skuljan}, L., {Sullivan}, D.~J., {Sumi}, T., {Sweatman}, W.~L., {Yock},
  P.~C.~M., {Albrow}, M.~D., {Allan}, A., {Beaulieu}, J.-P., {Burgdorf}, M.~J.,
  {Cook}, K.~H., {Coutures}, C., {Dominik}, M., {Dieters}, S., {Fouqu{\'e}},
  P., {Greenhill}, J., {Horne}, K., {Steele}, I., {Tsapras}, Y., {Chaboyer},
  B., {Crocker}, A., {Frank}, S., \& {Macintosh}, B. 2008, Science, 319, 927--

\bibitem[{Goldreich} \& {Sari}(2003){Goldreich} and {Sari}]{Goldreich03}
{Goldreich}, P., \& {Sari}, R. 2003, \apj, 585, 1024--1037

\bibitem[{Gregory}(2007){Gregory}]{Gregory07}
{Gregory}, P.~C. 2007, \mnras, 381, 1607--1616

\bibitem[{Ida} \& {Lin}(2004){Ida} and {Lin}]{Ida04b}
{Ida}, S., \& {Lin}, D.~N.~C. 2004, \apj, 616, 567--572

\bibitem[{Ida} \& {Lin}(2008){Ida} and {Lin}]{Ida08}
{Ida}, S., \& {Lin}, D.~N.~C. 2008, \apj, 685, 584

\bibitem[{Johnson} {et~al.}(2007a){Johnson}, {Butler}, {Marcy}, {Fischer},
  {Vogt}, {Wright}, and {Peek}]{Johnson07b}
{Johnson}, J.~A., {Butler}, R.~P., {Marcy}, G.~W., {Fischer}, D.~A., {Vogt},
  S.~S., {Wright}, J.~T., \& {Peek}, K.~M.~G. 2007a, \apj, 670, 833--840

\bibitem[{Johnson} {et~al.}(2007b){Johnson}, {Marcy}, {Wright}, Driscoll,
  {Butler}, {Hekker}, {Reffert}, and {Vogt}]{Johnson07}
{Johnson}, J.~A., {Marcy}, G.~W., {Wright}, J.~T., Driscoll, P., {Butler},
  R.~P., {Hekker}, S., {Reffert}, S., \& {Vogt}, S.~S. 2007b, \apj,
  665, 785

\bibitem[{Kley}, {Peitz}, \& {Bryden}(2004){Kley}, {Peitz}, and
  {Bryden}]{KleyResonance}
{Kley}, W., {Peitz}, J., \& {Bryden}, G. 2004, \aap, 414, 735--747

\bibitem[{Kley} {et~al.}(2005){Kley}, {Lee}, {Murray}, and {Peale}]{Kley05b}
{Kley}, W., {Lee}, M.~H., {Murray}, N., \& {Peale}, S.~J. 2005, \aap, 437,
  727--742

\bibitem[{Knuth}(1997){Knuth}]{computerprogramming}
{Knuth}, D.~E. 1997.
\newblock {The art of computer programming}, volume~2, Addison-Wesley

\bibitem[{Kokubo} \& {Ida}(2002){Kokubo} and {Ida}]{Kokubo02}
{Kokubo}, E., \& {Ida}, S. 2002, \apj, 581, 666--680

\bibitem[{Konacki} {et~al.}(2003){Konacki}, {Torres}, {Sasselov}, and
  {Jha}]{Konacki03}
{Konacki}, M., {Torres}, G., {Sasselov}, D.~D., \& {Jha}, S. 2003, \apj, 597,
  1076--1091

\bibitem[{Laughlin} \& {Chambers}(2002){Laughlin} and {Chambers}]{Laughlin02}
{Laughlin}, G., \& {Chambers}, J.~E. 2002, \aj, 124, 592--600

\bibitem[{Lee} \& {Peale}(2003){Lee} and {Peale}]{Lee03}
{Lee}, M.~H., \& {Peale}, S.~J. 2003, \apj, 592, 1201--1216

\bibitem[{Lee} {et~al.}(2006){Lee}, {Butler}, {Fischer}, {Marcy}, and
  {Vogt}]{Lee06}
{Lee}, M.~H., {Butler}, R.~P., {Fischer}, D.~A., {Marcy}, G.~W., \& {Vogt},
  S.~S. 2006, \apj, 641, 1178--1187

\bibitem[{Levison}, {Lissauer}, \& {Duncan}(1998){Levison}, {Lissauer}, and
  {Duncan}]{Levison98}
{Levison}, H.~F., {Lissauer}, J.~J., \& {Duncan}, M.~J. 1998, \aj, 116,
  1998--2014

\bibitem[{Lovis} {et~al.}(2006){Lovis}, {Mayor}, {Pepe}, {Alibert}, {Benz},
  {Bouchy}, {Correia}, {Laskar}, {Mordasini}, {Queloz}, {Santos}, {Udry},
  {Bertaux}, and {Sivan}]{Lovis06}
{Lovis}, C., {Mayor}, M., {Pepe}, F., {Alibert}, Y., {Benz}, W., {Bouchy}, F.,
  {Correia}, A.~C.~M., {Laskar}, J., {Mordasini}, C., {Queloz}, D., {Santos},
  N.~C., {Udry}, S., {Bertaux}, J.-L., \& {Sivan}, J.-P. 2006, \nat, 441,
  305--309

\bibitem[{Marcy} {et~al.}(2001){Marcy}, {Butler}, {Fischer}, {Vogt},
  {Lissauer}, and {Rivera}]{Marcy_876}
{Marcy}, G.~W., {Butler}, R.~P., {Fischer}, D., {Vogt}, S.~S., {Lissauer},
  J.~J., \& {Rivera}, E.~J. 2001, \apj, 556, 296--301

\bibitem[{Marcy} {et~al.}(2005){Marcy}, {Butler}, {Vogt}, {Fischer}, {Henry},
  {Laughlin}, {Wright}, and {Johnson}]{Marcy05}
{Marcy}, G.~W., {Butler}, R.~P., {Vogt}, S.~S., {Fischer}, D.~A., {Henry},
  G.~W., {Laughlin}, G., {Wright}, J.~T., \& {Johnson}, J.~A. 2005, \apj, 619,
  570--584

\bibitem[{Marzari}, {Scholl}, \& {Tricarico}(2005){Marzari}, {Scholl}, and
  {Tricarico}]{Marzari05}
{Marzari}, F., {Scholl}, H., \& {Tricarico}, P. 2005, In 36th Annual Lunar and
  Planetary Science Conference, S.~{Mackwell} and E.~{Stansbery}, eds., pp.
  1289--+

\bibitem[{Mayor} \& {Queloz}(1995){Mayor} and {Queloz}]{Mayor_queloz}
{Mayor}, M., \& {Queloz}, D. 1995, \nat, 378, 355--+

\bibitem[{Mayor} {et~al.}(2004){Mayor}, {Udry}, {Naef}, {Pepe}, {Queloz},
  {Santos}, and {Burnet}]{Mayor04}
{Mayor}, M., {Udry}, S., {Naef}, D., {Pepe}, F., {Queloz}, D., {Santos}, N.~C.,
  \& {Burnet}, M. 2004, \aap, 415, 391--402

\bibitem[{Mayor} {et~al.}(2008){Mayor}, {Udry}, {Lovis}, {Pepe}, {Queloz},
  {Benz}, {Bertaux}, {Bouchy}, {Mordasini}, and {Segransan}]{Mayor08}
{Mayor}, M., {Udry}, S., {Lovis}, C., {Pepe}, F., {Queloz}, D., {Benz}, W.,
  {Bertaux}, J.~., {Bouchy}, F., {Mordasini}, C., \& {Segransan}, D. 2008,
  \aap, 493, 639

\bibitem[{McArthur} {et~al.}(2004){McArthur}, {Endl}, {Cochran}, {Benedict},
  {Fischer}, {Marcy}, {Butler}, {Naef}, {Mayor}, {Queloz}, {Udry}, and
  {Harrison}]{McArthur04}
{McArthur}, B.~E., {Endl}, M., {Cochran}, W.~D., {Benedict}, G.~F., {Fischer},
  D.~A., {Marcy}, G.~W., {Butler}, R.~P., {Naef}, D., {Mayor}, M., {Queloz},
  D., {Udry}, S., \& {Harrison}, T.~E. 2004, \apjl, 614, L81--L84

\bibitem[{Murray}(2003){Murray}]{Murray2003}
{Murray}, N. 2003, In Scientific Frontiers in Research on Extrasolar
Planets, {\em Astronomical Society of the Pacific Conference Series}
  pp. 165--172

\bibitem[{Naef} {et~al.}(2004){Naef}, {Mayor}, {Beuzit}, {Perrier}, {Queloz},
  {Sivan}, and {Udry}]{Naef04}
{Naef}, D., {Mayor}, M., {Beuzit}, J.~L., {Perrier}, C., {Queloz}, D., {Sivan},
  J.~P., \& {Udry}, S. 2004, \aap, 414, 351--359

\bibitem[{Nelson} \& {Papaloizou}(2002){Nelson} and {Papaloizou}]{Nelson02}
{Nelson}, R.~P., \& {Papaloizou}, J.~C.~B. 2002, \mnras, 333, L26--L30

\bibitem[{Niedzielski} {et~al.}(2008){Niedzielski}, {Go\'zdziewski}, {Wolszczan}, {Konacki}, {Nowak}, and
  {Zieli{\'n}ski}]{Niedzielski08b}
{Niedzielski}, A., {Go$\backslash$' zdziewski}, K., {Wolszczan}, A., {Konacki},
  M., {Nowak}, G., \& {Zieli{\'n}ski}, P. 2008, ArXiv 0810.1710

\bibitem[{Papaloizou}(2005){Papaloizou}]{Paploizou05}
{Papaloizou}, J.~C.~B. 2005, Celestial Mechanics and Dynamical Astronomy, 91,
  33--57

\bibitem[{Pepe} {et~al.}(2003){Pepe}, {Bouchy}, {Queloz}, and {Mayor}]{Pepe03}
{Pepe}, F., {Bouchy}, F., {Queloz}, D., \& {Mayor}, M. 2003, In Scientific
  Frontiers in Research on Extrasolar Planets, D.~{Deming} and S.~{Seager},
  eds., volume 294 of {\em Astronomical Society of the Pacific Conference
  Series\/}, pp. 39--42

\bibitem[{Pepe} {et~al.}(2007){Pepe}, {Correia}, {Mayor}, {Tamuz}, {Couetdic},
  {Benz}, {Bertaux}, {Bouchy}, {Laskar}, {Lovis}, {Naef}, {Queloz}, {Santos},
  {Sivan}, {Sosnowska}, and {Udry}]{Pepe07}
{Pepe}, F., {Correia}, A.~C.~M., {Mayor}, M., {Tamuz}, O., {Couetdic}, J.,
  {Benz}, W., {Bertaux}, J.-L., {Bouchy}, F., {Laskar}, J., {Lovis}, C.,
  {Naef}, D., {Queloz}, D., {Santos}, N.~C., {Sivan}, J.-P., {Sosnowska}, D.,
  \& {Udry}, S. 2007, \aap, 462, 769--776

\bibitem[{Pollack} {et~al.}(1996){Pollack}, {Hubickyj}, {Bodenheimer},
  {Lissauer}, {Podolak}, and {Greenzweig}]{1996Icar..124...62P}
{Pollack}, J.~B., {Hubickyj}, O., {Bodenheimer}, P., {Lissauer}, J.~J.,
  {Podolak}, M., \& {Greenzweig}, Y. 1996, Icarus, 124, 62--85

\bibitem[{Raghavan} {et~al.}(2006){Raghavan}, {Henry}, {Mason}, {Subasavage},
  {Jao}, {Beaulieu}, and {Hambly}]{Raghavan06}
{Raghavan}, D., {Henry}, T.~J., {Mason}, B.~D., {Subasavage}, J.~P., {Jao},
  W.-C., {Beaulieu}, T.~D., \& {Hambly}, N.~C. 2006, \apj, 646, 523--542

\bibitem[{Rasio} {et~al.}(1996){Rasio}, {Tout}, {Lubow}, and {Livio}]{Rasio96b}
{Rasio}, F.~A., {Tout}, C.~A., {Lubow}, S.~H., \& {Livio}, M. 1996, \apj, 470,
  1187--+

\bibitem[{Rivera} {et~al.}(2005){Rivera}, {Lissauer}, {Butler}, {Marcy},
  {Vogt}, {Fischer}, {Brown}, {Laughlin}, and {Henry}]{Rivera05}
{Rivera}, E.~J., {Lissauer}, J.~J., {Butler}, R.~P., {Marcy}, G.~W., {Vogt},
  S.~S., {Fischer}, D.~A., {Brown}, T.~M., {Laughlin}, G., \& {Henry}, G.~W.
  2005, \apj, 634, 625--640

\bibitem[{S\'{a}ndor} \& {Kley}(2006){S\'{a}ndor} and {Kley}]{SandorScatter}
{S\'{a}ndor}, Z., \& {Kley}, W. 2006, \aap, 451, L31--L34

\bibitem[{Santos}, {Israelian}, \& {Mayor}(2001){Santos}, {Israelian}, and
  {Mayor}]{Santos01b}
{Santos}, N.~C., {Israelian}, G., \& {Mayor}, M. 2001, \aap, 373, 1019--1031

\bibitem[{Santos} {et~al.}(2004){Santos}, {Bouchy}, {Mayor}, {Pepe}, {Queloz},
  {Udry}, {Lovis}, {Bazot}, {Benz}, {Bertaux}, {Lo Curto}, {Delfosse},
  {Mordasini}, {Naef}, {Sivan}, and {Vauclair}]{Santos04b}
{Santos}, N.~C., {Bouchy}, F., {Mayor}, M., {Pepe}, F., {Queloz}, D., {Udry},
  S., {Lovis}, C., {Bazot}, M., {Benz}, W., {Bertaux}, J.-L., {Lo Curto}, G.,
  {Delfosse}, X., {Mordasini}, C., {Naef}, D., {Sivan}, J.-P., \& {Vauclair},
  S. 2004, \aap, 426, L19--L23

\bibitem[{Takeda} {et~al.}(2007){Takeda}, {Ford}, {Sills}, {Rasio}, {Fischer},
  and {Valenti}]{Takeda07}
{Takeda}, G., {Ford}, E.~B., {Sills}, A., {Rasio}, F.~A., {Fischer}, D.~A., \&
  {Valenti}, J.~A. 2007, \apjs, 168, 297--318

\bibitem[{Tanaka} \& {Ward}(2004){Tanaka} and {Ward}]{Tanaka04}
{Tanaka}, H., \& {Ward}, W.~R. 2004, \apj, 602, 388--395

\bibitem[{Tinney} {et~al.}(2006){Tinney}, {Butler}, {Marcy}, {Jones},
  {Laughlin}, {Carter}, and {Bailey}]{Tinney06}
{Tinney}, C.~G., {Butler}, R.~P., {Marcy}, G.~W., {Jones}, H. R.~A.,
  {Laughlin}, G., {Carter}, B., \& {Bailey}, J. 2006, \apj\ submitted

\bibitem[{Trilling}, {Lunine}, \& {Benz}(2002){Trilling}, {Lunine}, and
  {Benz}]{Trilling02}
{Trilling}, D.~E., {Lunine}, J.~I., \& {Benz}, W. 2002, \aap, 394, 241--251

\bibitem[{Udry} {et~al.}(2007){Udry}, {Bonfils}, {Delfosse}, {Forveille},
  {Mayor}, {Perrier}, {Bouchy}, {Lovis}, {Pepe}, {Queloz}, and
  {Bertaux}]{Udry07}
{Udry}, S., {Bonfils}, X., {Delfosse}, X., {Forveille}, T., {Mayor}, M.,
  {Perrier}, C., {Bouchy}, F., {Lovis}, C., {Pepe}, F., {Queloz}, D., \&
  {Bertaux}, J.-L. 2007, \aap, 469, L43--L47

\bibitem[{Valenti} \& {Fischer}(2005){Valenti} and {Fischer}]{SPOCS}
{Valenti}, J.~A., \& {Fischer}, D.~A. 2005, \apjs, 159, 141--166

\bibitem[{Vogt} {et~al.}(2005){Vogt}, {Butler}, {Marcy}, {Fischer}, {Henry},
  {Laughlin}, {Wright}, and {Johnson}]{Vogt05}
{Vogt}, S.~S., {Butler}, R.~P., {Marcy}, G.~W., {Fischer}, D.~A., {Henry},
  G.~W., {Laughlin}, G., {Wright}, J.~T., \& {Johnson}, J.~A. 2005, \apj, 632,
  638--658

\bibitem[{Wisdom} \& {Holman}(1991){Wisdom} and {Holman}]{Wisdom91}
{Wisdom}, J., \& {Holman}, M. 1991, \aj, 102, 1528--1538

\bibitem[{Wittenmyer}, {Endl}, \& {Cochran}(2007){Wittenmyer}, {Endl}, and
  {Cochran}]{Wittenmyer07}
{Wittenmyer}, R.~A., {Endl}, M., \& {Cochran}, W.~D. 2007, \apj, 654, 625--632

\bibitem[{Wolszczan} \& {Frail}(1992){Wolszczan} and {Frail}]{Wolszczan92}
{Wolszczan}, A., \& {Frail}, D.~A. 1992, \nat, 355, 145--147

\bibitem[{Wright}(2005){Wright}]{Wright05}
{Wright}, J.~T. 2005, \pasp, 117, 657--664

\bibitem[{Wright} {et~al.}(2007){Wright}, {Marcy}, {Fischer}, {Butler}, {Vogt},
  {Tinney}, {Jones}, {Carter}, {Johnson}, {McCarthy}, and {Apps}]{Wright07}
{Wright}, J.~T., {Marcy}, G.~W., {Fischer}, D.~A., {Butler}, R.~P., {Vogt},
  S.~S., {Tinney}, C.~G., {Jones}, H.~R.~A., {Carter}, B.~D., {Johnson}, J.~A.,
  {McCarthy}, C., \& {Apps}, K. 2007, \apj, 657, 533--545

\end{thebibliography}
